\documentclass[a4paper,11pt]{article}
\usepackage{amssymb}
\usepackage{aaskaiid}
\usepackage{orcidlink}
\usepackage[dvipsnames]{xcolor}

\newcommand{\HI}{\mbox{H\,{\sc i}}} 
\newcommand{\CII}{\mbox{C\,{\sc ii}}} 
\newcommand{\CI}{\mbox{C\,{\sc i}}} 

\title{Tracing Cold Gas in Absorption Across Cosmic Time with the SKA}

\ShortTitle{\HI\ absorption with the SKA}

\author[1]{Elizabeth K. Mahony\orcidlink{0000-0002-5053-2828}}
\ShortName{Mahony et al.} 
\author[2]{Neeraj Gupta\orcidlink{0000-0001-7547-4241}}
\author[3]{Sergei A. Balashev}
\author[4]{Yogesh Chandola\orcidlink{0000-0002-3698-3294}}
\author[5]{Francoise Combes\orcidlink{0000-0003-2658-7893}}
\author[6]{Rebecca Davies\orcidlink{0000-0002-3324-4824}}
\author[7,8]{Jens-Kristian Krogager}
\author[9]{Wenkai Hu}
\author[10,11]{Filippo M. Maccagni\orcidlink{0000-0002-9930-1844}}
\author[12]{Pasquier Noterdaeme\orcidlink{0000-0002-5777-1629}}
\author[13]{Mamta Pandey-Pommier\orcidlink{0000-0001-5829-1099}}
\author[14,1]{Elaine M. Sadler\orcidlink{0000-0002-1136-2555}}
\author[15]{Rasha M. Samir\orcidlink{0000-0003-2716-8332}}
\author[16]{Nick Seymour\orcidlink{0000-0003-3506-5536}}
\author[17]{Hyein Yoon\orcidlink{0000-0003-4048-2203}}

\affiliation[1]{ATNF, CSIRO Space and Astronomy, PO Box 76, Epping, NSW 1710, Australia}
\emailAdd{elizabeth.mahony@csiro.au}
\affiliation[2]{Inter-University Centre for Astronomy and Astrophysics, 
Post Bag 4, Ganeshkhind, Pune 411 007, India}
\emailAdd{ngupta@iucaa.in}
\affiliation[3]{Ioffe Institute, 26 Politeknicheskaya st., St. Petersburg, 194021, Russia}
\emailAdd{s.balashev@gmail.com}
\affiliation[4]{Indian Institute of Astrophysics (IIA), 2nd Block, Koramangala, Bengaluru 560034, India}
\emailAdd{yogesh.chandola@gmail.com}
\affiliation[5]{Observatoire de Paris, LUX, Coll\`ege de France, PSL University, Sorbonne University, CNRS, Paris, France}
\emailAdd{francoise.combes@obspm.fr}
\affiliation[6]{Centre for Astrophysics and Supercomputing, Swinburne University of Technology, John Street, Hawthorn, 3122, VIC, Australia}
\emailAdd{rdavies@swin.edu.au}
\affiliation[7]{Universit\'e Lyon1, ENS de Lyon, CNRS, Centre de Recherche Astrophysique de Lyon UMR5574, F-69230 Saint-Genis-Laval, France}
\affiliation[8]{French-Chilean Laboratory for Astronomy, IRL 3386, CNRS and Universidad de Chile, Santiago, Chile}
\emailAdd{jens-kristian.krogager@univ-lyon1.fr}
\affiliation[9]{National Astronomical Observatories, Chinese Academy of Sciences, Beijing 100101, China}
\emailAdd{wkhu@nao.cas.cn}
\affiliation[10]{INAF -- Osservatorio Astronomico di Cagliari, via della Scienza 5, Selargius (CA), Italy}
\affiliation[11]{Wits Centre for Astrophysics, School of Physics, University of the Witwatersrand, 1 Jan Smuts Avenue, 2000, Johannesburg, South Africa}
\emailAdd{filippo.maccagni@inaf.it}
\affiliation[12]{Institut d'Astrophysique de Paris, CNRS, 98bis bd Arago, 75014, Paris, France}
\emailAdd{noterdaeme@iap.fr}
\affiliation[13]{Pole Scientific, University Catholic of Lyon, 10 place des Archives 69288, Lyon, France}
\emailAdd{mamtapommier@gmail.com}
\affiliation[14]{Sydney Institute for Astronomy, School of Physics A28, University of Sydney, NSW 2006, Australia}
\emailAdd{elaine.sadler@sydney.edu.au}
\affiliation[15]{Astronomy Department, National Research Institute of Astronomy and Geophysics (NRIAG), EL Marsad Street 1, Helwan, Cairo, Egypt}
\emailAdd{rasha.samir@nriag.sci.eg}
\affiliation[16]{International Centre for Radio Astronomy Research, Curtin University, GPO Box U1987, Bentley, WA 6845, Australia}
\emailAdd{nick.seymour@curtin.edu.au }
\affiliation[17]{Korea Astronomy and Space Science Institute, 776 Daedeok-daero, Yuseong-gu, Daejeon 34055, Republic of Korea}
\emailAdd{hiyoon.astro@gmail.com}

\abstract{

Observing the 21-cm \HI\ line in absorption provides a powerful means of tracing the cold neutral gas in normal and active galaxies across cosmic time. The frequency coverage and sensitivity of SKAO will allow us to detect \HI\ in absorption from $z=0$ to beyond $z=6$, enabling the characterisation of the properties of cold gas in and around galaxies at all epochs. This chapter summarises recent advances in absorption-line studies, lessons learned from precursor surveys, and updates the science case presented in \citet{Kanekar2004} and \citet{Morganti2015}, focusing on the capabilities enabled by the SKA design baseline, Array Assembly 4 (AA4). We expand on these earlier works by presenting new opportunities to simultaneously search for OH 18-cm absorption, an efficient tracer of diffuse molecular gas that complements the atomic gas traced by \HI\ absorption, as well as the need for sub-arcsecond scale spectroscopic imaging and multi-wavelength data from large surveys. These advances will allow SKAO absorption surveys to address key questions surrounding the fuelling and feedback cycles of AGN and the evolution of the cold neutral gas across cosmic time.
}

\begin{document}

\maketitle

\section{Introduction}

Neutral atomic hydrogen (\HI) is a fundamental ingredient in cosmic star formation and galaxy evolution, and plays a key role in the baryon cycle by tracing how gas is accreted, processed to form molecular gas (H$_2$) and stars, and expelled in and around galaxies via various feedback mechanisms \citep[e.g.][]{Peroux2020}. Observations of the 21-cm line of \HI\ in emission and absorption provide a unique probe of the interstellar medium (ISM), enabling measurements of atomic gas content, kinematics, structure, and cold atomic gas ($\sim$100\,K) fraction across different environments and epochs. \HI\ gas observed in 21-cm absorption also serves as a probe of a crucial reservoir fueling central active galactic nuclei \citep[AGN;][]{Morganti2018}, thereby regulating black hole growth and linking the availability of neutral gas (\HI\ + H$_2$) to feedback processes that shape galaxy evolution. A key scientific goal for the SKAO will be to chart the evolution of \HI\ over cosmic time, building a direct link between the neutral gas reservoirs of galaxies, their star formation histories, and AGN fuelling and feedback processes.

Large-area surveys of the 21-cm \HI\ line in emission using single dish telescopes, such as HIPASS \citep{Koribalski2004, Meyer2004} and ALFALFA \citep{Haynes2011}, and spatially resolved radio interferometric \HI\ images from THINGS \citep[][]{Walter2008}, and more recently WALLABY \citep[][]{Koribalski2020} and MHONGOOSE \citep[][]{deBlok2024} have built a detailed picture of the \HI\ content of galaxies in the local Universe and its connection to environment and star formation. 
At higher redshifts ($z \gtrsim 1.7$), Ly$\alpha$ absorption studies using ground based optical telescopes provide constraints on the neutral gas reservoir \citep[e.g.][]{Wolfe2005, Prochaska2009, Noterdaeme2012}, while 21-cm emission line stacking analyses have been employed to estimate the average \HI\ content of galaxy populations out to $z\sim1$ \citep[e.g.][]{Lah2007, Delhaize2013, Rhee2018, Chowdhury2020, Chowdhury2022, Bera2023, dePalma2025}. 

Despite these advancements through \HI\ 21-cm emission and Ly$\alpha$ absorption line techniques, significant gaps remain in our knowledge of the \HI\ properties of galaxies especially at $z \sim 0.1-3$, a period when galaxies undergo massive transformation as the star-formation rate of galaxies peaks and rapidly declines \citep[][]{Madau2014}. An alternative, and complementary, approach is to detect the \HI\ 21-cm line in absorption against bright background radio continuum sources. 
Since the \HI\ 21-cm absorption strength depends on the brightness of the background radio continuum source rather than the distance to the absorber, and is inversely sensitive to the spin temperature of the gas, it is an excellent probe of the cold atomic gas \citep{Kulkarni08}.
Therefore, \HI\ 21-cm absorption-line observations offer a powerful and redshift-independent probe for tracing cold gas in normal and active galaxies.

Due to its radio-quiet sites, broad frequency coverage and sensitivity, SKAO will be the first observatory capable of detecting the 21-cm \HI\ absorption line across all redshifts. This capability opens the door to systematic studies of cold gas from the local Universe out to the epoch of reionisation, offering a direct window into the neutral gas properties of galaxies across cosmic time. 
The wide instantaneous bandwidth also makes OH an inseparable part of the \HI\ absorption search, since both lines fall within the same observing band at overlapping redshifts. OH forms efficiently in cold atomic gas alongside H$_2$ and serves as a tracer of diffuse molecular gas — the so-called 'CO-dark' gas — that is neither detectable in CO nor \HI\ emission \citep[e.g.,][]{Liszt1999, Balashev2021}. The prospects for OH detection are therefore an integral part of the \HI\ absorption science case presented here.
This chapter builds on the \HI\ absorption case presented in \citet{Kanekar2004} and \citet[][]{Morganti2015}, and includes discussion of OH throughout. H$_2$ is discussed where necessary to contextualise the role of OH as a bridge between the atomic and molecular ISM phases.

\subsection{Key Observables}

\begin{figure}[ht]
    \vspace{0.2cm}
    \includegraphics[clip,width=0.5\textwidth]{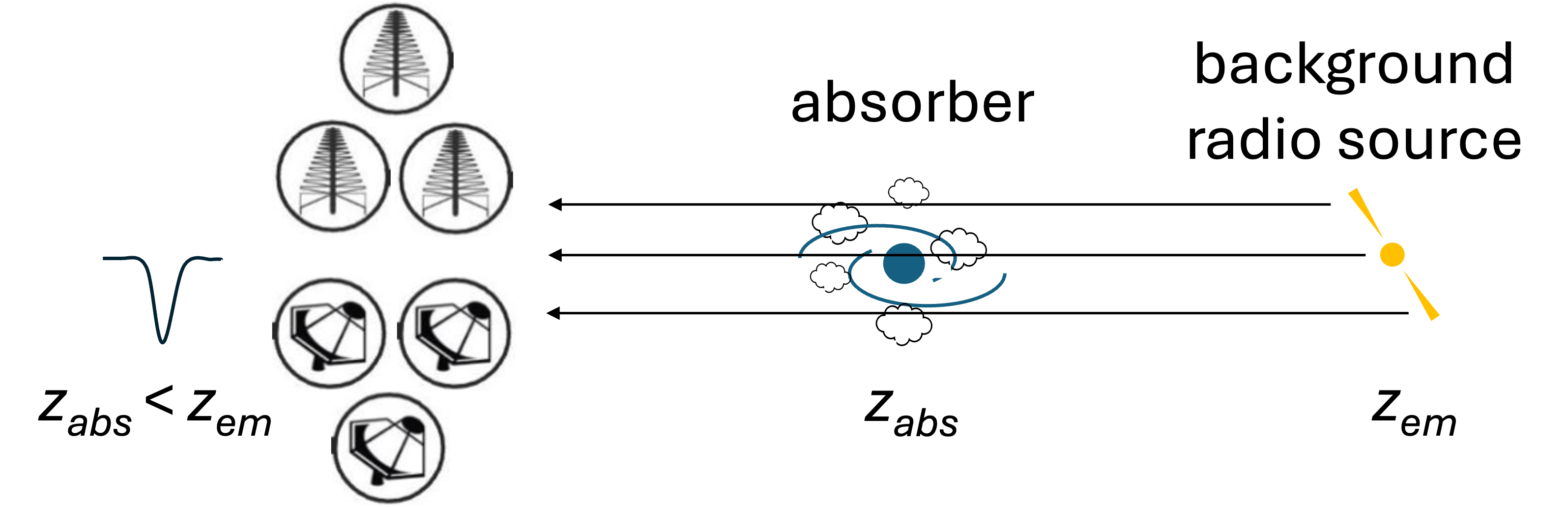}
    \includegraphics[clip,width=0.5\textwidth]{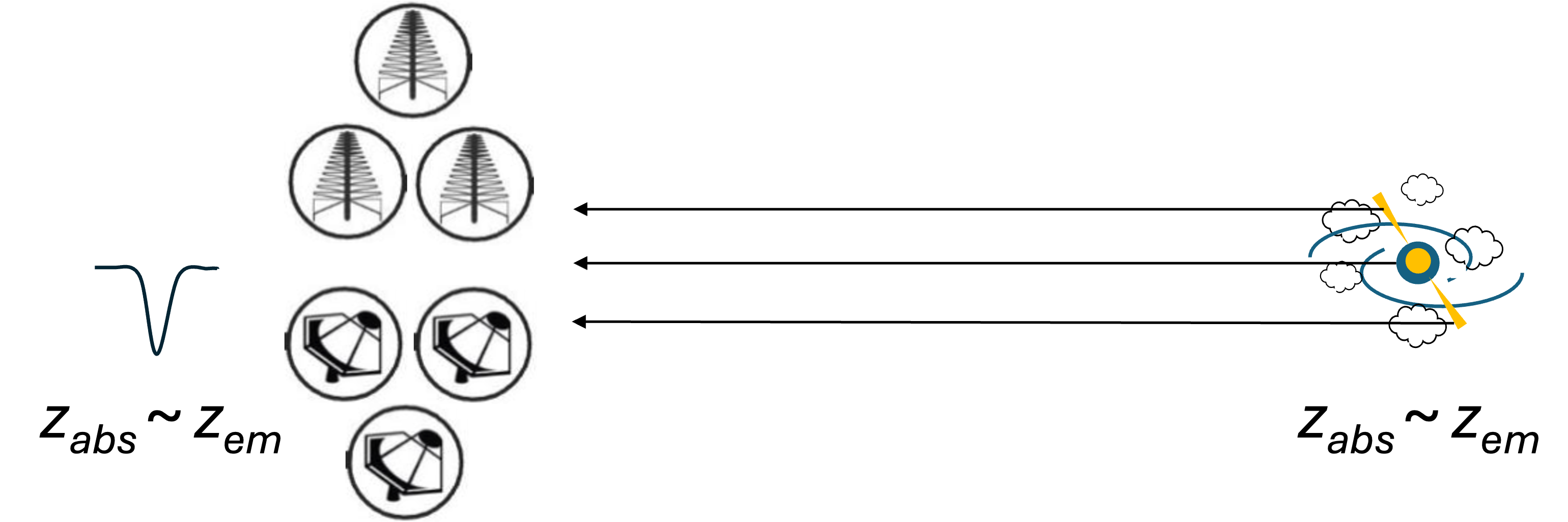}
    \vspace{0.2cm}
    \caption{Schematic of intervening \HI\ absorption (left) and associated \HI\ absorption (right). }
    \label{fig:cartoon}
\end{figure}

Absorption-line studies fall in two categories: {\it (i)} intervening absorption - when absorbing gas ($z_{\rm abs}$) is unrelated to the background source at $z_{\rm em}$, with a working definition that the difference in $z_{\rm em}$ and $z_{\rm abs}$, expressed in Doppler shift, is higher than 3000\,kms$^{-1}$ \citep[e.g.,][]{Wolfe1986, Ellison2002}.  The absorbing gas in these cases correspond to the interstellar or circumgalactic medium of an intervening galaxy or intragroup medium (see Fig.~\ref{fig:cartoon}). 
{\it (ii)} Associated absorption: when $z_{\rm em}$ $\approx$ $z_{\rm abs}$, the absorbing gas may be associated with the AGN, its host galaxy, a nearby companion galaxy, outflows driven by its feedback, or infalling material.

The principal observable in \HI\ absorption studies is the line optical depth, $\tau(v)$, which describes the fractional attenuation of background radio continuum as a function of velocity:

\begin{equation}
\tau(v) = -\ln \left( 1 + \frac{\Delta S(v)}{c_{\rm f} S_{\rm c}(v)} \right),
\end{equation}

where $\Delta S(v)$ is the continuum-subtracted absorption line depth (in Jy), 
$S_{\rm c}(v)$ is the continuum flux density (in Jy), and $c_{\rm f}$ is the covering factor representing the fraction of the radio continuum flux covered by the absorbing gas. In the optically thin regime, the \HI\ column density is given by:
\begin{equation}
N_{\rm HI} = 1.823 \times 10^{18} T_{\rm s}  \int \tau(v) \, dv \ \ {\rm cm}^{-2},
\label{eq21}
\end{equation}
where $T_{\rm s}$ is the spin temperature of the gas (in Kelvin), and the integral is over velocity in km\,s$^{-1}$ \citep[][]{Draine2011}. Accurate derivation of $N_{\rm HI}$ thus depends on independent constraints or assumptions about both $T_{\rm s}$ and $c_{\rm f}$. 
In addition, the full width at half maximum (FWHM) of 21-cm absorption line detected in a high-resolution spectrum yields a direct measurement of (or stringent upper limit on) the kinetic temperature (in Kelvin), $T_k$ = 21.866$\times\Delta v^2$ \citep[][]{Mcclure-Griffiths2023}.

The spin temperature $T_{\rm s}$ is a measure of the excitation of the 21-cm transition, and depends on a combination of collisional excitation, Ly$\alpha$ pumping, and coupling to the local radiation field \citep{Field1958,Liszt2001}. In the cold neutral medium (CNM), where densities are higher ($n \sim 100$ cm$^{-3}$), $T_{\rm s} \approx T_{\rm k} \sim 100$~K. In contrast, the warm neutral medium (WNM) has lower densities and higher temperatures, leading to $T_{\rm s} \sim 1000$–10,000~K. 
If the total \HI\ column density associated with the absorbing gas is independently known from 21-cm emission or Ly$\alpha$ absorption line, then $T_{\rm s}$ can be inferred under suitable assumptions of $c_{\rm f}$ 
\citep[e.g.][]{Kanekar2014b,allison2021}. 
The covering factor in principle is determined by the parsec-scale structure of the gas, which in turn depends sensitively on the stellar feedback processes, though \cite{Braun2012} suggests that cosmological intervening absorbers with high \HI\ column densities are likely to arise in atomic clouds of $\sim100$\,pc linear size which are
self-opaque in the 21 cm transition. 

Instantaneous wideband coverages of modern radio telescopes enables simultaneous searches of OH main and satellite lines in absorption. OH lines can provide independent constraints on H$_2$ fraction in the CNM.
OH is most commonly observed in the 18-cm ground state transitions which occur at rest frequencies of 1665.402 and 1667.359 (main lines), and 1612.231 and 1720.530\,MHz (satellite lines). The relative strengths of these lines in the local thermodynamic equilibrium (LTE) are 1612:1665:1667:1720 MHz = 1:5:9:1. The typical abundance ratios observed in the Galaxy are $\sim10^{-7}$ \citep[e.g.][]{Li2018}.
They often exhibit maser emission in regions associated with high-density and far-infrared (FIR) radiation, and can provide additional constraints on the neutral gas conditions in normal and active galaxies. 
For an optically thin cloud, the integrated OH optical depth of the 1667\,MHz line is related to the OH column density $N$(OH) through
\begin{equation}
N{({\rm OH})}=2.24\times10^{14}~{T_{\rm ex}}\int~\tau_{1667}(v)~{\rm d}v~{\rm cm^{-2}}, 
\label{eqoh}
\end{equation}
where $T_{\rm ex}$ is the excitation temperature in Kelvin and $\tau_{1667}$($v$) is the optical depth of the 1667\,MHz line at velocity $v$ (in km\,s$^{-1}$).

Note that the $c_{\rm f}$ of the \HI\ 21-cm and OH 18-cm absorbing gas in Equations~\ref{eq21} and \ref{eqoh} even for the same absorber could be different \citep[][]{Gupta2018oh}. 
Although these various parameters introduce degeneracy and uncertainty, \HI\ and OH absorption lines have proven to be an excellent probe of the properties of the cold neutral gas in the Milky Way \citep[e.g.][]{Li2018, Dawson2022, Rugel2025}, and deliver some of the most stringent ($<10^{-6}$) constraints on the fractional variations of fundamental constants of physics \citep[][]{Kanekar2005, Uzan2011}. 
Large sensitive searches with SKAO, aided by sub-arcsecond scale spectroscopy with SKA-VLBI, can potentially alleviate these limitations and help build a more complete dust-unbiased view of extragalactic ISM.
%

\subsection{\HI\ absorption studies in the pre-SKA era}
\label{sec:preska-hi}

The past decade has seen the emergence of \HI\ absorption line surveys enabled by wideband receivers and correlators on modern radio facilities. These surveys mark a transition from searches based on targets selected from optical spectroscopic catalogues to large untargeted surveys capable of obtaining unbiased estimates of cold gas associated with normal and active galaxies.  This includes dedicated large absorption line surveys at ASKAP \citep[$0.4<z<1.0$;][]{Allison2022} and MeerKAT \citep[$0<z<1.4$;][]{MALS}, as well as large \HI\ 21-cm emission line surveys with FAST \citep[$z<0.4$;][]{FASHI} and intensity mapping experiments ($0.8<z<2.5$) such as Canadian Hydrogen Intensity Mapping Experiment \citep[CHIME;][]{CHIME2025arXiv250611269C} and the Hydrogen Intensity and Real-time Analysis eXperiment \citep[HIRAX;][]{Newburgh2016}.

Typically, prior to the advent of these new wideband receivers, AGN with known redshift and compact morphology were observed for associated absorption line studies \citep[e.g.][]{Vermeulen2003, Gupta2006sur, Gereb2015, aditya2017MNRAS.465.5011A, Maccagni2017, Grasha2019}.
For intervening absorption line searches the targets have been sight lines with indications of large \HI\ column densities, $N$(\HI), suggested by the presence of damped Ly$\alpha$ absorber \citep[DLA; $N$(\HI) $> 2 \times 10^{20}$\,cm$^{-2}$;][]{Kanekar2003, Srianand2012, Kanekar2014b} or Mg~{\sc ii} absorption \citep[][]{Briggs1983, Lane2000, Gupta2009, Curran2010, Dutta2017} or a galaxy at a low impact parameter \citep[][]{Carilli1992qgp, Gupta2010, Reeves2016, Dutta2017}. 
Only a handful of OH absorbers are known at present \citep[][]{Chengalur2003, Kanekar2005, Combes2023}, and these have rarely been systematically searched in galaxies \citep[][]{Gupta2018oh}.

Although targeted studies can provide an estimate of the cold gas fraction in a specific class of radio loud AGN or galaxies probed by DLA or Mg~{\sc ii} absorption, they often lack the statistical power to explore trends and correlations across diverse galaxy populations. The interpretation is also complicated by selection methods used to define optical spectroscopic samples and the possibility of an associated dust-bias \citep[e.g.][]{Krogager2015}. 
The ongoing precursor surveys offer an opportunity to fill this gap and also identify rare or extreme systems for detailed follow-ups.  These untargeted radio surveys have the potential to uncover populations of absorbers that may have been missed in previous targeted searches. 
The ongoing searches have already revealed new absorption-line systems that would have been missed in traditional targeted searches. These include absorption detected against the radio lobes of galaxies rather than the central core \citep{Murthy2022, Mahony2022} and absorbers with no signatures of high $N$(\HI) in their optical spectra \citep{Gupta2022, deka2024A&A...687A..50D}, and very high optical-depth lines that require wide-area coverage to uncover uncommon systems \citep{Su2022, aditya2024MNRAS.527.8511A,Yoon2025}.

The next two sections explore primary science goals for associated and intervening absorbers, and synergies with multi-wavelength observations. 

\section{Absorption line studies as a probe of AGN fuelling and feedback} 

Understanding the physics of AGN fuelling and feedback is a key science goal for the SKAO, and observations of the cold gas provides crucial insight into the processes that regulate the growth and activity of AGN. Associated \HI\ absorption lines offer a powerful means of tracing the cold neutral gas in the immediate environments of radio AGN, shedding light on the complex interactions between the central engine and its surrounding interstellar medium (ISM). 

\subsection{Detailed studies at low-redshift} 

\begin{figure}[ht]
    \includegraphics[clip,width=0.45\textwidth]{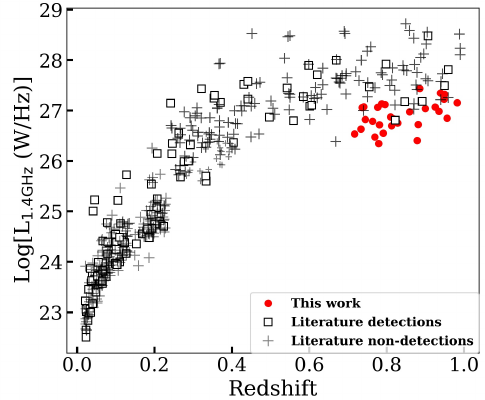}
    \includegraphics[clip,width=0.54\textwidth]{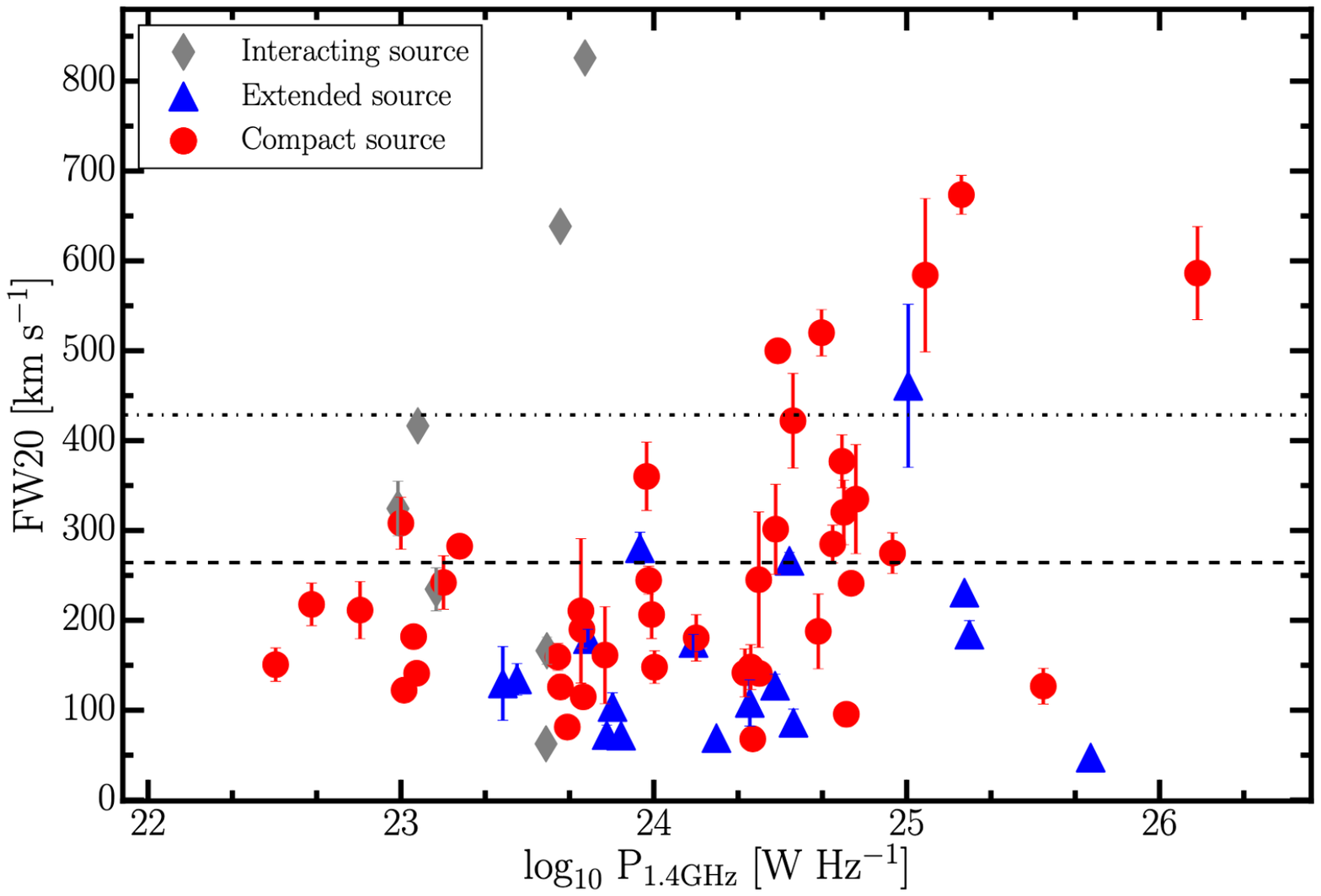}
    \caption{{\em Left panel}: Distribution of radio luminosities of sources searched for absorption as a function of redshift~\citep[Fig. 1 of ~][]{murthy2022A&A...659A.185M}. The JVLA high redshift sample of ~\citet{murthy2022A&A...659A.185M} is shown in red while other surveys are shown in black. In particular, at $z < 0.25$ is shown the sample of~\citet{Maccagni2017}. At $z > 0.25$ the searches are from~\citet{Carilli1992,Carilli1998,Pihlstrom2003,Vermeulen2003,Gupta2006,Orienti2006,Curran2006,Curran2011,Allison2015,Yan2016,Curran2017,Ostorero2017,aditya2018MNRAS.473...59A,Aditya2019,Curran2019,Grasha2019,Mhaskey2020,Murthy2021}. {\em Right panel}: Full width at $20\%$ of the intensity of the \HI\ detections  vs. the radio power of the sources~\citep[Fig. 4 of][]{Maccagni2017}. Sources are classified in compact (red) and extended (blue) according to the extent of their radio continuum. The dashed line indicates the mean of the distribution of rotational velocities of the sources of the sample. The fine dashed line shows the $3\sigma$ upper limit of the distribution of line-widths.}
    \label{fig:lowz_survey}
\end{figure}

At low-redshifts ($z\lesssim 0.25$) \HI\ absorption studies against AGN allow us to probe the cold gas in the very proximity of their SMBH ($\lesssim 1$ kpc). Over time, different surveys at low-redshifts~\citep[$\lesssim 0.25$, e.g.][]{vanGorkom1989,Gallimore1999, Gupta2006sur,Maccagni2017,chandola2020MNRAS.494.5161C} discovered hundreds of \HI\ absorption lines associated to the circumnuclear regions of AGN~\citep{Morganti2018}. 
In the nearby Universe the detection rate of associated \HI\ absorption is $\sim30\%$~\citep{Maccagni2017} above a peak optical depth of $\tau_{\rm pk} \sim 0.08$ and is independent of redshift and radio power (Fig.~\ref{fig:lowz_survey} left panel). Compact AGN with sizes typically less than 15\,kpc (i.e. sources where the jets are embedded within the host galaxy) show not only higher detection rates but also broader lines indicative of more perturbed kinematics (Fig.~\ref{fig:lowz_survey} right panel).  The occurrence of \HI\ absorption may vary across different classes of AGN, with extended ($>$15\,kpc) radio sources showing much lower \citep[$\sim$15\%;][]{Chandola2013} and those associated with merging galaxy pairs showing much higher ($\sim$85\%) detection rates \citep[][]{Dutta2019}.

Most \HI\ absorption lines trace the neutral atomic phase of the circumnuclear disk, revealing the structure and dynamics of gas that may be directly feeding the central black hole. However, the detailed kinematics and profiles of these lines can also unveil other processes at play. Absorption lines with a more complex morphology not explained by a rotating disk ~\citep[$5\%$ of the detections, see][]{Morganti2005,Gereb2015,Maccagni2017} can be classified into two groups: narrow redshifted lines and broad blue-shifted wings. Narrow redshifted lines have been mostly found in bright early type galaxies~\citep[][]{vanGorkom1989} and likely trace gas that is falling towards the AGN because of turbulence, as often seen in bright cluster galaxies in the molecular gas phase~\citep[see, for example,][]{Tremblay2016,Tremblay2018,Tamhane2022}, and also detected by \HI\ absorption lines~\citep[see, for example,][]{jaffe90,taylor1999,Saraf2023}.

\begin{figure}[ht]
    \includegraphics[clip,width=0.35\textwidth]{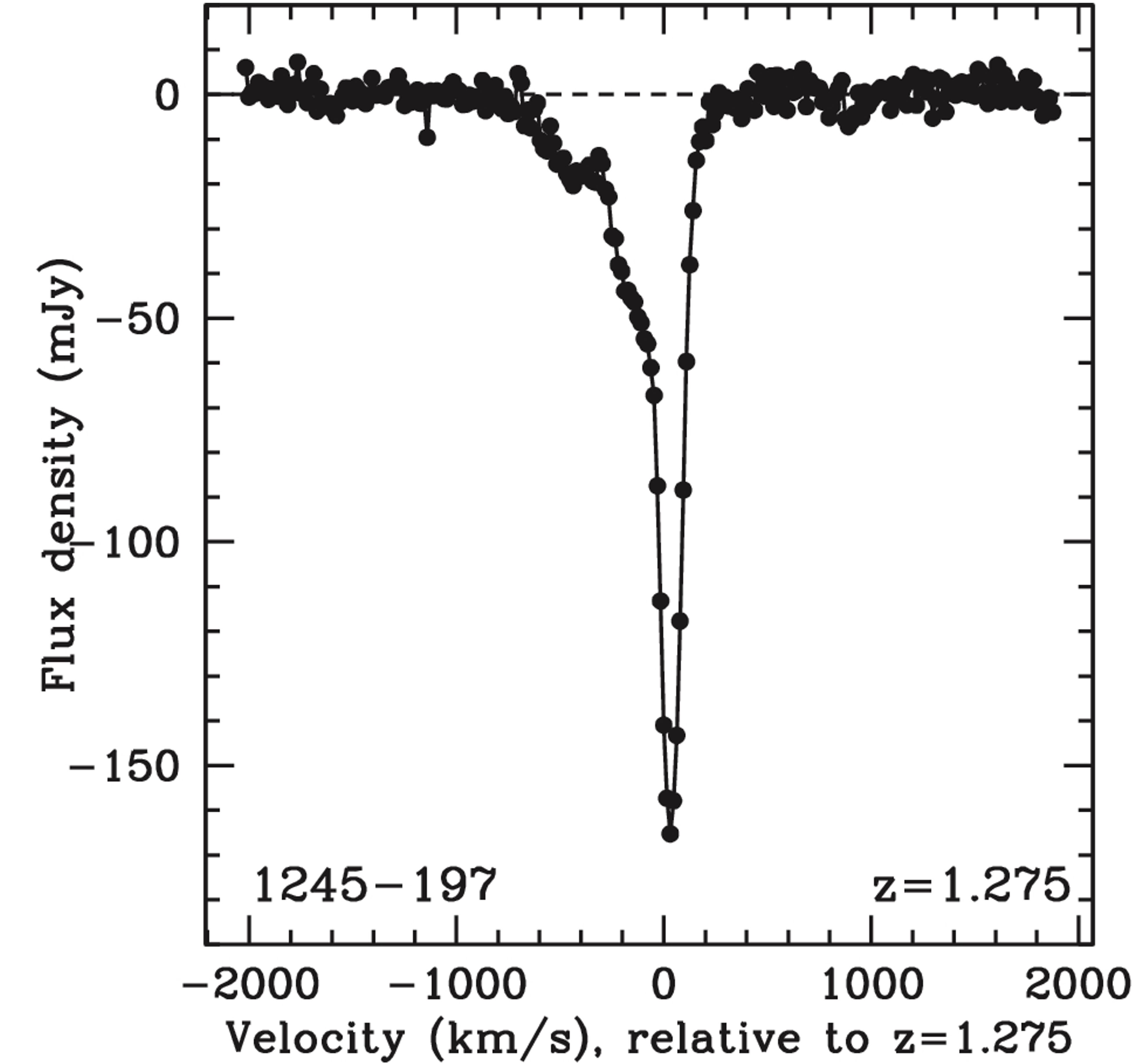}
    \includegraphics[clip,width=0.68\textwidth]{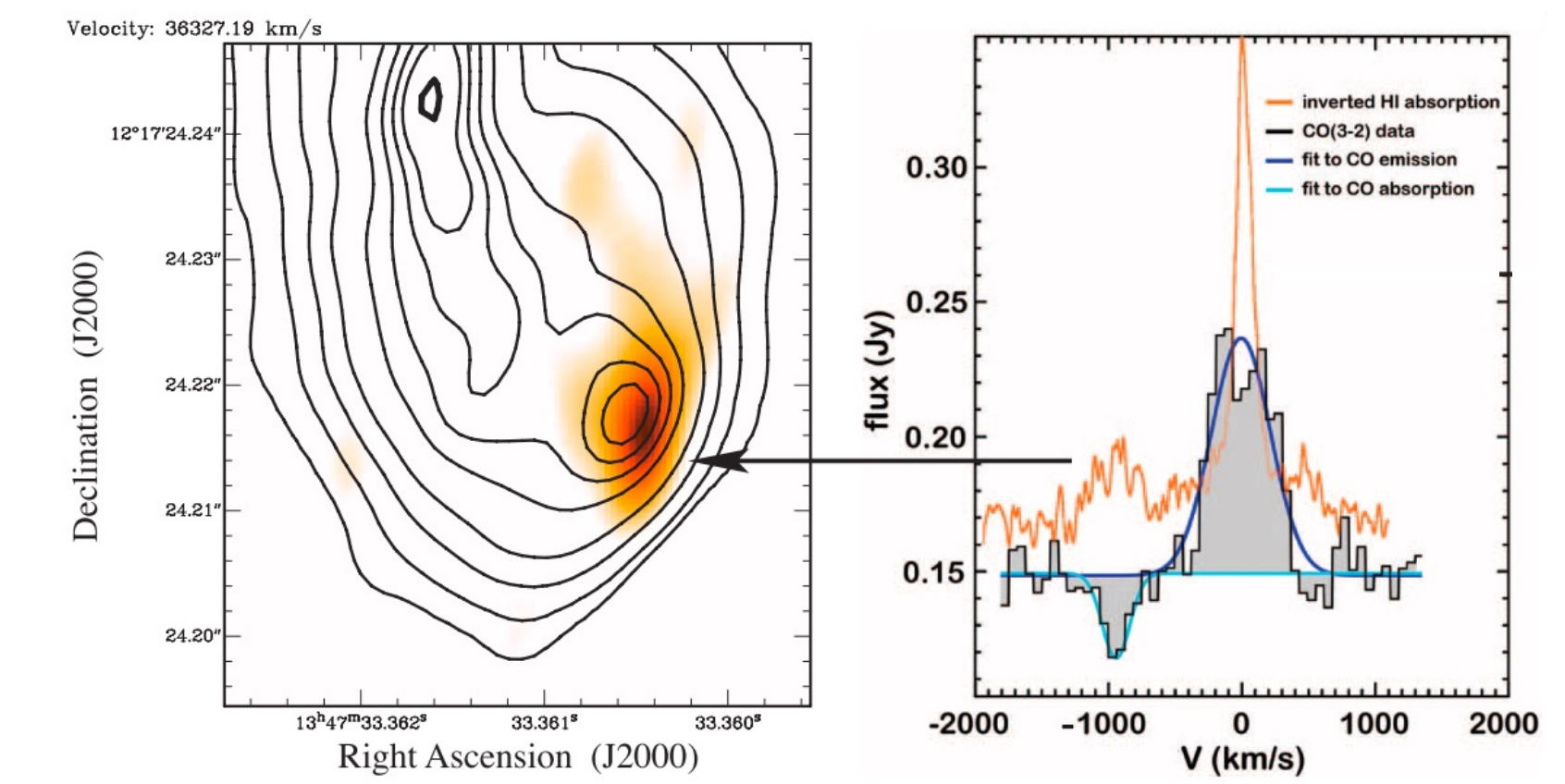}    
    \caption{Some examples of fast outflows detected via \HI\ absorption studies: TXS 1245-197 ({\it left panel}) at $z=1.25$ is one of the highest redshift detections of a distinctive blue-shifted wing indicative of a fast outflow \citep{aditya2018MNRAS.473...59A}; and 4C12.50 ({\it middle and right panels}), where VLBI observations pinpoint the location of the \HI\ outflow to the hotspot of the radio lobe \citep{Morganti2013}.}
    \label{fig:outflows}
\end{figure}

Broad ($> 100$ km s$^{-1}$) blue-shifted wings can trace \HI\ outflows entrained by the AGN radio jets (most commonly), but also radiative winds (see Fig. \ref{fig:outflows} for some examples). Given that most gas involved in feeding and feedback mechanisms is in the cold phase~\citep[e.g.][]{Fluetsch2019,Murthy2022}, \HI\ absorption studies can provide some of the strongest evidence for AGN jets accelerating and clearing gas from a galaxy. 
Some well-studied examples include IC5063~\citep{Morganti1998}, 4C12.50~\citep{Morganti2013}, 3C293~\citep{Mahony2013} and NCG1266~\citep{Alatalo2011}); see \citet{Morganti2018} and \citet{Morganti2024} for a comprehensive overview. These sources typically have outflow rates ranging from a few solar masses to tens of solar masses per year in atomic gas, often far higher than the outflow rates detected in ionised gas \citep{Mahony2016}. However, the presence of a blue-shifted wing alone should be treated with caution. For example, even though the \HI\ absorbing system of IC5063 has extreme velocities of $\approx 700$ km s$^{-1}$, final evidence that the \HI\ outflow is caused by expansion of the radio jets has been provided only by follow-up sub-arcsecond resolution VLBI observations~\citep{Oosterloo2000}. 

High resolution observations which resolve the \HI\ line against the radio continuum can reveal the true nature of the \HI\ gas, if it is outflowing and what is causing its acceleration \citep[see e.g.][]{Morganti2013, Schulz2018, Murthy2024}. SKA-VLBI will enable direct resolution of \HI\ absorption lines, while also allowing simultaneous study of the radio continuum properties of background sources. This combination of sensitivity and pc-scale resolution will facilitate spatially resolved studies of circumnuclear disks, jet–ISM interactions, and pc-scale outflows, vastly expanding the statistical sample of AGN and providing a more complete view of cold gas kinematics and feedback across cosmic time. Further details on the potential of SKA VLBI \HI\ absorption studies are given in the VLBI papers in this collection.

\subsection{Building statistical samples of \HI\ absorbers}

In addition to studies of individual systems, statistical analyses of samples of associated absorbers can reveal how the presence and properties of cold gas relate to AGN characteristics such as accretion mode and AGN type. Previous studies have found a notable fraction of detections occur toward compact, peaked-spectrum sources \citep{Pihlstrom2003, Gupta2006,Yoon2025}, which are typically interpreted as young or confined radio AGN \citep{ODea2021}. These sources may be embedded in dense circumnuclear gas, increasing the likelihood of detecting absorption. Studies have also highlighted potential links between \HI\ absorption and the IR galaxy colours \citep{chandola2017MNRAS.465..997C, Glowacki2017, Curran2018} or accretion mode of the central black hole, with cold-mode accretors more likely to show higher instances of absorption compared to hot-mode accretors \citep{chandola2017MNRAS.465..997C,chandola2020MNRAS.494.5161C}. These findings suggest that \HI\ absorption can be used as a potential diagnostic of AGN evolutionary stage and feedback processes. 

The sample shown in Fig. \ref{fig:lowz_survey} represents the combination of many individual targeted surveys carried out with multiple telescopes over the past two decades—a global, long-term effort required to build even a modest census of \HI\ absorbers. The emergence of large, untargeted surveys with SKA precursors has transformed the search for \HI\ absorption. The First Large Absorption Survey in \HI\ (FLASH; \citealt{Allison2022}) takes advantage of ASKAP's large field of view afforded by its phased array feeds to carry out a wide-area survey searching for \HI\ in absorption in the redshift range $0.42<z<1$. FLASH searches for \HI\ absorption towards all radio sources brighter than 30\,mJy south of declination $+15^\circ$, making it the first `all-sky' search for \HI\ absorption. The MeerKAT Absorption Line Survey (MALS; \citealt{MALS}) takes a different approach; observing $\sim$400 pointings centered at radio sources brighter than 200\,mJy at 1\,GHz to search for \HI\ absorption out to $z=1.5$ towards all the radio sources ($\gtrsim$5\,mJy) within the MeerKAT primary beam. In addition to SKA-precursor surveys, recent drift-scan surveys with FAST -- FASHI and CRAFTS -- have conducted untargeted searches over large areas of the sky, detecting 51 \HI\ absorption systems at $z<0.09$ (FASHI; \citealt{FASHI}) and 34 systems out to $z=0.35$ (CRAFTS; \citealt{Hu2025}). These surveys maximize on the sensitivity of a large single-dish to probe absorption towards lower luminosity radio sources, complementing the precursor surveys FLASH and MALS.

Early results from FLASH already demonstrate how this new observing approach is shedding fresh light on the properties of \HI\ absorbers. As expected, a large fraction of detections are associated with young, compact radio sources, consistent with previous targeted studies. However, FLASH is also uncovering rarer populations of absorbers, including systems with exceptionally high optical depths and those detected against the extended radio lobes of galaxies \citep{Su2022, Mahony2022, aditya2024MNRAS.527.8511A, Yoon2025}. 
 \citet{Yoon2025} show examples of \HI\ absorption lines detected in the FLASH pilot survey, illustrating both the survey’s large field of view and its wide instantaneous bandwidth.

\subsection{The evolution of \HI\ absorption detection rates}

The enhanced low-frequency coverage of radio telescopes over the last two decades has also enabled us to push targeted absorption line searches out to higher redshifts, usually with prior knowledge of the galaxy redshift \citep[e.g.][]{Vermeulen2003, Gupta2006sur, murthy2022A&A...659A.185M, aditya2024MNRAS.527.8511A}. 
Despite efforts over the past three decades, only a handful of absorption systems are known at $z > 1$: nine systems at $1 < z < 2$, and five systems at $2 < z < 3.5$ \citep[see][for the latest detections]{deka2024A&A...687A..50D, CHIME2025arXiv250611269C}. 

Even though affected by low number statistics and by the fact at higher redshifts only the most powerful radio sources ($\gtrsim 10^{26}$ W/Hz) have been observed, 
the detection rate of associated \HI\ absorption decreases with redshift ~\citep[see also][]{Curran2008,aditya2018MNRAS.473...59A, Su2022, aditya2024MNRAS.527.8511A}. 
At $2<z<3.5$, the detection rate is only $1.6^{+3.8}_{-1.4}$\% \citep[][]{gupta2021ApJS..255...28G}.

The evolution of the detection rate of \HI\ absorption is still unclear and remains a subject of debate. With its combination of wide frequency coverage and spectral-line sensitivity, the SKAO will directly test whether this evolution reflects intrinsic changes in the AGN population with redshift or arises from observational biases, such as the orientation of the radio source, UV emission photoionising the neutral gas or higher $T_{\rm spin}$ of the gas which would decrease the observed optical depth of the lines for the same column density of the gas \citep[]{Gupta2006, CurranWhiting2012, aditya2018MNRAS.481.1578A}.

\subsection{Feedback processes at cosmic noon}

Galaxy-scale outflows are ubiquitous in massive galaxies at cosmic noon (\mbox{1~$\lesssim z \lesssim$~3}), but their role in quenching star-formation is strongly debated \citep[e.g.][]{ForsterSchreiber2020}. The SKAO will significantly advance our understanding of the mechanisms powering AGN-driven outflows, the amount of mass ejected by these outflows and quantify the impact that these outflows have on star-formation processes.

Outflows driven by luminous AGN at $z\sim$~2 are often assumed to be radiation-driven because the energy injection rates correlate with the AGN bolometric luminosity and are sufficiently large to drive the observed outflows \citep[e.g.][]{Fiore2017}. However, compact radio jets have been shown to be a crucial feedback mechanism in radio-quiet quasars at $z<$~0.2 \citep[e.g.][]{Jarvis2019}. Furthermore, radio jets may become increasingly important outflow drivers as galaxies start to exhaust their cold gas reservoirs \citep[e.g.][]{Roy2021}, and \cite{heckman2024} propose that radio jets are the main drivers behind the quenching of star formation in all massive galaxies across cosmic time.  

As discussed in the previous section, at low redshifts there are a number of radio galaxies that show evidence for jet-driven outflows \citep{Morganti2005, Mahony2013, Morganti2013, Schulz2018}. Additionally, optical studies of the ionised gas show that these jet-driven outflows can impact the gas throughout the host galaxy via turbulence and heating processes created by the jets as they propagate through the clumpy ISM \citep{Mahony2016, Mukherjee2018}. \citet{aditya2018MNRAS.473...59A} and \citet{Aditya2019} reported the first detections of cold gas outflows detected in \HI\ absorption beyond the local Universe. Using GMRT to search for \HI\ absorption in peaked spectrum sources at $z>1$, they reported that five out of the six sources detected in \HI\ absorption also show evidence for broad blue-shifted features indicative of fast-outflows, the most extreme of which exhibits a mass outflow rate of 78 $M_\odot$\,yr$^{-1}$. Although still small numbers at present, the higher detection rate of outflowing material, and the higher outflow rates than that seen in the local Universe, indicates that outflows could be more prevalent, and carry more mass and energy at intermediate redshifts.

Both simulations and observations of AGN-driven outflows at $z\sim$~2 suggest that most outflowing mass is likely to be found in the  neutral and molecular phases \citep[e.g.][]{HerreraCamus2019,Belli2024}, but these phases are poorly characterised observationally at $z\sim$~2 compared to the warmer ionised phase. Ly$\alpha$ absorption has been used for this purpose \citep[e.g.][]{Moretti2025} but generally traces more extended, lower column density \HI\ \citep[e.g.][]{Zwaan2005} and may therefore be a less direct probe of the dense outflowing gas. Recent works showed that cold gas in the AGN environment (including outflows) can be efficiently selected and studied in details using molecular hydrogen UV transitions that are conveniently shifted in the optical band for $z\gtrsim 2.5$ \citep{Noterdaeme2019,Noterdaeme2021,Noterdaeme2023}. Additionally, JWST has enabled the use of Na~\textsc{i} 5891,5897\AA\ absorption as a tracer of neutral gas outflows at cosmic noon \citep[e.g.][]{Davies2024}. However, the relationship between Na~\textsc{i} column density and H~\textsc{i} column density has only been measured for a few sources outside the Milky Way. Detections of \HI\ 21-cm absorbers at $z\sim$~2 will provide large samples of galaxies to accurately calibrate optical and UV tracers of neutral outflows. %

Combining deep, high resolution multi-band radio continuum observations from SKAO with optical spectroscopy will enable us to measure radio spectral indices and jet powers for $z\sim$~2 galaxies and explore the role of radio jets in powering outflows, as well as the impact of radio jets on the interstellar medium of the host galaxy. Using broadband radio spectra, it may be possible to infer both the jet kinetic power and ISM properties, providing a direct link between jet energetics and their impact on the host galaxy \citep{Young2025}.

\subsection{\HI\ absorption at high redshift} \label{sec:highz} 

Much of the science discussed above, particularly examination of outflows and AGN feedback, is possible at higher redshifts with SKA-Low. The frequency range and sensitivity of SKA-Low will provide the ability to observe \HI\ absorption at unprecedented redshifts ($z>3$). 

SKA-Low AA4 will not be sensitive enough for wide-area, untargeted surveys, especially as the observed flux densities of even the most luminous radio galaxies rapidly become faint at $z>3$. Naturally, any wide area continuum surveys conducted with SKA-Low could be used to search for \HI\ absorption, but this science case is unlikely to be enough justification for such a survey on its own. Hence, above $z=3$, where only SKA-Low can probe, targeted observations of known compact, bright radio sources would provide the best scientific return. However, the wide field-of-view of SKA-Low will enable absorption searches against all sources in the field, providing the opportunity for serendipitous detections beyond the primary target.

A key question is what impact do radio jets have on their host galaxies in the early Universe? Can short ($\le 100\,$kyr) bursts of powerful radio jets successfully quench star formation in massive high-z radio galaxies as seen in deep {\it JWST} surveys \citep{Nanayakkara2024}? If caused by radio jets with $L_{\rm 500MHz}>10^{26}$ they may be too rare to be detected in the deep, narrow surveys probed by {\it JWST} and can only be studied by selecting these sources from wide-area surveys when they are in the short-lived radio-loud phase. %

\subsection{Tracing the multi-phase gas} \label{sec:molecular}

The cold atomic gas traced by \HI\ absorption represents just one phase of the complex interstellar and circumgalactic medium. Understanding how this atomic component connects to the molecular and ionised phases is critical for interpreting the physical conditions, kinematics, and fueling mechanisms in galaxies. The combination of \HI\ absorption with OH 18-cm absorption lines and high-resolution ionised and molecular gas observations is therefore essential for a complete view of the gas cycle in galaxies, particularly around AGN and in dense environments. 

Recent studies have demonstrated the diagnostic power of combining \HI\ and CO data in the same systems~\citep[see for example, PKS 1718-649 and PKS B1740-517, ][]{Maccagni2014,Maccagni2016,Maccagni2018,Allison2015,Allison2019}.  Such comparisons reveal the physical and dynamical transition between atomic and molecular phases, and the balance between inflow and feedback. 
Currently, the main limitation is the small number of known molecular absorption line systems in OH and CO \citep[][]{Kanekar2008, Combes2024}.
SKAO observations will extend these studies to much larger and more diverse samples, enabling systematic investigation of how the cold gas phases co-exist and evolve across cosmic time.

\subsubsection{Multi-phase Gas in Cluster Environments}

\begin{figure}[ht]

\begin{minipage}[h]{0.6\textwidth}
    \centering
    \includegraphics[clip,width=\linewidth]{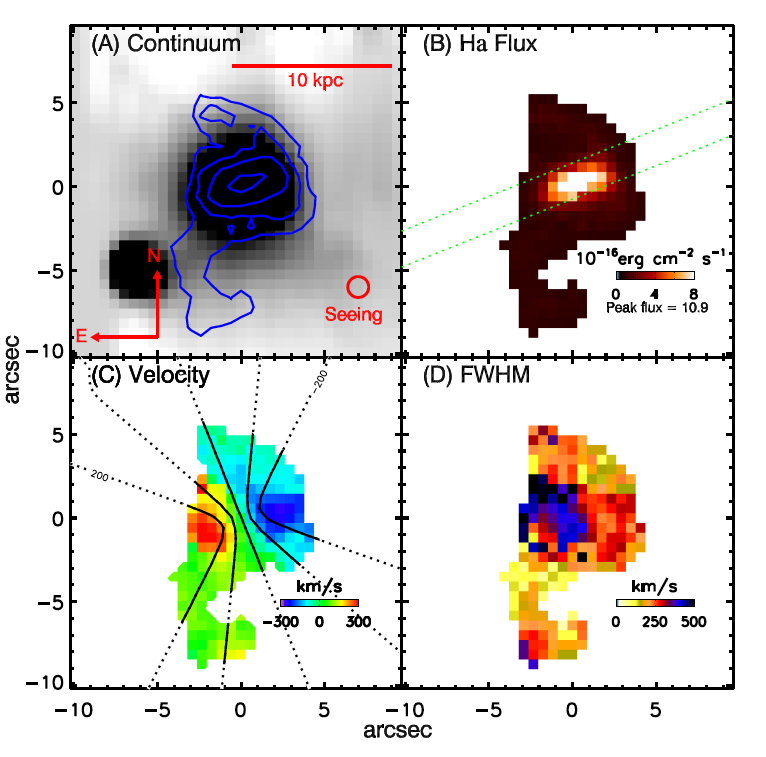}
\end{minipage}
\begin{minipage}[h]{0.35\textwidth}
    \centering
    \includegraphics[clip,width=\linewidth]{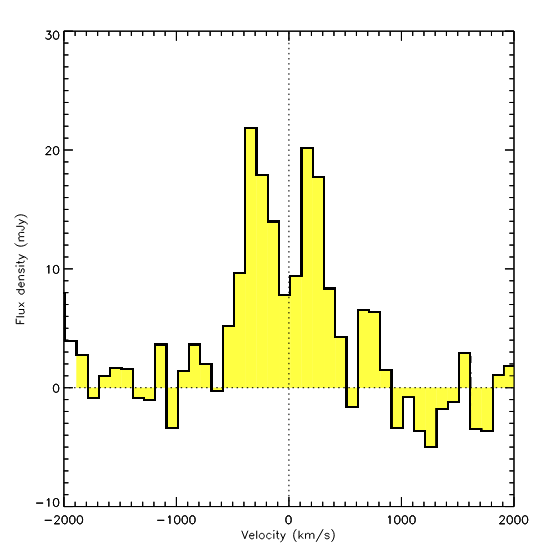}
    \includegraphics[clip,width=\linewidth]{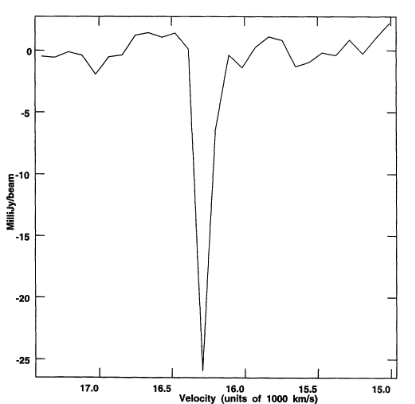}
\end{minipage}

    \caption{Molecular gas in BCGs. From left to right; IFU observations of 3C\,218 (Hydra A) from \citet{Hamer2014HydraA} showing the continuum image made by collapsing the H$\alpha$ cube (panel A), flux map (B), velocity field relative to the galaxy redshift (C), and FWHM which broadens at the centre of the velocity gradient (D). CO(2–1) emission detected with IRAM \citep{Hamer2014HydraA} and VLA \HI\ absorption toward the Hydra~A core \citep{dwarakanath1994}.} 
    \label{fig:moheg}
\end{figure}

In cluster and group environments, the detection of \HI\ 21-cm absorption provides a powerful tool to investigate the evolution of cold atomic gas in dense environments, where molecular and ionized gases are commonly observed. This is particularly significant in the case of interacting galaxies within cluster environments, where filamentary structures of cooling gas flows fuel the central galaxies. Molecular Hydrogen Emission Galaxies (MOHEGs), particularly those found in cluster environments, can be ideal targets for studying the relationship between these different gas phases due to their abundance of molecular hydrogen \citep{Ogle2010}. One such example is 3C\,218, the radio galaxy at the center of the Hydra\,A cluster, where multi-wavelength observations reveal a complex interplay between various gas phases associated with the intra-cluster medium (ICM). As seen in Fig. \ref{fig:moheg}, gas filaments in this system are detected across multiple phases, offering valuable insights into the transition between neutral, molecular, and ionized gas, enhancing our understanding of the complex dynamics of AGN fuelling and feedback and provide direct evidence for cold gas accretion in cluster cores. 

\subsubsection{Molecular absorption in front of Brightest Cluster Galaxies (BCG)} 

At low redshift, molecular absorption is easier to detect, since the high spatial resolution achievable is crucial to separate it from molecular emission: the continuum in the millimeter domain is in general limited to the tiny core. The local radio sources rich in gas are frequently detected, such as Centaurus A, and they are also frequent in cool-core clusters, in front of the BCG \citep{Rose2024}.

In these cool-core clusters, cooling gas fuels the central AGN while feedback processes expel some gas, to prevent excessive star formation. Radio jets are sculpting cavities in the hot X-ray gas, in virial equilibrium with the cluster potential \citep[e.g.][]{Fabian2012}, but it can be difficult to know whether the emitting gas towards the BCG is infalling or outflowing. Absorption line detections can be used to determine the direction of motion. In a vast majority, the gas is infalling, supporting the cooling \citep{Bera2023}. When detected, the \HI\ 21-cm absorption tends to be broader, tracing gas not only in front of the radio core, but also the radio lobes. An example of infalling molecular gas is displayed in Fig.~\ref{fig:moheg}. SKAO surveys will uncover many more such absorbers, providing an unprecedented view of cooling gas filaments in clusters and the fueling of central AGN.

With the sensitivity and sky coverage of SKAO, it will be possible to identify many more such systems through large \HI\ absorption surveys, providing a statistical view of cold gas in BCGs and their surroundings. Follow-up with ALMA and JWST will enable detailed comparisons between the atomic, molecular, and ionized components, offering new insights into how AGN feedback regulates cooling and star formation in dense environments.

\section{Absorption line studies as a probe of cold atomic gas in and around galaxies}

The star formation history of the Universe is primarily driven by the gas content of galaxies and their star formation efficiency.  \HI\ 21-cm line absorption is an excellent tracer of the CNM and can reveal the processes leading to the conversion of atomic gas into molecular gas and eventually into stars. 
Primarily due to the technical limitations imposed by narrow bandwidths and a hostile radio frequency environment the earlier searches of intervening 21-cm 
absorption lines, summarized in Section~\ref{sec:preska-hi}, focused on quasar sight lines from optical surveys with indications of high \HI\ column density.
Here we present key results, complications due to different pre-selections methods and advancements possible with the SKAO.

\subsection{Cosmic evolution of CNM fraction}

DLAs detected in the optical and ultraviolet spectra of distant quasars trace the bulk of neutral \HI\ in the Universe \citep[][]{Noterdaeme2009}.  Consequently, these provide a unique laboratory to measure physical conditions in neutral gas and build-up of metals associated with galaxies out to $z\sim6$.  
Typical ionization structure of DLAs at $z\gtrsim1.8$, studied at high-resolution at large optical telescopes, indicates that DLAs predominantly trace WNM. While \CII\ fine-structure lines are ubiquitously detected in DLAs \citep{Wolfe2003} they may arise from both WNM and CNM \citep{Balashev2022}. The more suitable CNM tracers, such as \CI\ and H$_2$ have been detected in DLAs with a fairly low ($\lesssim 10$\%) detection rate  \cite[][]{Noterdaeme2008, Jorgenson2010, Balashev2018}.
While detailed analysis of excitation of their levels provided precise constraints on the physical conditions \citep{Balashev2019, Klimenko2020}, they trace, unlike the 21-cm absorption line, only the part of CNM associated with molecular hydrogen \citep{Balashev2024}.

The \HI\ 21-cm absorption line measurements of DLAs towards radio bright quasars have yielded detection rates (10-20\%) similar to the H$_2$ searches \citep[][]{Srianand2012, Kanekar2014b}.  
The lack of 21-cm absorption in DLAs, even when H$_2$ is detected, suggests that the H$_2$ components seen in DLAs are compact ($\le$15\,pc) and contain only a small fraction ($\le$10 per cent) of the DLA's total $N$(\HI). 
An anti-correlation observed between the spin temperature, which represents the column-density weighted mean of the spin temperatures along the sight line,  and gas metallicity  suggests that the higher WNM fraction implied by higher $T_{\rm s}$ values is related to fewer gas cooling routes \citep[][]{Kanekar2009Ts}.  The spin temperature measurements from these observations independently confirm that DLAs at $z>2$ are primarily tracing the WNM.  DLAs at lower redshifts show lower spin temperatures implying an evolution in CNM fraction with cosmic time \citep[][]{Kanekar2014b}.       
%

At $z\lesssim2$ Ly$\alpha$ falls below the atmospheric cut-off and the DLA samples are thus small \citep[][]{Neeleman2016}, owing to the need for space-based UV observations and the lower incidence of DLAs at these redshifts. It has been shown using Hubble Space Telescope (HST) spectroscopic observations that about 36\% of Mg~{\sc ii} absorbers with rest equivalent widths, W$_r$(Mg~{\sc ii}$\lambda$2796) $>$ 0.5\AA\ and W$_r$(Fe~{\sc ii}$\lambda$2600) $>$ 0.5\AA\ are DLAs \citep[][]{Rao2006}.  Specifically, absorbers with W$_r$(Mg~{\sc ii}$\lambda$2796) $>$ 1\AA\ typically have $N$(\HI) $\sim$ 10$^{19-20}$\,cm$^{-2}$.  
Consequently, at $0.2<z<1.8$ primarily Mg~{\sc ii} absorbers with large equivalent widths have been used to pre-select sight lines to search for \HI\ 21-cm absorption. 
A 21-cm absorption detection rate of 10-20\% is achieved in these studies \citep[][]{Gupta2009, Kanekar2009, Curran2010}, with higher detection rates for absorbers with metal line ratios implying higher \HI\ column density \citep[][]{Gupta2012, Dutta2017}.  As further discussed below, there is little evidence of the evolution of CNM fraction with redshift revealed through the studies based on Mg~{\sc ii} absorption.  
%

The major boost in 21-cm absorption line studies described above has been due to the large spectroscopic catalogues from the SDSS \citep[][]{York2000}, and the availability of suitable low-frequency coverage ($<$1\,GHz) at the Green Bank Telescope (GBT), the Giant Metrewave Telescope (GMRT) and the Westerbork Synthesis Radio Telescope (WSRT) in the 2000s.  
However, cold gas as traced by \HI\ 21-cm absorption is likely to be metal-rich and associated with dust, as metals and dust promote efficient cooling into the cold neutral phase.  The dust in the absorbing gas introduces a bias against quasars with dust-rich absorption systems in surveys such as SDSS \citep[][]{Krogager2015, Krogager2016, Fall1989, Ellison2008, Pontzen2009, Fynbo2013}.  While, in principle, dust-bias can be corrected statistically \citep[][]{Krogager2019}, the inhomogeneous optical and infrared colour criteria used to select optical spectroscopic targets still make this practically impossible.
Different pre-selection methods based on the presence of a DLA or a Mg~{\sc ii} absorber, depending on the redshift range, further complicate the interpretation of the 21-cm absorption detection rate.  This limitation can be naturally overcome through large, untargetted 21-cm absorption line surveys enabled by wide frequency coverage and large instantaneous bandwidth offered by SKAO. An early ASKAP search at $0.4<z<1$ reported in \citet[][]{Sadler2020} and \citet[][]{Yoon2025} provides a first glimpse at the population of intervening \HI\ absorption lines that future surveys should reveal in large numbers \citep[see also][for initial FAST results at $z<0.4$]{Hu2025}.

\begin{figure}[ht]
    \includegraphics[clip,width=1.0\textwidth]{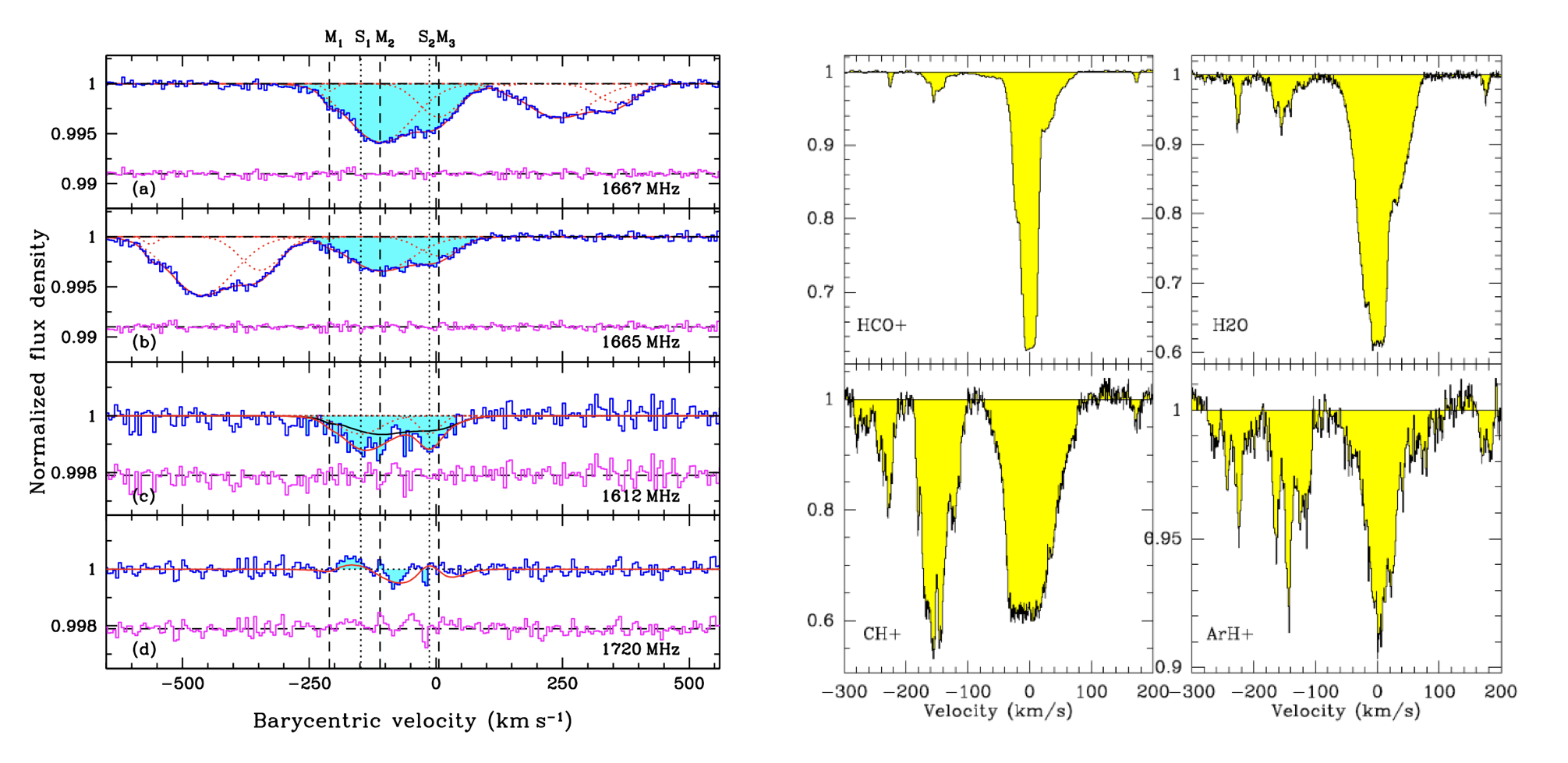}
    
    \caption{{\it Left:} Zoomed-in plot of OH-main and satellite lines detected in the MeerKAT UHF-band spectrum  {\it Right:} Corresponding molecular absorption lines obtained with ALMA tracing denser gas.
    The differences in the absorption line profiles of different species are due to gas physics, structure and the frequency-dependent structure of the background radio source over 0.5 - 200\,GHz \citep[details and full L- and UHF-band spectra covering 580 -- 1670\,MHz are given in][]{Gupta2021artip, Combes2021}.
    }
    \label{fig:pks1830}
\end{figure}

In the near future, FASHI, FLASH and MALS will provide first constraints on the evolution of CNM covering factor at $z<1.4$ \citep[e.g.][]{allison2021}. Various ongoing intensity mapping experiments may extend these to higher redshifts ($z\sim2.5$).
Fig.~\ref{fig:pks1830} shows zoomed-in MeerKAT UHF-band spectra of the quasar PKS \,1830-211 ($z_{em}$ = 2.1). Instantaneous wideband coverage, covering 580 - 1015\,MHz as shown in \citet{Combes2021}, also enables us to simultaneously search for the OH main and satellite lines in absorption. 

OH lines can provide independent constraints on the evolution of H$_2$-bearing CNM fraction, with first constraints coming from MALS, and eventually SKA \citep[][]{Balashev2021}. Moreover, ISM models and observations indicate that OH provides a complementary view between \HI\ and CO, as tracers of CNM. In particular, OH traces diffuse molecular gas \citep[CO-dark gas, i.e. not traced by CO;][]{Rugel2025}, and therefore does not suffer from the \HI\ depletion towards the center of the galaxies caused by \HI/H$_2$ transition at high pressure. Consequently, in disk models, the OH radical remains distributed in a thick plane, like the \HI, that flares with galactocentric radius. This is reflected in the broad velocity extent of the \HI\ and OH absorption lines in comparison to dense gas tracers detected at mm wavelengths \citep[e.g.][]{Combes2021}.

\begin{figure}[ht]
    \includegraphics[clip,width=1.0\textwidth]{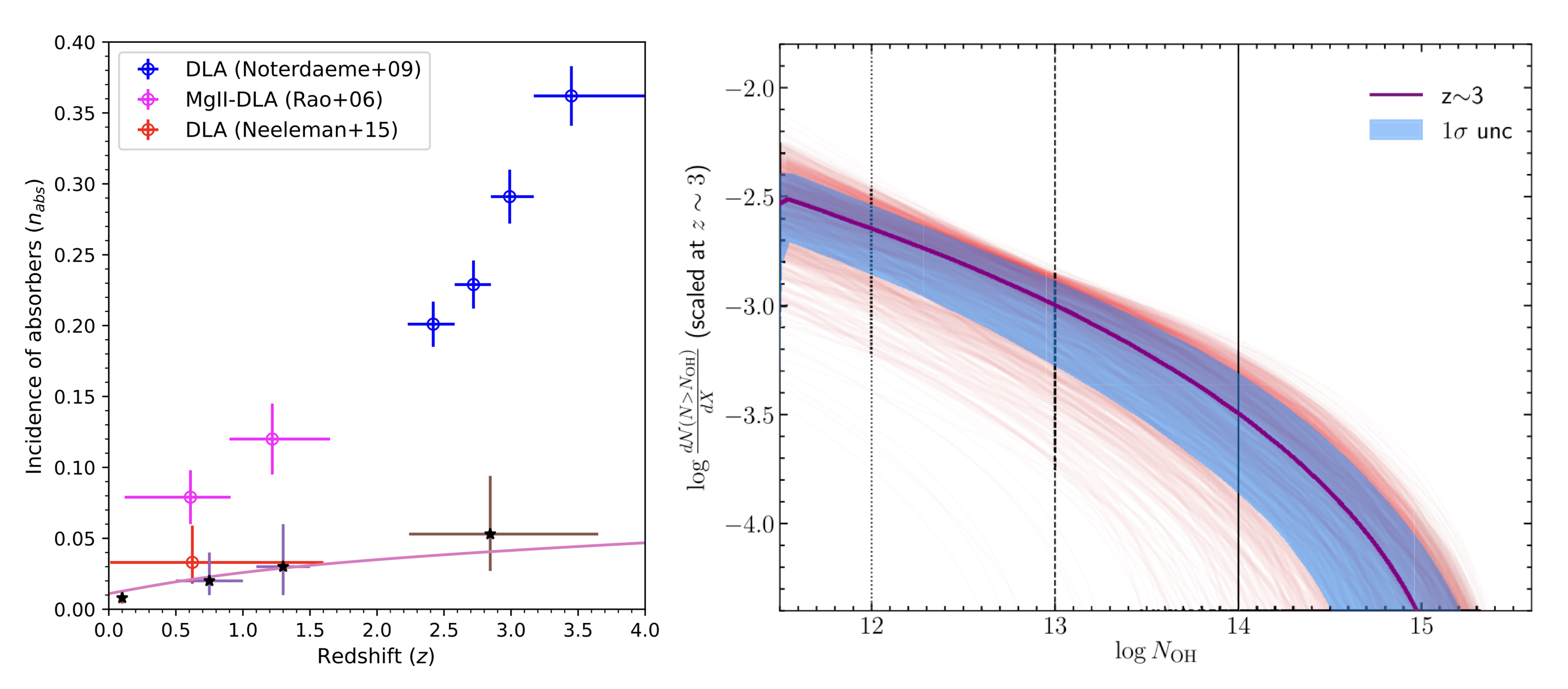}    
    \caption{{\it Left:} Number of DLAs (open circles) and 21-cm absorbers (stars) per unit redshift range as a function of redshift.  The incidences of 21-cm absorbers are estimated for optical depth cut-off, ${\cal{T}_{\rm o}}$ = 0.3\,km\,s$^{-1}$ based on DLAs and Mg~{\sc ii} systems at $2<z<3.5$ and $0.5<z<1.3$, respectively \citep[][]{Gupta2012} and nearby galaxies at $z\sim0.1$ \citep[][]{Dutta2017qgp}. The curve for non-evolving population of 21-cm absorbers normalized at $z$ = 1.3 is also plotted. Note that ${\cal{T}_{\rm o}}$ = 0.3\,km\,s$^{-1}$ corresponds to $N(\HI)$ $\approx 5\times10^{19}$ for an assumed T$_{\rm s}$ = 100\,K.
    {\it Right:} The expected incidence rate of absorption systems with $N > N(\rm OH)$ as a function of $N(\rm OH)$. The violet line indicates the averaged incidence rate, while the blue region represents 0.68 confidence region, based on the distribution (shown by red lines), calculated assuming a range of expected physical conditions and constrained H$_2$ column density distribution, reproduced from \citet[][]{Balashev2021}.
    }
    \label{fig:dndz}
\end{figure}

The most crucial metric for any absorption line survey is the sensitivity function, $g({\cal{T}}, z)$, representing the number of spectra in which it is possible to detect absorption of a given strength i.e. integrated optical depth (${\cal{T}}$)  at a given redshift ($z$).
The total or integrated completeness-corrected redshift path of the survey at the $j$th ${\cal{T}}$-interval and $k$th redshift-interval considering all sight lines, is then given by
\begin{equation}
	\Delta z ({\cal{T}}_j) \equiv g({\cal{T}}_j) = \sum_k g({\cal{T}}_j, z_k)\Delta z_k. 
\label{eqgz3}
\end{equation}
The incidence or number of absorbers per unit redshift ($n_{abs}$) with integrated optical depth greater than some threshold (${\cal{T}}_j$) can then be estimated as 
\begin{equation}
	n_{abs} = \sum_{i=1}^{N_{abs}} \frac{1}{\Delta z({\cal{T}}_i)}, 
\label{eqgz4}
\end{equation}
where the sum extends over all the absorbers ($N_{abs}$) with ${\cal{T}}_i \ge {\cal{T}}_j$, and $g({\cal{T}}_i)$ is the redshift path over which the $i$th absorber could be detected.
This readily explains the direct dependence of absorption line survey yield on the survey speed to obtain large number of sensitive wideband spectra. 
Fig.~\ref{fig:dndz} (left panel) shows incidences of DLAs ($n_{\rm DLA}$) -- the discrepancy at $0.1<z<2$ between the measurements from DLAs from unbiased surveys \citep[][]{Neeleman2016} and Mg~{\sc ii}-based selection \citep[][]{Rao2006} is apparent. 

Fig.~\ref{fig:dndz} (left panel) also shows incidences of 21-cm absorbers ($n_{21}$) based on samples of nearby galaxies ($\sim$0.1), Mg~{\sc ii} systems ($0.5<z<1.5$) and DLAs ($2<z<3.5$).  These different pre-selection methods and large uncertainties limit the constraints on the evolution of CNM fraction.  Larger samples and uniform redshift coverage from FLASH and MALS will improve these over $0<z<1.5$, and the low-frequency coverage of the upgraded GMRT may enhance constraints at $2<z<3.5$ \citep[][]{gupta2021ApJS..255...28G}.  Independent constraints on CNM fraction at intermediate redshifts ($0.8<z<2.5$) may also come from  the absorption line searches based on intensity mapping experiments with CHIME \citep[][]{CHIME2025arXiv250611269C} and HIRAX \citep[][]{Newburgh2016}.

Currently, only four intervening OH absorbers are known at present.  These sightlines have OH column densities, log\,$N$(OH) $>$ 15.  In the Galaxy, such column densities are associated with the dense molecular ISM.  The estimated incidence rate of OH at $z\sim3$ based on constrained H$_2$ column density distribution is shown in Fig.~\ref{fig:dndz} \citep[see][for details]{Balashev2021}. Based on this, one can see that at $z\sim3$ to detect a single OH absorption at the sensitivity limits corresponding to log\,$N$(OH) $<$ 14 total absorption path lengths in excess of 1000 will be needed.  
While ongoing surveys may be able to achieve this, significant enhancement in the number of known OH absorbers will come from SKAO, especially in intermediate redshifts $z\sim 1-2$, where the CNM incidence rate is expected to be larger.
%

In addition to probing physical conditions in gas at high-$z$, the absorption lines seen in the spectra of distant QSOs can be used to place constraints on the space and time variations of different dimensionless fundamental constants of physics \citep[see][for a review]{Uzan2011}.
As \HI\ and OH absorption line frequencies depend differently on $\alpha$ -- the fine structure constant, $\mu$ -- the electron-proton mass ratio and $g_p$ -- the proton gyromagnetic ratio, relative shifts between the observed frequencies of these lines can be used to constrain the variations of these fundamental constants of physics \citep[][]{Chengalur2003}.
Because the frequency scales at radio telescopes are known to be well-defined and compared to optical / ultraviolet lines the radio lines are more sensitive to the variation of fundamental constants \citep[e.g.][]{Darling2004, Kanekar2005, Rahmani2012, Gupta2018oh, Shu2025}.  
A major limitation is the availability of suitable \HI\ and OH absorbers required for this purpose.  The next major improvement (1 part in $10^{7}$) in these constraints may come from the large samples of \HI\ and OH absorbers from wide area, untargetted surveys.

\subsection{Origin of absorbing gas}

\begin{figure}[ht]
    \includegraphics[clip,width=1.0\textwidth]{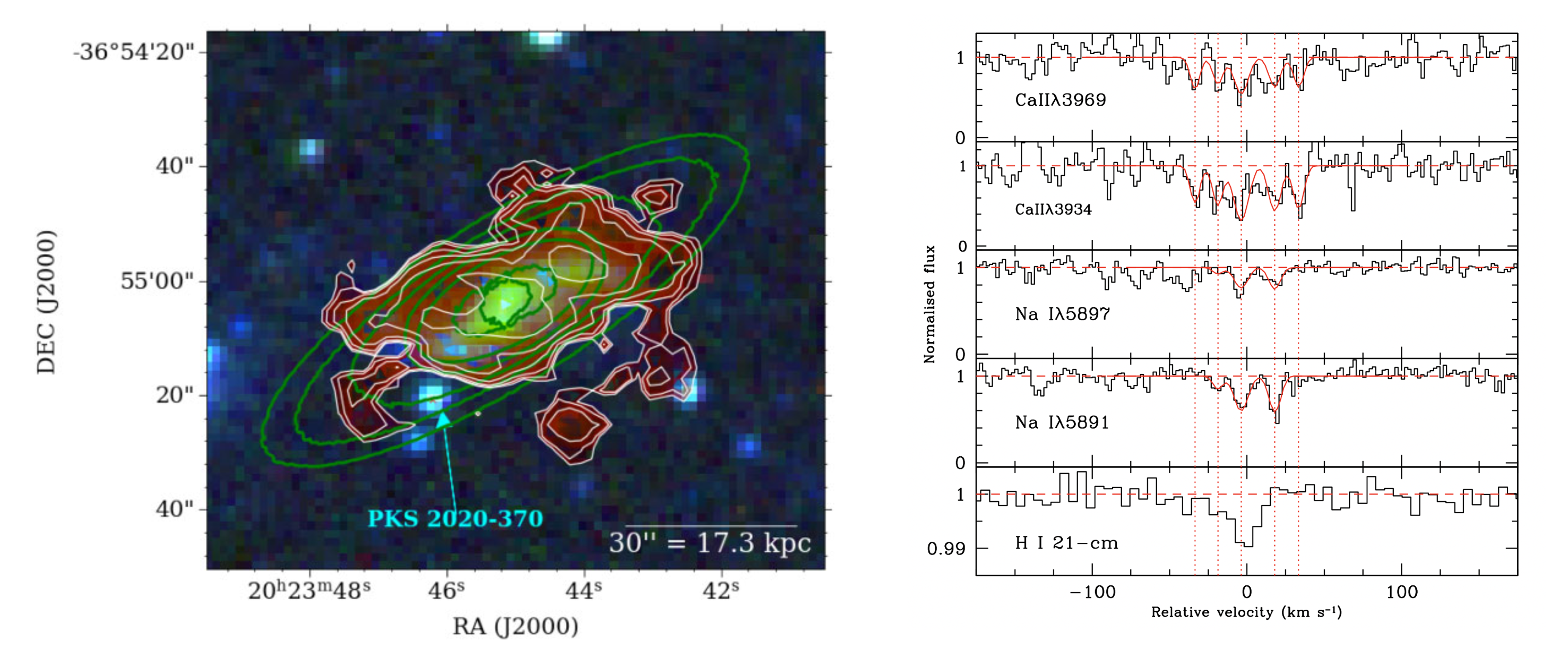}    
    \caption{
    {\it Left:} Total intensity \HI\ 21-cm emission map of Klemola31A, member of a galaxy group at $z = 0.029$, in autumn colour gradient (increasing $N$(\HI) from brown to yellow) overlaid on the RGB cutout of DSS image. Also shown are the observed (white) and TIRIFIC tilted- ring model (green) moment-0 contours, and the location of background quasar PKS\,2020-370 ($z_{em}$ = 1.048) probing the outer disk.
    {\it Right:} Comparison of the absorption profiles Ca~{\sc ii}, Na~{\sc i}, and \HI\ 21-cm absorption associated with Klemola\,31A towards the background quasar PKS\,2020-370. The best-fitting Voigt profiles (components identified with red vertical dotted lines) are overplotted. The zero velocity scale is defined at $z_{abs}$ = 0.028725. 
    The joint \HI\ emission and absorption line analysis providing constraints on the kinematics and spin temperature of absorbing, along with depletion pattern inferred from metal absorption lines, suggest an extra-planar origin of the absorbing gas \citep[see][for details]{Maina2022}.
    }
    \label{fig:pks2020}
\end{figure}

While it is well established that strong Mg~{\sc ii} absorption lines and DLAs typically arise from the gas within galaxies, the exact nature and the underlying physical process driving the gas seen in absorption is still a matter of debate and may evolve significantly with redshift \citep[e.g.][]{Fynbo2018, Neeleman2018, Kanekar2020}. 
These studies demonstrate that mergers and interactions with other galaxies, and ram pressure striping due to hot gas in group and cluster environments, also play an important role in defining the gas structures in the circumgalactic medium.
A weak anti-correlation between the \HI\ 21-cm absorption optical depth and impact parameter from galaxies has been reported based on searches in the vicinity of galaxy at $z<0.4$ \citep[][]{Gupta2013, Dutta2017qgp}.
Wide-area, untargetted surveys will provide a more complete view of the cold gas structures in and around galaxies.  Of particular interest are investigations of 21-cm absorbers and DLAs at low-redshift where \HI\ gas is also detected in emission \citep[e.g.][]{Dutta2016, Boettcher2022, Maina2022}.  Such investigations help in recognizing the galactic processes that contribute to the absorbing gas in different environments (Fig.~\ref{fig:pks2020}), and will be useful for interpreting the origin of high-$z$ absorbers.
Currently, only a handful of such investigations have been reported.  The ongoing emission-absorption surveys such as FASHI, MALS and WALLABY may provide samples for systematic exploration.

Besides enabling first steps towards understanding the absorber-galaxy relationship, low-$z$ absorption line studies have also facilitated mapping of parsec-scale structure of the absorbing gas through very long baseline interferometry (VLBI) which offers milliarcsecond (mas) resolution. The structure and the size distribution of neutral gas are required for determining the true 21-cm absorption optical depth and spin temperature \citep[][]{Briggs1983}, and relevant for understanding the processes that determine the stability of these clouds \citep[][]{Maclow2004}. 
However, the lack of suitable low-frequency receivers at existing VLBI facilities has severely limited such studies, particularly at higher redshifts. 
Presently, mas-scale spectroscopy has been possible only for a handful of cases at $z<0.1$, implying coherent gas structures larger than $>$30\,pc \citep[e.g.][]{Keeney2005, Borthakur2010, Srianand2013, Gupta2018vlbi}.
At higher redshifts, in the absence of VLBI spectroscopic capabilities, mas-scale continuum imaging has been used to put constraints on the covering factor of absorbing gas and infer spin temperatures \citep[][]{Kanekar2009vlbi, Gupta2012, Curran2013}.
The typical upper limit on the extent of radio emission from these observations is $\sim$300 pc, consistent with the lower limits on the sizes of absorbing clouds inferred from low-$z$ VLBI spectroscopic observations. While this justifies the practice of using a single covering factor to correct for the partial coverage of radio emission, large scale studies enabled by suitable low-frequency ($<$1 GHz) receivers are essential to test this explicitly.
Constraints on the parsec-scale structure of the ISM will also come from variability studies of absorption lines \citep[][]{Kanekar2001, Allison2017, Srianand2022, Combes2023}.
 Variability studies over several decades may also produce the first constraints on cosmic acceleration \citep[][]{Darling2012}. 

\subsection{The 21-cm forest}

The search for the `21-cm forest', i.e. multiple weak 21cm absorption features from the cold neutral intergalactic and circumgalactic medium along sightlines to powerful high-redshift radio sources, has been a unique and long-standing science case for the SKA \citep{carilli2002,ciardi2015}. 
Simulations suggest that a 21cm-forest signal should be detectable out to redshifts as high as $z\sim10$ \citep{ma2020}, and the frequency coverage of SKA-Low allows the \HI\ line to be traced to $z\sim27$. 
However, the most distant radio sources currently known are only at  $z\sim 7$ \citep{Banados2023}. At slightly lower redshifts, i.e. $5<z<6$, there are numerous bright compact targets with $S_{\rm 150}\gg 100\,$mJy \citep[e.g.][]{vanBreugel1999,Drouart:20,Capettie2024} which could be considered prime targets for in-situ and intervening absorption searches. We note that the fainter radio sources lying at $z>6.5$ are within the Epoch of Reionisation (EoR). As implied by observations of the Gunn-Peterson effect \citep{gunn1965} in quasar spectra at $z>6$ \citep[e.g.][]{becker2001,fan2006}, these high-redshift radio sources  could probe intervening neutral hydrogen before it is completely reionised. 
For a further discussion of this topic, we refer the reader to the EoR papers in this collection.

\subsection{Probing the multi-phase ISM}

The physical conditions in the multi-phase ISM are influenced by a number of physical processes such as metallicity, dust, in situ star formation, cosmic rays and various feedback processes.  Multi-wavelength data constraining these are crucial to understand the occurrence and the origin of gas detected in \HI\ and OH absorption.
At high redshifts, molecular absorption lines have been challenging to detect since the majority of synchrotron radio sources generally exhibit steep spectra, and the K-correction further reduces their millimeter flux, thereby reducing the number of suitable background sources.  Currently only 14 high-redshift ($0.1 < z <$ 3.5) molecular absorption line systems with \HI\ 21-cm absorption are known, evenly split between associated and intervening detections. While the derived \HI\ column densities vary from 10$^{19}$ cm$^{-2}$ to 10$^{22}$ cm$^{-2}$, the H$_2$ ones vary from 10$^{20}$ cm$^{-2}$ to 10$^{23}$ cm$^{-2}$ \citep{Combes2024}.
Naturally, larger samples of \HI\ and OH absorbers from ongoing surveys will expand these studies to investigate the chemical composition of the cold gas at high redshifts \citep[e.g.,][]{Combes2021}.

In turn, as previously noted, H$_2$ can be efficiently detected at high-$z$ in absorption using electronic transitions (UV in rest-frame), providing a complementary and detailed view of the CNM \citep{Noterdaeme2008, Srianand2010}. Interestingly, H$_2$ is found in at least the same proportion in proximate DLAs as in intervening DLAs \citep{Noterdaeme2019}, despite the efficient dissociation of H$_2$ by AGN UV radiation. The most efficient way to find H$_2$ was found to use spectroscopic survey facilities, such as SDSS, DESI and 4MOST facilities to preselect H$_2$ \citep{Balashev2014} and subsequently follow-up at high-resolution at large optical telescopes e.g. VLT and KECK, or ELT in near future. The statistical analysis of the emergent samples \citep{Noterdaeme2008, Noterdaeme2018, Balashev2019}, as well as detailed analysis of individual systems \citep{Balashev2017, Noterdaeme2017, Noterdaeme2021,Balashev2025} indicate a wide diversity of physical properties, environment and nature of the CNM gas associated with detected H$_2$. Despite the usual mismatch between the sightlines and the sizes of the optical and radio emitting regions of AGNs, 21-cm absorption provides a powerful synergy with UV–optical studies of the CNM, as it traces a larger fraction of the cold neutral gas—not only the H$_2$-bearing component.  
Multi-wavelength follow-up of large samples of \HI\ and OH absorbers to detect corresponding metal and molecular absorption lines will be essential to understand the interplay of different phases of ISM. 

%

\section{\HI\ absorption observations with SKA-AA4}

The SKAO’s sensitivity and frequency coverage will open a new discovery space for both associated and intervening \HI\ and OH absorption systems, allowing us to trace the evolution of cold gas from the nearby Universe to $z>6$, connecting the era of peak star formation and AGN activity to the earliest stages of galaxy assembly. 

\subsection{Wide-area absorption-line surveys with the SKA}

\begin{figure}[ht]
    \includegraphics[clip,width=0.9\textwidth]{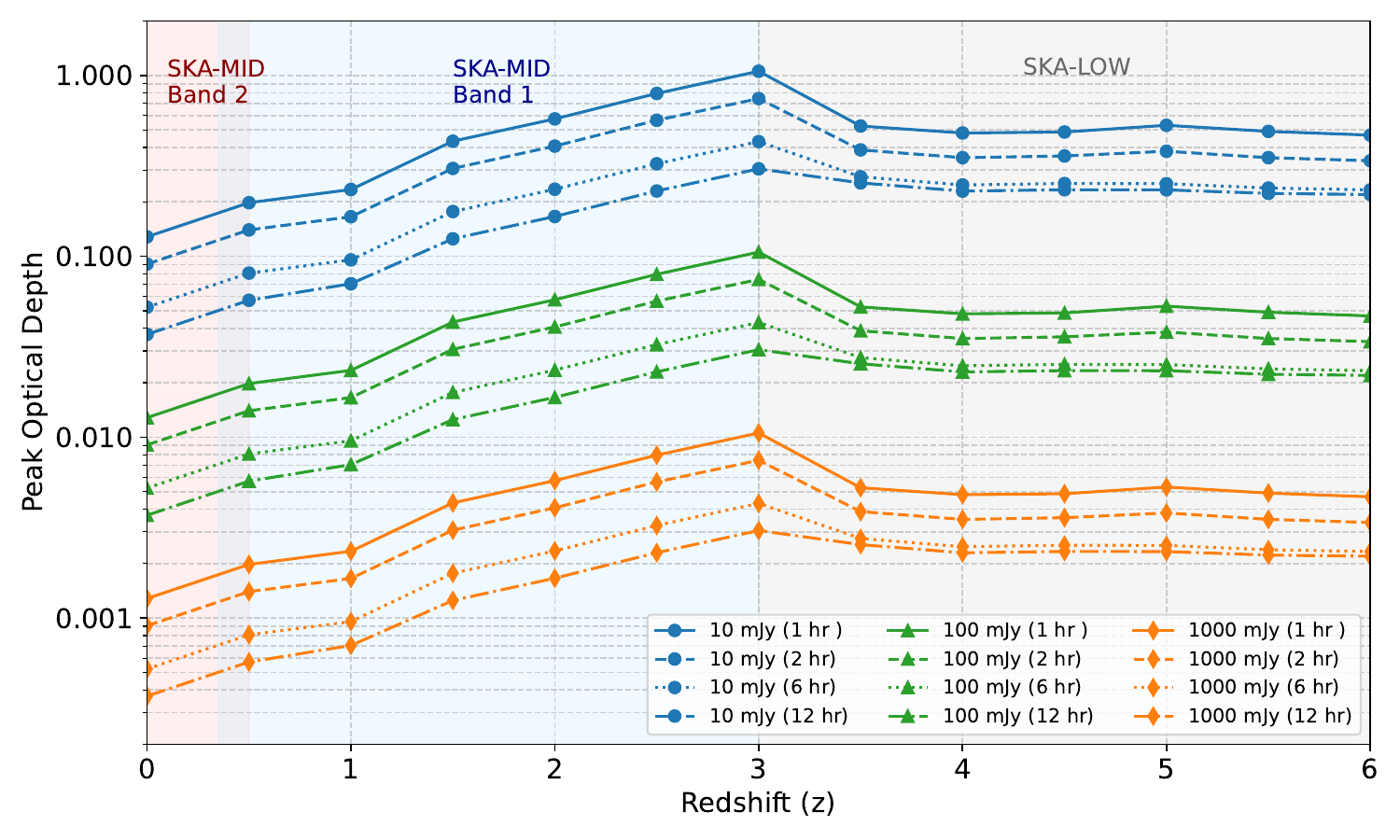}
    \caption{Peak optical depth limits reached for integration times ranging from 1 - 12hr, against a 10\,mJy, 100\,mJy and 1\,Jy background continuum source. Assumes a $5\sigma$ detection threshold and no spectral binning.} 
    \label{fig:SKA_tau}
\end{figure}

Fig. \ref{fig:SKA_tau} shows the peak optical depths achievable for a range of integration times (1–12 hrs per pointing) against background sources with flux densities of 10\,mJy, 100\,mJy, and 1\,Jy. These calculations assume a $5\sigma$ detection (per channel) and Briggs weighting with robustness 0. No spectral binning was used for these estimates meaning that binning spectrally by a factor of 2-4 could reduce these limits slightly with a trade-off in spectral resolution. These estimates assume use of the continuum mode at the default spectral resolution, however zoom modes could be used to achieve higher spectral resolution depending on the specific science aims. Sensitivity is poorest in the range $1.5<z<3$ due to the 13.5m MeerKAT antennas not being equipped with receivers that operate below 580\,MHz. Nevertheless, 1\,hr per pointing would probe peak optical depths of $\tau_{pk}>0.02$–0.1, and the wide bandwidth of SKA allows coverage of the redshift range $0.5<z<3$ in a single observation using the full 700\,MHz bandwidth.

Table~\ref{tab:ska_aa4_surveys} summarises possible SKAO absorption-line surveys, updated from \citet{Morganti2015}. Observations with SKA-Mid will explore the redshift range $0<z\lesssim3$, providing large samples of both associated and intervening 21-cm absorbers suitable for statistical studies of cold gas evolution and AGN feedback.  For OH absorbers, redshift paths in excess of a few 1000 will be essential.  At higher redshifts, SKA-Low will extend these studies into the largely unexplored regime $3<z<6$, probing cold gas in the very early stages of galaxy evolution.

\begin{table*}
\centering
\caption{Approximate peak optical depth sensitivity and spectral resolution for representative SKA \HI\ absorption surveys using the default spectral resolution in continuum mode. This assumes a 5$\sigma$ detection against background sources of 10, 30 and 100\,mJy for comparison. Sightline numbers assume a $10^4\,{\rm deg^2}$ survey and are estimated from the RACS-low and LOTSS radio source counts.}
\label{tab:ska_aa4_surveys}
\begin{tabular}{ccccccc}
\hline
Survey & Redshift & Source flux & $\tau_{\rm pk}$ limit & Spatial res. & Spectral res. & Sightlines \\
 &  & (mJy) &  & (arcsec) & (km\,s$^{-1}$) & ($10^4$ deg$^2$) \\
\hline

SKA-Mid & $0<z<3$ & 100 & $\sim0.02$--0.1 & $\sim1$--2 & 3--11 & $\sim28000$ \\
AA*             &         & 30  & $\sim0.07$--0.5 &            &       & $\sim90000$ \\

SKA-Mid & $0<z<3$ & 100 & $\sim0.01$--0.07 & $\sim1$--2 & 3--11 & $\sim28000$ \\
AA4              &         & 30  & $\sim0.05$--0.3  &         &       & $\sim90000$ \\
              &         & 10  & $\sim0.2$--0.7   &         &       & $\sim230000$ \\
\hline

SKA-Low & $z>3$ & 100 & $\sim0.07$ & $\sim5$--8 & 5--8 & $\sim1000$ \\
AA*             &       & 30  & $\sim0.5$ &           &     & $\sim2600$ \\

SKA-Low  & $z>3$ & 100 & $\sim0.04$ & $\sim5$--8 & 5--8 & $\sim1000$ \\
AA4          &       & 30  & $\sim0.2$   &           &     & $\sim2600$ \\
             &       & 10  & $\sim0.4$       &           &     & $\sim5500$ \\

\hline
\end{tabular}
\end{table*}

\subsection{Expected detection rates for a wide-area survey with the SKA}

Estimating the number of 21-cm absorption systems that future SKA surveys will detect remains highly uncertain, largely due to the poorly constrained redshift distribution of radio-loud AGN at high redshifts, the uncertainity on the cold-gas fraction and how this evolves with redshift. Ongoing large-area surveys with ASKAP and MeerKAT will provide critical constraints on these quantities, enabling more reliable predictions over the coming years.

Nonetheless, the optical depth limits shown in Fig.~\ref{fig:SKA_tau} allow a first-order estimate of the number of absorption systems that may be detectable, summarised in Table~\ref{tab:ska_aa4_surveys}. As a representative case, we consider a shallow, wide-area survey covering $\sim10,000~{\rm deg^2}$ with $\sim2$\,hours of integration per pointing. For SKA-Mid (Band~1, centred near 700\,MHz), this would require roughly $5,000$ pointings, given an effective field of view of $\sim2.0~{\rm deg^2}$. At these integration times, peak optical depth limits range from $\tau_{\rm pk} \approx 0.07$ at low redshift to $\tau_{\rm pk} \approx 0.5$ at $z\sim3$, using the AA* array at the native spectral resolution ($\sim4$--$11~{\rm km,s^{-1}}$). Based on radio source counts from the \emph{Rapid ASKAP Continuum Survey} (RACS-low; \citealt{Hale2021}), such a survey would include roughly 90,000 background sources with peak flux densities above 30\,mJy. The increased sensitivity of AA4 would enable searches for absorption against fainter background sources, increasing the number of sightlines to 230\,000 for sources with flux densities greater than 10\,mJy. Naturally, the strong absorber population will be detectable towards even fainter sources.

At higher redshift, a comparable $10,000~{\rm deg^2}$ survey with SKA-Low ($z>3$) would require only $\sim500$ pointings due to its much larger field of view ($\sim20~{\rm deg^2}$), corresponding to $\sim1,000$ observing hours in total at 2\,hours per pointing. Estimating the number of detectable absorption lines remains very uncertain, as the redshift distribution of radio sources at $z>3$ remains poorly constrained. However, recent results from JWST and LOFAR indicate the existence of a substantial population of high-redshift massive galaxies, many of which are known to host radio jets \citep[e.g.][]{Gloudemans2022, Nanayakkara2024, Roy2024}. Using source counts from the \emph{LOFAR Two-metre Sky Survey} (LoTSS; \citealt{Shimwell2026, Hardcastle2025}) combined with radio luminosity functions indicating that only $\sim1\%$ of radio-loud AGN lie at $z>3$, this yields roughly 5,000 sightlines over the survey area. These observations would reach optical depth limits comparable to the SKA-Mid survey at similar spectral resolution. 

The detection rate in previous \HI\ absorption surveys varies significantly, depending on the sample selection and optical depth limits reached. Adopting a conservative detection rate of 1–3\% from similar untargeted searches \citep[e.g.][]{gupta2021ApJS..255...28G,Su2022,Hu2025}, a $10,000~{\rm deg^2}$ SKA-Mid survey could yield $\sim2000$–$7000$ absorbers, while SKA-Low might detect a few hundred systems at $z>3$.  Even under conservative assumptions, SKA will increase \HI\ 21-cm absorber samples from the current few hundred to thousands across $0<z<6$, enabling statistically robust subdivision by redshift, radio luminosity, AGN type as well as studies of spin-temperature evolution, and the changing cold-gas fraction over more than 12\,Gyr of cosmic history.
As previously noted, with only a handful of known OH absorbers, the OH detection rate is even more uncertain. However, based on the incidence rates estimated by \citet{Balashev2021}, SKA-Mid surveys could plausibly detect tens of OH absorbers, providing the first statistical constraints on diffuse molecular gas across cosmic time.

\subsection{Targeted follow-up observations}

Wide-area, untargeted surveys with SKA-AA4 will identify large samples of \HI\ and OH absorption systems, revealing galaxies selected purely by their cold gas content. These surveys, combined with other selection criteria, can provide candidates for deeper, targeted follow-up observations designed to probe weaker absorption features and provide detailed information on the kinematics, structure, and origin of the absorbing gas, and even test invariance of fundamental constants of physics. Signatures of outflows, for instance, are often seen at optical depths of $\tau<0.01$, requiring longer integration times to reach suitable limits, particularly at redshifts $1.5<z<3$. 

\begin{figure}[ht]
    \includegraphics[clip,width=0.9\textwidth]{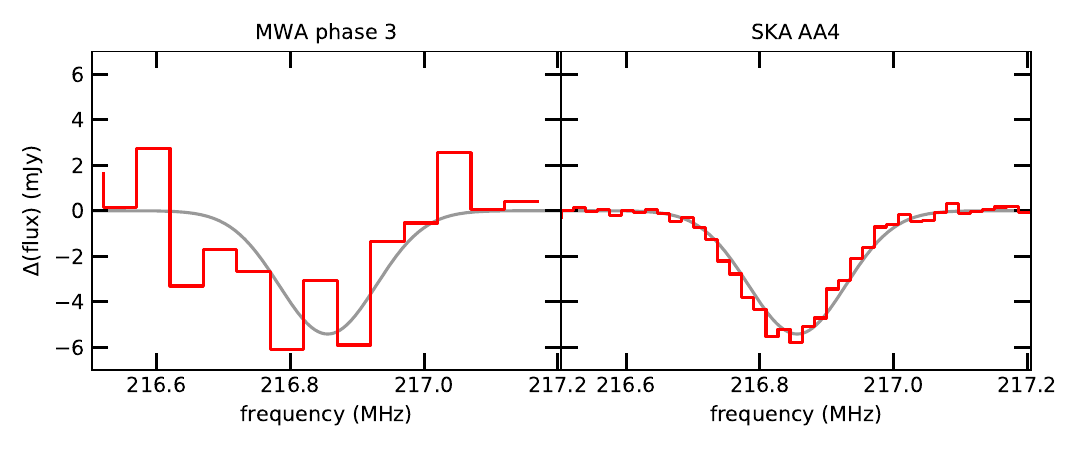}
    \caption{Simulations of 12\,hour observations of associated \HI\ absorption in GLEAM J0856$+$0224 at $z=5.55$ for a Gaussian profile line of width 100\,km\,s$^{-1}$ and maximum depth of $\tau=0.01$ (input spectrum shown by the grey line). The left panel shows what would be observed with MWA Phase 3 with an RMS of 13.4\,mJy per 200\,kHz channel, and the right panel shows SKA-Low AA4 with an RMS of 2.2\,mJy per 18\,kHz channel.} 
    \label{fig:sim}
\end{figure}

At the highest redshifts ($z>3$), the 21-cm \HI\ line will shift into the SKA-Low frequency range. As an example we look at the detectability of associated \HI\ absorption in one of the most luminous radio galaxies at $5<z<6$; GLEAM J0856$+$0224 \citep[at $z=5.55$ and with $S_{\rm 150MHz}\approx 0.9\,$Jy, ][]{Drouart:20}. Fig. ~\ref{fig:sim} shows a simulated \HI\ absorption line observation with MWA phase 3 and SKA-AA4. This assumes a Gaussian absorption profile of width 100\,km\,s$^{-1}$ and a maximum peak optical depth of $\tau_{pk}=0.01$. For the MWA we assume 256 tiles, overlapping coarse channels to provide contiguous spectral coverage and an rms noise of 13.4\,mJy per 200\,kHz channel from 12\,hours integration. For SKA-Low we use the same exposure time and assume all observations are at high elevation with SKA-Low AA4 providing us with an RMS of 2.2\,mJy per 18\,kHz channel. We see a clear detection with SKA-Low and a marginal one at best with MWA.   

The balance between wide-area, untargeted surveys and deeper, targeted follow-up observations will be key to advancing our understanding of \HI\ absorption in the SKA era. Untargeted surveys will provide large, unbiased samples of \HI\ -rich galaxies and identify candidates that can then be followed up with deeper, pointed observations. The optimal balance between these approaches will be informed by the ongoing SKA precursor surveys, which are already demonstrating the complementary nature of large survey programmes that efficiently uncover new absorbers, and deeper targeted observations that reveal more detailed gas kinematics. The frequency coverage of SKAO will enable us to push the study of \HI\ to redshifts beyond $z>3$, tracing the evolution of cold neutral gas over more than 10 billion years of cosmic history.

\section{Conclusions}

Over the past decade, \HI\ absorption studies have progressed from small, heterogeneous samples to large, systematic surveys with SKA precursors, revealing new absorber populations and expanding our view of cold gas in galaxies.  Instantaneous wide bandwidths for the first time are enabling simultaneous searches of OH absorption, capable of delivering independent constraints on H$_2$ gas fraction and excitation conditions. 

\subsection{Science outcomes}

Observations of the \HI\ 21-cm and OH 18-cm lines in absorption with the SKAO will provide a statistical census of these absorbers across cosmic time, tracing both associated and intervening systems from $z=0$ to $z>6$, including the 21-cm forest. These surveys will allow us to explore the following key science goals and will provide crucial observational constraints that can inform galaxy evolution simulations.

\begin{enumerate}
    \item Identification and characterisation of neutral gas outflows at all redshifts, but especially at ``cosmic noon'' ($z\sim1$--2), providing crucial constraints on AGN feedback and its impact on galaxy evolution. 
    \item Measurements of the CNM and H$_2$ (OH) fraction revealing processes associated with conversion of \HI\ gas to cold gas and eventually stars over cosmic time. 
    \item Discovery of new populations of gas-rich, optically faint or dust-obscured associated and intervening absorbers, to resolve the long-standing question of expected dust-bias in optically-selected samples.
    \item The most stringent constraints on variations of fundamental constants of physics, and first constraints on cosmic acceleration.
    \item Through sub-arcsecond scale spectroscopy (SKA-VLBI), constraints on parsec-scale structure of cold neutral gas in normal and active galaxies, and its relationship with stellar and AGN feedback processes. 
\end{enumerate}

\subsection{Challenges and lessons learned from SKA precursors}

Experience from precursor surveys provides essential guidance for planning SKA-AA4 observations. These surveys have highlighted, and in many cases helped overcome, several challenges that must be considered when designing absorption-line surveys with the SKA. Radio-frequency interference (RFI) remains a key concern. Even though both SKA telescopes are built on radio-quiet sites, the presence of RFI can still impact observations. This can either be due to aircraft or satellites overhead, or via tropospheric ducting which can transport radio waves across large distances. While affected data or contaminated frequency ranges can be flagged, this results in a loss of redshift coverage for spectral-line surveys that cannot be recovered. 

Bandpass stability is another key issue for spectral line surveys. The need for a flat, stable bandpass over wide frequency range is critical to achieve the science goals discussed in this chapter, particular for the detection of weak, broad absorption features, and exploration of line variability. Lastly, spectral-line observations produce extremely large datasets, making automated pipelines for line extraction, data verification and quality control, and stacking essential to fully exploit the data from SKAO absorption surveys. 
These challenges are already being addressed by the large teams conducting current absorption-line surveys, but more advanced and efficient methods will be needed to tackle more sensitive and larger data volumes from SKAO.

\subsection{Synergies with multi-wavelength data}

Maximising the scientific return of \HI\ absorption studies with the SKAO will depend on complementary multi-wavelength data and high-resolution follow-up observations. While the SKAO will provide unparalleled sensitivity for detecting the 21-cm and OH lines across cosmic time, interpreting these signals requires optical and infrared photometry and spectroscopy to secure redshifts for distinguishing between associated and intervening absorbers, and performing stacking analyses.  
Large optical and infrared spectroscopic surveys such as DESI \citep[][]{Desi2024} and 4MOST \citep[][]{4most2021} will provide large number of spectra to distinguish between associated and intervening absorbers, as well as information on stellar populations, AGN diagnostics and gas physics in absorber host galaxies.
Detailed follow-up observations with existing and planned facilities such as ALMA and ngVLA in radio, and JWST, VLT, KECK and ELTs in optical / IR will allow characterization of individual systems and their multi-phase ISM in greater depth. %
Together, these facilities will provide a comprehensive, multi-phase picture of the gas, dust, AGN and star-formation processes shaping galaxy evolution. 

\newpage

\bibliographystyle{abbrvnat-maxbibnames4}
\bibliography{chapter} 

@ARTICLE{Kanekar2003,
       author = {{Kanekar}, Nissim and {Chengalur}, Jayaram N.},
        title = "{A deep search for 21-cm absorption in high redshift damped Lyman-alpha  systems}",
      journal = {\aap},
         year = 2003,
        month = mar,
       volume = {399},
        pages = {857-868},
          doi = {10.1051/0004-6361:20021922},
archivePrefix = {arXiv},
       eprint = {astro-ph/0211637},
 primaryClass = {astro-ph},
       adsurl = {https://ui.adsabs.harvard.edu/abs/2003A&A...399..857K}
}

@ARTICLE{Mcclure-Griffiths2023,
       author = {{McClure-Griffiths}, Naomi M. and {Stanimirovi{\'c}}, Sne{\v{z}}ana and {Rybarczyk}, Daniel R.},
        title = "{Atomic Hydrogen in the Milky Way: A Stepping Stone in the Evolution of Galaxies}",
      journal = {\araa},
         year = 2023,
        month = aug,
       volume = {61},
        pages = {19-63},
          doi = {10.1146/annurev-astro-052920-104851},
archivePrefix = {arXiv},
       eprint = {2307.08464},
 primaryClass = {astro-ph.GA},
       adsurl = {https://ui.adsabs.harvard.edu/abs/2023ARA&A..61...19M}
}

@ARTICLE{Dawson2022,
       author = {{Dawson}, J.~R. and {Jones}, P.~A. and {Purcell}, C. and {Walsh}, A.~J. and {Breen}, S.~L. and {Brown}, C. and {Carretti}, E. and {Cunningham}, M.~R. and {Dickey}, J.~M. and {Ellingsen}, S.~P. and {Gibson}, S.~J. and {G{\'o}mez}, J.~F. and {Green}, J.~A. and {Imai}, H. and {Krishnan}, V. and {Lo}, N. and {Lowe}, V. and {Marquarding}, M. and {McClure-Griffiths}, N.~M.},
        title = "{SPLASH: the Southern Parkes Large-Area Survey in Hydroxyl - data description and release}",
      journal = {\mnras},
         year = 2022,
        month = may,
       volume = {512},
       number = {3},
        pages = {3345-3364},
          doi = {10.1093/mnras/stac636},
archivePrefix = {arXiv},
       eprint = {2203.05131},
 primaryClass = {astro-ph.GA},
       adsurl = {https://ui.adsabs.harvard.edu/abs/2022MNRAS.512.3345D}
}

@BOOK{Draine2011,
       author = {{Draine}, Bruce T.},
        title = "{Physics of the Interstellar and Intergalactic Medium}",
         year = 2011,
       adsurl = {https://ui.adsabs.harvard.edu/abs/2011piim.book.....D}
}

@ARTICLE{Noterdaeme2008,
       author = {{Noterdaeme}, P. and {Ledoux}, C. and {Petitjean}, P. and {Srianand}, R.},
        title = "{Molecular hydrogen in high-redshift damped Lyman-{\ensuremath{\alpha}} systems: the VLT/UVES database}",
      journal = {\aap},
         year = 2008,
        month = apr,
       volume = {481},
       number = {2},
        pages = {327-336},
          doi = {10.1051/0004-6361:20078780},
archivePrefix = {arXiv},
       eprint = {0801.3682},
 primaryClass = {astro-ph},
       adsurl = {https://ui.adsabs.harvard.edu/abs/2008A&A...481..327N}
}

@ARTICLE{Briggs1983,
       author = {{Briggs}, F.~H. and {Wolfe}, A.~M.},
        title = "{The incidence of 21 centimeter absorption in QSO redshift systems selected for MG II absorption : evidence for a two-phase nature of the absorbing gas.}",
      journal = {\apj},
         year = 1983,
        month = may,
       volume = {268},
        pages = {76-89},
          doi = {10.1086/160931},
       adsurl = {https://ui.adsabs.harvard.edu/abs/1983ApJ...268...76B}
}

@PHDTHESIS{Lane2000,
       author = {{Lane}, Wendy Meredith},
        title = "{HI 21cm absorbers at moderate redshifts}",
       school = {University of Groningen, Netherlands},
         year = 2000,
        month = sep,
       adsurl = {https://ui.adsabs.harvard.edu/abs/2000PhDT.......283L}
}

@ARTICLE{Curran2010,
       author = {{Curran}, S.~J.},
        title = "{On the detectability of HI 21cm in MgII absorption systems}",
      journal = {\mnras},
         year = 2010,
        month = mar,
       volume = {402},
       number = {4},
        pages = {2657-2665},
          doi = {10.1111/j.1365-2966.2009.16080.x},
archivePrefix = {arXiv},
       eprint = {0910.3998},
 primaryClass = {astro-ph.CO},
       adsurl = {https://ui.adsabs.harvard.edu/abs/2010MNRAS.402.2657C}
}

@ARTICLE{Gupta2009,
       author = {{Gupta}, N. and {Srianand}, R. and {Petitjean}, P. and {Noterdaeme}, P. and {Saikia}, D.~J.},
        title = "{A complete sample of 21-cm absorbers at z \raisebox{-0.5ex}\textasciitilde 1.3: Giant Metrewave Radio Telescope survey using MgII systems}",
      journal = {\mnras},
         year = 2009,
        month = sep,
       volume = {398},
       number = {1},
        pages = {201-220},
          doi = {10.1111/j.1365-2966.2009.14933.x},
archivePrefix = {arXiv},
       eprint = {0904.2878},
 primaryClass = {astro-ph.CO},
       adsurl = {https://ui.adsabs.harvard.edu/abs/2009MNRAS.398..201G}
}

@ARTICLE{Kanekar2009,
       author = {{Kanekar}, N. and {Prochaska}, J.~X. and {Ellison}, S.~L. and {Chengalur}, J.~N.},
        title = "{A search for HI 21cm absorption in strong MgII absorbers in the redshift desert}",
      journal = {\mnras},
         year = 2009,
        month = jun,
       volume = {396},
       number = {1},
        pages = {385-401},
          doi = {10.1111/j.1365-2966.2009.14661.x},
archivePrefix = {arXiv},
       eprint = {0903.4487},
 primaryClass = {astro-ph.CO},
       adsurl = {https://ui.adsabs.harvard.edu/abs/2009MNRAS.396..385K}
}

@ARTICLE{Gupta2012,
       author = {{Gupta}, N. and {Srianand}, R. and {Petitjean}, P. and {Bergeron}, J. and {Noterdaeme}, P. and {Muzahid}, S.},
        title = "{Search for cold gas in strong Mg II absorbers at 0.5 < z < 1.5: nature and evolution of 21-cm absorbers}",
      journal = {\aap},
         year = 2012,
        month = aug,
       volume = {544},
          eid = {A21},
        pages = {A21},
          doi = {10.1051/0004-6361/201219159},
archivePrefix = {arXiv},
       eprint = {1205.4029},
 primaryClass = {astro-ph.CO},
       adsurl = {https://ui.adsabs.harvard.edu/abs/2012A&A...544A..21G}
}

@ARTICLE{Dutta2017,
       author = {{Dutta}, R. and {Srianand}, R. and {Gupta}, N. and {Joshi}, R.},
        title = "{H I 21-cm absorption from z {\ensuremath{\sim}} 0.35 strong Mg II absorbers}",
      journal = {\mnras},
         year = 2017,
        month = jun,
       volume = {468},
       number = {1},
        pages = {1029-1037},
          doi = {10.1093/mnras/stx538},
archivePrefix = {arXiv},
       eprint = {1703.00457},
 primaryClass = {astro-ph.GA},
       adsurl = {https://ui.adsabs.harvard.edu/abs/2017MNRAS.468.1029D}
}

@ARTICLE{Gupta2021artip,
       author = {{Gupta}, N. and {Jagannathan}, P. and {Srianand}, R. and {Bhatnagar}, S. and {Noterdaeme}, P. and {Combes}, F. and {Petitjean}, P. and {Jose}, J. and {Pandey}, S. and {Kaski}, C. and {Baker}, A.~J. and {Balashev}, S.~A. and {Boettcher}, E. and {Chen}, H.-W. and {Cress}, C. and {Dutta}, R. and {Goedhart}, S. and {Heald}, G. and {J{\'o}zsa}, G.~I.~G. and {Kamau}, E. and {Kamphuis}, P. and {Kerp}, J. and {Kl{\"o}ckner}, H.-R. and {Knowles}, K. and {Krishnan}, V. and {Krogager}, J.-. K. and {Kulkarni}, V.~P. and {Momjian}, E. and {Moodley}, K. and {Passmoor}, S. and {Schr{\"o}eder}, A. and {Sekhar}, S. and {Sikhosana}, S. and {Wagenveld}, J. and {Wong}, O.~I.},
        title = "{Blind H I and OH Absorption Line Search: First Results with MALS and uGMRT Processed Using ARTIP}",
      journal = {\apj},
         year = 2021,
        month = jan,
       volume = {907},
       number = {1},
          eid = {11},
        pages = {11},
          doi = {10.3847/1538-4357/abcb85},
archivePrefix = {arXiv},
       eprint = {2007.04347},
 primaryClass = {astro-ph.GA},
       adsurl = {https://ui.adsabs.harvard.edu/abs/2021ApJ...907...11G}
}

@ARTICLE{Darling2004,
       author = {{Darling}, Jeremy},
        title = "{A Laboratory for Constraining Cosmic Evolution of the Fine-Structure Constant: Conjugate 18 Centimeter OH Lines toward PKS 1413+135 at z = 0.24671}",
      journal = {\apj},
         year = 2004,
        month = sep,
       volume = {612},
       number = {1},
        pages = {58-63},
          doi = {10.1086/422450},
archivePrefix = {arXiv},
       eprint = {astro-ph/0405240},
 primaryClass = {astro-ph},
       adsurl = {https://ui.adsabs.harvard.edu/abs/2004ApJ...612...58D}
}

@ARTICLE{Kanekar2005,
       author = {{Kanekar}, N. and {Carilli}, C.~L. and {Langston}, G.~I. and {Rocha}, G. and {Combes}, F. and {Subrahmanyan}, R. and {Stocke}, J.~T. and {Menten}, K.~M. and {Briggs}, F.~H. and {Wiklind}, T.},
        title = "{Constraints on Changes in Fundamental Constants from a Cosmologically Distant OH Absorber or Emitter}",
      journal = {\prl},
         year = 2005,
        month = dec,
       volume = {95},
       number = {26},
          eid = {261301},
        pages = {261301},
          doi = {10.1103/PhysRevLett.95.261301},
archivePrefix = {arXiv},
       eprint = {astro-ph/0510760},
 primaryClass = {astro-ph},
       adsurl = {https://ui.adsabs.harvard.edu/abs/2005PhRvL..95z1301K}
}

@ARTICLE{Rahmani2012,
       author = {{Rahmani}, H. and {Srianand}, R. and {Gupta}, N. and {Petitjean}, P. and {Noterdaeme}, P. and {V{\'a}squez}, D. Albornoz},
        title = "{Constraining the variation of fundamental constants at z {\ensuremath{\sim}} 1.3 using 21-cm absorbers}",
      journal = {\mnras},
         year = 2012,
        month = sep,
       volume = {425},
       number = {1},
        pages = {556-576},
          doi = {10.1111/j.1365-2966.2012.21503.x},
archivePrefix = {arXiv},
       eprint = {1206.2653},
 primaryClass = {astro-ph.CO},
       adsurl = {https://ui.adsabs.harvard.edu/abs/2012MNRAS.425..556R}
}

@ARTICLE{Darling2012,
       author = {{Darling}, Jeremy},
        title = "{Toward a Direct Measurement of the Cosmic Acceleration}",
      journal = {\apjl},
         year = 2012,
        month = dec,
       volume = {761},
       number = {2},
          eid = {L26},
        pages = {L26},
          doi = {10.1088/2041-8205/761/2/L26},
archivePrefix = {arXiv},
       eprint = {1211.4585},
 primaryClass = {astro-ph.CO},
       adsurl = {https://ui.adsabs.harvard.edu/abs/2012ApJ...761L..26D}
}

@article{Maclow2004,
  title = {Control of star formation by supersonic turbulence},
  author = {Mac Low, Mordecai-Mark and Klessen, Ralf S.},
  journal = {Rev. Mod. Phys.},
  volume = {76},
  issue = {1},
  pages = {125--194},
  numpages = {0},
  year = {2004},
  month = {Jan},
  publisher = {American Physical Society},
  doi = {10.1103/RevModPhys.76.125},
  url = {https://link.aps.org/doi/10.1103/RevModPhys.76.125}
}

@ARTICLE{Srianand2013,
       author = {{Srianand}, R. and {Gupta}, N. and {Rahmani}, H. and {Momjian}, E. and {Petitjean}, P. and {Noterdaeme}, P.},
        title = "{Parsec-scale structures and diffuse bands in a translucent interstellar medium at z≃ 0.079}",
      journal = {\mnras},
         year = 2013,
        month = jan,
       volume = {428},
       number = {3},
        pages = {2198-2206},
          doi = {10.1093/mnras/sts190},
archivePrefix = {arXiv},
       eprint = {1210.3036},
 primaryClass = {astro-ph.CO},
       adsurl = {https://ui.adsabs.harvard.edu/abs/2013MNRAS.428.2198S}
}

@ARTICLE{Gupta2006sur,
       author = {{Gupta}, Neeraj and {Salter}, C.~J. and {Saikia}, D.~J. and {Ghosh}, T. and {Jeyakumar}, S.},
        title = "{Probing radio source environments via HI and OH absorption}",
      journal = {\mnras},
         year = 2006,
        month = dec,
       volume = {373},
       number = {3},
        pages = {972-992},
          doi = {10.1111/j.1365-2966.2006.11064.x},
archivePrefix = {arXiv},
       eprint = {astro-ph/0605423},
 primaryClass = {astro-ph},
       adsurl = {https://ui.adsabs.harvard.edu/abs/2006MNRAS.373..972G}
}

@ARTICLE{Gupta2018oh,
       author = {{Gupta}, N. and {Momjian}, E. and {Srianand}, R. and {Petitjean}, P. and {Noterdaeme}, P. and {Gyanchandani}, D. and {Sharma}, R. and {Kulkarni}, S.},
        title = "{Discovery of OH Absorption from a Galaxy at z {\ensuremath{\sim}} 0.05: Implications for Large Surveys with SKA Pathfinders}",
      journal = {\apjl},
         year = 2018,
        month = jun,
       volume = {860},
       number = {2},
          eid = {L22},
        pages = {L22},
          doi = {10.3847/2041-8213/aac9cd},
archivePrefix = {arXiv},
       eprint = {1806.00172},
 primaryClass = {astro-ph.GA},
       adsurl = {https://ui.adsabs.harvard.edu/abs/2018ApJ...860L..22G}
}

@ARTICLE{Madau2014,
       author = {{Madau}, Piero and {Dickinson}, Mark},
        title = "{Cosmic Star-Formation History}",
      journal = {\araa},
         year = 2014,
        month = aug,
       volume = {52},
        pages = {415-486},
          doi = {10.1146/annurev-astro-081811-125615},
archivePrefix = {arXiv},
       eprint = {1403.0007},
 primaryClass = {astro-ph.CO},
       adsurl = {https://ui.adsabs.harvard.edu/abs/2014ARA&A..52..415M}
}

@ARTICLE{Morganti2018,
       author = {{Morganti}, Raffaella and {Oosterloo}, Tom},
        title = "{The interstellar and circumnuclear medium of active nuclei traced by H i 21 cm absorption}",
      journal = {\aapr},
         year = 2018,
        month = jul,
       volume = {26},
       number = {1},
          eid = {4},
        pages = {4},
          doi = {10.1007/s00159-018-0109-x},
archivePrefix = {arXiv},
       eprint = {1807.01475},
 primaryClass = {astro-ph.GA},
       adsurl = {https://ui.adsabs.harvard.edu/abs/2018A&ARv..26....4M}
}

@ARTICLE{Gupta2018vlbi,
       author = {{Gupta}, N. and {Srianand}, R. and {Farnes}, J.~S. and {Pidopryhora}, Y. and {Vivek}, M. and {Paragi}, Z. and {Noterdaeme}, P. and {Oosterloo}, T. and {Petitjean}, P.},
        title = "{Revealing H I gas in emission and absorption on pc to kpc scales in a galaxy at z {\ensuremath{\sim}} 0.017}",
      journal = {\mnras},
         year = 2018,
        month = may,
       volume = {476},
       number = {2},
        pages = {2432-2445},
          doi = {10.1093/mnras/sty384},
archivePrefix = {arXiv},
       eprint = {1712.03511},
 primaryClass = {astro-ph.GA},
       adsurl = {https://ui.adsabs.harvard.edu/abs/2018MNRAS.476.2432G}
}

@ARTICLE{Carilli1992qgp,
       author = {{Carilli}, C.~L. and {van Gorkom}, J.~H.},
        title = "{H i Imaging of Four Quasar-Galaxy Pairs: The Parent Galaxies of Low-Redshift Quasar Absorption Systems}",
      journal = {\apj},
         year = 1992,
        month = nov,
       volume = {399},
        pages = {373},
          doi = {10.1086/171934},
       adsurl = {https://ui.adsabs.harvard.edu/abs/1992ApJ...399..373C}
}

@ARTICLE{Gupta2010,
       author = {{Gupta}, N. and {Srianand}, R. and {Bowen}, D.~V. and {York}, D.~G. and {Wadadekar}, Y.},
        title = "{GMRT mini-survey to search for 21-cm absorption in quasar-galaxy pairs at z \raisebox{-0.5ex}\textasciitilde 0.1}",
      journal = {\mnras},
         year = 2010,
        month = oct,
       volume = {408},
       number = {2},
        pages = {849-864},
          doi = {10.1111/j.1365-2966.2010.17198.x},
archivePrefix = {arXiv},
       eprint = {1007.0288},
 primaryClass = {astro-ph.CO},
       adsurl = {https://ui.adsabs.harvard.edu/abs/2010MNRAS.408..849G}
}

@ARTICLE{dePalma2025,
       author = {{DePalma}, David and {Gupta}, Neeraj and {Chen}, Hsiao-Wen and {Simcoe}, Robert A. and {Balashev}, Sergei and {Boettcher}, Erin and {Cantalupo}, Sebastiano and {Chen}, Mandy C. and {Combes}, Fran{\c{c}}oise and {Faucher-Gigu{\`e}re}, Claude-Andr{\'e} and {Johnson}, Sean D. and {Kl{\"o}ckner}, Hans-Rainer and {Krogager}, Jens-Kristian and {Li}, Jennifer I-Hsiu and {L{\'o}pez}, Sebasti{\'a}n and {Noterdaeme}, Pasquier and {Petitjean}, Patrick and {Qu}, Zhijie and {Rudie}, Gwen C. and {Schaye}, Joop and {Zahedy}, Fakhri},
        title = "{H I Properties of Field Galaxies at $\boldsymbol{z\approx 0.2}$-0.6: Insights into Declining Cosmic Star Formation}",
      journal = {arXiv e-prints},
         year = 2025,
        month = oct,
          eid = {arXiv:2510.03400},
        pages = {arXiv:2510.03400},
          doi = {10.48550/arXiv.2510.03400},
archivePrefix = {arXiv},
       eprint = {2510.03400},
 primaryClass = {astro-ph.GA},
       adsurl = {https://ui.adsabs.harvard.edu/abs/2025arXiv251003400D}
}

@INCOLLECTION{Kulkarni08,
       author = {{Kulkarni}, Shrinivas R. and {Heiles}, Carl},
        title = "{Neutral hydrogen and the diffuse interstellar medium.}",
    booktitle = {Galactic and Extragalactic Radio Astronomy},
         year = 1988,
       editor = {{Kellermann}, K.~I. and {Verschuur}, G.~L.},
        pages = {95-153},
       adsurl = {https://ui.adsabs.harvard.edu/abs/1988gera.book...95K}
}

@ARTICLE{Gupta2022,
       author = {{Gupta}, N. and {Srianand}, R. and {Momjian}, E. and {Shukla}, G. and {Combes}, F. and {Krogager}, J.-K. and {Noterdaeme}, P. and {Petitjean}, P.},
        title = "{H I Gas Playing Hide-and-seek around a Powerful FRI-type Quasar at z   2.1}",
      journal = {\apjl},
         year = 2022,
        month = mar,
       volume = {927},
       number = {2},
          eid = {L24},
        pages = {L24},
          doi = {10.3847/2041-8213/ac589f},
archivePrefix = {arXiv},
       eprint = {2202.11755},
 primaryClass = {astro-ph.GA},
       adsurl = {https://ui.adsabs.harvard.edu/abs/2022ApJ...927L..24G}
}

@ARTICLE{York2000,
       author = {{York}, Donald G. and {Adelman}, J. and {Anderson}, Jr., John E. and {Anderson}, Scott F. and {Annis}, James and {Bahcall}, Neta A. and {Bakken}, J.~A. and {Barkhouser}, Robert and {Bastian}, Steven and {Berman}, Eileen and {Boroski}, William N. and {Bracker}, Steve and {Briegel}, Charlie and {Briggs}, John W. and {Brinkmann}, J. and {Brunner}, Robert and {Burles}, Scott and {Carey}, Larry and {Carr}, Michael A. and {Castander}, Francisco J. and {Chen}, Bing and {Colestock}, Patrick L. and {Connolly}, A.~J. and {Crocker}, J.~H. and {Csabai}, Istv{\'a}n and {Czarapata}, Paul C. and {Davis}, John Eric and {Doi}, Mamoru and {Dombeck}, Tom and {Eisenstein}, Daniel and {Ellman}, Nancy and {Elms}, Brian R. and {Evans}, Michael L. and {Fan}, Xiaohui and {Federwitz}, Glenn R. and {Fiscelli}, Larry and {Friedman}, Scott and {Frieman}, Joshua A. and {Fukugita}, Masataka and {Gillespie}, Bruce and {Gunn}, James E. and {Gurbani}, Vijay K. and {de Haas}, Ernst and {Haldeman}, Merle and {Harris}, Frederick H. and {Hayes}, J. and {Heckman}, Timothy M. and {Hennessy}, G.~S. and {Hindsley}, Robert B. and {Holm}, Scott and {Holmgren}, Donald J. and {Huang}, Chi-hao and {Hull}, Charles and {Husby}, Don and {Ichikawa}, Shin-Ichi and {Ichikawa}, Takashi and {Ivezi{\'c}}, {\v{Z}}eljko and {Kent}, Stephen and {Kim}, Rita S.~J. and {Kinney}, E. and {Klaene}, Mark and {Kleinman}, A.~N. and {Kleinman}, S. and {Knapp}, G.~R. and {Korienek}, John and {Kron}, Richard G. and {Kunszt}, Peter Z. and {Lamb}, D.~Q. and {Lee}, B. and {Leger}, R. French and {Limmongkol}, Siriluk and {Lindenmeyer}, Carl and {Long}, Daniel C. and {Loomis}, Craig and {Loveday}, Jon and {Lucinio}, Rich and {Lupton}, Robert H. and {MacKinnon}, Bryan and {Mannery}, Edward J. and {Mantsch}, P.~M. and {Margon}, Bruce and {McGehee}, Peregrine and {McKay}, Timothy A. and {Meiksin}, Avery and {Merelli}, Aronne and {Monet}, David G. and {Munn}, Jeffrey A. and {Narayanan}, Vijay K. and {Nash}, Thomas and {Neilsen}, Eric and {Neswold}, Rich and {Newberg}, Heidi Jo and {Nichol}, R.~C. and {Nicinski}, Tom and {Nonino}, Mario and {Okada}, Norio and {Okamura}, Sadanori and {Ostriker}, Jeremiah P. and {Owen}, Russell and {Pauls}, A. George and {Peoples}, John and {Peterson}, R.~L. and {Petravick}, Donald and {Pier}, Jeffrey R. and {Pope}, Adrian and {Pordes}, Ruth and {Prosapio}, Angela and {Rechenmacher}, Ron and {Quinn}, Thomas R. and {Richards}, Gordon T. and {Richmond}, Michael W. and {Rivetta}, Claudio H. and {Rockosi}, Constance M. and {Ruthmansdorfer}, Kurt and {Sandford}, Dale and {Schlegel}, David J. and {Schneider}, Donald P. and {Sekiguchi}, Maki and {Sergey}, Gary and {Shimasaku}, Kazuhiro and {Siegmund}, Walter A. and {Smee}, Stephen and {Smith}, J. Allyn and {Snedden}, S. and {Stone}, R. and {Stoughton}, Chris and {Strauss}, Michael A. and {Stubbs}, Christopher and {SubbaRao}, Mark and {Szalay}, Alexander S. and {Szapudi}, Istvan and {Szokoly}, Gyula P. and {Thakar}, Anirudda R. and {Tremonti}, Christy and {Tucker}, Douglas L. and {Uomoto}, Alan and {Vanden Berk}, Dan and {Vogeley}, Michael S. and {Waddell}, Patrick and {Wang}, Shu-i. and {Watanabe}, Masaru and {Weinberg}, David H. and {Yanny}, Brian and {Yasuda}, Naoki and {SDSS Collaboration}},
        title = "{The Sloan Digital Sky Survey: Technical Summary}",
      journal = {\aj},
         year = 2000,
        month = sep,
       volume = {120},
       number = {3},
        pages = {1579-1587},
          doi = {10.1086/301513},
archivePrefix = {arXiv},
       eprint = {astro-ph/0006396},
 primaryClass = {astro-ph},
       adsurl = {https://ui.adsabs.harvard.edu/abs/2000AJ....120.1579Y}
}

@ARTICLE{Ellison2002,
       author = {{Ellison}, S.~L. and {Yan}, L. and {Hook}, I.~M. and {Pettini}, M. and {Wall}, J.~V. and {Shaver}, P.},
        title = "{The CORALS survey. II. Clues to galaxy clustering around QSOs from z$_{abs}$ \raisebox{-0.5ex}\textasciitilde z$_{em}$ damped Lyman alpha systems}",
      journal = {\aap},
         year = 2002,
        month = jan,
       volume = {383},
        pages = {91-97},
          doi = {10.1051/0004-6361:20011738},
archivePrefix = {arXiv},
       eprint = {astro-ph/0112135},
 primaryClass = {astro-ph},
       adsurl = {https://ui.adsabs.harvard.edu/abs/2002A&A...383...91E}
}

@ARTICLE{Wolfe2003,
       author = {{Wolfe}, Arthur M. and {Gawiser}, Eric and {Prochaska}, Jason X.},
        title = "{C II* Absorption in Damped Ly{\ensuremath{\alpha}} Systems. II. A New Window on the Star Formation History of the Universe}",
      journal = {\apj},
         year = 2003,
        month = aug,
       volume = {593},
       number = {1},
        pages = {235-257},
          doi = {10.1086/376521},
archivePrefix = {arXiv},
       eprint = {astro-ph/0304042},
 primaryClass = {astro-ph},
       adsurl = {https://ui.adsabs.harvard.edu/abs/2003ApJ...593..235W}
}

@ARTICLE{Combes2021,
       author = {{Combes}, F. and {Gupta}, N. and {Muller}, S. and {Balashev}, S. and {J{\'o}zsa}, G.~I.~G. and {Srianand}, R. and {Momjian}, E. and {Noterdaeme}, P. and {Kl{\"o}ckner}, H.-R. and {Baker}, A.~J. and {Boettcher}, E. and {Bosma}, A. and {Chen}, H.-W. and {Dutta}, R. and {Jagannathan}, P. and {Jose}, J. and {Knowles}, K. and {Krogager}, J.-. K. and {Kulkarni}, V.~P. and {Moodley}, K. and {Pandey}, S. and {Petitjean}, P. and {Sekhar}, S.},
        title = "{PKS 1830-211: OH and H I at z = 0.89 and the first MeerKAT UHF spectrum}",
      journal = {\aap},
         year = 2021,
        month = apr,
       volume = {648},
          eid = {A116},
        pages = {A116},
          doi = {10.1051/0004-6361/202040167},
archivePrefix = {arXiv},
       eprint = {2101.00188},
 primaryClass = {astro-ph.GA},
       adsurl = {https://ui.adsabs.harvard.edu/abs/2021A&A...648A.116C}
}

@ARTICLE{Li2018,
       author = {{Li}, Di and {Tang}, Ningyu and {Nguyen}, Hiep and {Dawson}, J.~R. and {Heiles}, Carl and {Xu}, Duo and {Pan}, Zhichen and {Goldsmith}, Paul F. and {Gibson}, Steven J. and {Murray}, Claire E. and {Robishaw}, Tim and {McClure-Griffiths}, N.~M. and {Dickey}, John and {Pineda}, Jorge and {Stanimirovi{\'c}}, Sne{\v{z}}ana and {Bronfman}, L. and {Troland}, Thomas and {PRIMO Collaboration}},
        title = "{Where is OH and Does It Trace the Dark Molecular Gas (DMG)?}",
      journal = {\apjs},
         year = 2018,
        month = mar,
       volume = {235},
       number = {1},
          eid = {1},
        pages = {1},
          doi = {10.3847/1538-4365/aaa762},
archivePrefix = {arXiv},
       eprint = {1801.04373},
 primaryClass = {astro-ph.GA},
       adsurl = {https://ui.adsabs.harvard.edu/abs/2018ApJS..235....1L}
}

@INPROCEEDINGS{Liszt1999,
       author = {{Liszt}, H. and {Lucas}, R.},
        title = "{Molecular Absorption in the Local Diffuse Interstellar Medium}",
    booktitle = {Highly Redshifted Radio Lines},
         year = 1999,
       editor = {{Carilli}, C.~L. and {Radford}, S.~J.~E. and {Menten}, K.~M. and {Langston}, G.~I.},
       series = {Astronomical Society of the Pacific Conference Series},
       volume = {156},
        month = jan,
        pages = {188},
       adsurl = {https://ui.adsabs.harvard.edu/abs/1999ASPC..156..188L}
}

@ARTICLE{Rugel2025,
       author = {{Rugel}, M.~R. and {Beuther}, H. and {Soler}, J.~D. and {Goldsmith}, P. and {Anderson}, L. and {Hafner}, A. and {Dawson}, J.~R. and {Wang}, Y. and {Bihr}, S. and {Wiesemeyer}, H. and {G{\"u}sten}, R. and {Lee}, M.-Y. and {Riquelme}, D. and {Jacob}, A.~M. and {Kim}, W.-J. and {Busch}, M. and {Khan}, S. and {Brunthaler}, A.},
        title = "{Faint absorption of the ground state hyperfine-splitting transitions of hydroxyl at 18 cm in the Galactic disk}",
      journal = {\aap},
         year = 2025,
        month = aug,
       volume = {700},
          eid = {A171},
        pages = {A171},
          doi = {10.1051/0004-6361/202553998},
archivePrefix = {arXiv},
       eprint = {2506.06149},
 primaryClass = {astro-ph.GA},
       adsurl = {https://ui.adsabs.harvard.edu/abs/2025A&A...700A.171R}
}

@ARTICLE{Balashev2021,
       author = {{Balashev}, S.~A. and {Gupta}, N. and {Kosenko}, D.~N.},
        title = "{OH in the diffuse interstellar medium: physical modelling and prospects with upcoming SKA precursor/pathfinder surveys}",
      journal = {\mnras},
         year = 2021,
        month = jul,
       volume = {504},
       number = {3},
        pages = {3797-3811},
          doi = {10.1093/mnras/stab1122},
archivePrefix = {arXiv},
       eprint = {2012.12241},
 primaryClass = {astro-ph.GA},
       adsurl = {https://ui.adsabs.harvard.edu/abs/2021MNRAS.504.3797B}
}

@ARTICLE{Rao2006,
       author = {{Rao}, Sandhya M. and {Turnshek}, David A. and {Nestor}, Daniel B.},
        title = "{Damped Ly{\ensuremath{\alpha}} Systems at z<1.65: The Expanded Sloan Digital Sky Survey Hubble Space Telescope Sample}",
      journal = {\apj},
         year = 2006,
        month = jan,
       volume = {636},
       number = {2},
        pages = {610-630},
          doi = {10.1086/498132},
archivePrefix = {arXiv},
       eprint = {astro-ph/0509469},
 primaryClass = {astro-ph},
       adsurl = {https://ui.adsabs.harvard.edu/abs/2006ApJ...636..610R}
}

@ARTICLE{Krogager2015,
       author = {{Krogager}, J.-K. and {Geier}, S. and {Fynbo}, J.~P.~U. and {Venemans}, B.~P. and {Ledoux}, C. and {M{\o}ller}, P. and {Noterdaeme}, P. and {Vestergaard}, M. and {Kangas}, T. and {Pursimo}, T. and {Saturni}, F.~G. and {Smirnova}, O.},
        title = "{The High A$_{V}$ Quasar Survey: Reddened Quasi-Stellar Objects Selected from Optical/Near-Infrared Photometry{\textemdash}II}",
      journal = {\apjs},
         year = 2015,
        month = mar,
       volume = {217},
       number = {1},
          eid = {5},
        pages = {5},
          doi = {10.1088/0067-0049/217/1/5},
archivePrefix = {arXiv},
       eprint = {1410.7783},
 primaryClass = {astro-ph.GA},
       adsurl = {https://ui.adsabs.harvard.edu/abs/2015ApJS..217....5K}
}

@ARTICLE{Srianand2012,
       author = {{Srianand}, R. and {Gupta}, N. and {Petitjean}, P. and {Noterdaeme}, P. and {Ledoux}, C. and {Salter}, C.~J. and {Saikia}, D.~J.},
        title = "{Search for cold gas in z > 2 damped Ly{\ensuremath{\alpha}} systems: 21-cm and H$_{2}$ absorption}",
      journal = {\mnras},
         year = 2012,
        month = mar,
       volume = {421},
       number = {1},
        pages = {651-665},
          doi = {10.1111/j.1365-2966.2011.20342.x},
archivePrefix = {arXiv},
       eprint = {1112.1438},
 primaryClass = {astro-ph.CO},
       adsurl = {https://ui.adsabs.harvard.edu/abs/2012MNRAS.421..651S}
}

@ARTICLE{Taylor1999,
       author = {{Taylor}, G.~B. and {O'Dea}, C.~P. and {Peck}, A.~B. and {Koekemoer}, A.~M.},
        title = "{H I Absorption toward the Nucleus of the Radio Galaxy PKS 2322-123 in A2597}",
      journal = {\apjl},
         year = 1999,
        month = feb,
       volume = {512},
       number = {1},
        pages = {L27-L30},
          doi = {10.1086/311873},
archivePrefix = {arXiv},
       eprint = {astro-ph/9811343},
 primaryClass = {astro-ph},
       adsurl = {https://ui.adsabs.harvard.edu/abs/1999ApJ...512L..27T}
}

@ARTICLE{Saraf2023,
       author = {{Saraf}, Manasvee and {Wong}, O. Ivy and {Cortese}, Luca and {Koribalski}, B{\"a}rbel S.},
        title = "{H I absorption associated with Norma's brightest cluster galaxy}",
      journal = {\mnras},
         year = 2023,
        month = mar,
       volume = {519},
       number = {3},
        pages = {4128-4141},
          doi = {10.1093/mnras/stac3695},
archivePrefix = {arXiv},
       eprint = {2212.06680},
 primaryClass = {astro-ph.GA},
       adsurl = {https://ui.adsabs.harvard.edu/abs/2023MNRAS.519.4128S}
}

@ARTICLE{gupta2021ApJS..255...28G,
       author = {{Gupta}, N. and {Srianand}, R. and {Shukla}, G. and {Krogager}, J. -. K. and {Noterdaeme}, P. and {Combes}, F. and {Dutta}, R. and {Fynbo}, J.~P.~U. and {Hilton}, M. and {Momjian}, E. and {Moodley}, K. and {Petitjean}, P.},
        title = "{Evolution of Cold Gas at 2 < z < 5: A Blind Search for H I and OH Absorption Lines toward Mid-infrared Color-selected Radio-loud AGN}",
      journal = {\apjs},
         year = 2021,
        month = aug,
       volume = {255},
       number = {2},
          eid = {28},
        pages = {28},
          doi = {10.3847/1538-4365/ac03b5},
archivePrefix = {arXiv},
       eprint = {2103.09437},
 primaryClass = {astro-ph.GA},
       adsurl = {https://ui.adsabs.harvard.edu/abs/2021ApJS..255...28G}
}

@ARTICLE{aditya2017MNRAS.465.5011A,
       author = {{Aditya}, J.~N.~H.~S. and {Kanekar}, Nissim and {Prochaska}, J. Xavier and {Day}, Brandon and {Lynam}, Paul and {Cruz}, Jocelyn},
        title = "{Giant Metrewave Radio Telescope detection of associated H I 21-cm absorption at z = 1.2230 towards TXS 1954+513}",
      journal = {\mnras},
         year = 2017,
        month = mar,
       volume = {465},
       number = {4},
        pages = {5011-5015},
          doi = {10.1093/mnras/stw3105},
archivePrefix = {arXiv},
       eprint = {1612.01139},
 primaryClass = {astro-ph.GA},
       adsurl = {https://ui.adsabs.harvard.edu/abs/2017MNRAS.465.5011A}
}

@ARTICLE{deka2024A&A...687A..50D,
       author = {{Deka}, P.~P. and {Gupta}, N. and {Chen}, H.~W. and {Johnson}, S.~D. and {Noterdaeme}, P. and {Combes}, F. and {Boettcher}, E. and {Balashev}, S.~A. and {Emig}, K.~L. and {J{\'o}zsa}, G.~I.~G. and {Kl{\"o}ckner}, H. -R. and {Krogager}, J. -. K. and {Momjian}, E. and {Petitjean}, P. and {Rudie}, G.~C. and {Wagenveld}, J. and {Zahedy}, F.~S.},
        title = "{MALS discovery of a rare H I 21 cm absorber at z {\ensuremath{\sim}} 1.35: Origin of the absorbing gas in powerful active galactic nuclei}",
      journal = {\aap},
         year = 2024,
        month = jul,
       volume = {687},
          eid = {A50},
        pages = {A50},
          doi = {10.1051/0004-6361/202348464},
archivePrefix = {arXiv},
       eprint = {2311.00336},
 primaryClass = {astro-ph.GA},
       adsurl = {https://ui.adsabs.harvard.edu/abs/2024A&A...687A..50D}
}

@ARTICLE{aditya2018MNRAS.473...59A,
       author = {{Aditya}, J.~N.~H.~S. and {Kanekar}, Nissim},
        title = "{A Giant Metrewave Radio Telescope search for associated H I 21 cm absorption in GHz-peaked-spectrum sources}",
      journal = {\mnras},
         year = 2018,
        month = jan,
       volume = {473},
       number = {1},
        pages = {59-67},
          doi = {10.1093/mnras/stx2325},
archivePrefix = {arXiv},
       eprint = {1710.00711},
 primaryClass = {astro-ph.GA},
       adsurl = {https://ui.adsabs.harvard.edu/abs/2018MNRAS.473...59A}
}

@ARTICLE{aditya2024MNRAS.527.8511A,
       author = {{Aditya}, J.~N.~H.~S. and {Yoon}, Hyein and {Allison}, James R. and {An}, Tao and {Chhetri}, Rajan and {Curran}, Stephen J. and {Darling}, Jeremy and {Emig}, Kimberly L. and {Glowacki}, Marcin and {Kerrison}, Emily and {Koribalski}, B{\"a}rbel S. and {Mahony}, Elizabeth K. and {Moss}, Vanessa A. and {Morgan}, John and {Sadler}, Elaine M. and {Soria}, Roberto and {Su}, Renzhi and {Weng}, Simon and {Whiting}, Matthew},
        title = "{The FLASH pilot survey: an H I absorption search against MRC 1-Jy radio sources}",
      journal = {\mnras},
         year = 2024,
        month = jan,
       volume = {527},
       number = {3},
        pages = {8511-8534},
          doi = {10.1093/mnras/stad3722},
archivePrefix = {arXiv},
       eprint = {2310.14571},
 primaryClass = {astro-ph.GA},
       adsurl = {https://ui.adsabs.harvard.edu/abs/2024MNRAS.527.8511A}
}

@ARTICLE{aditya2018MNRAS.481.1578A,
       author = {{Aditya}, J.~N.~H.~S. and {Kanekar}, Nissim},
        title = "{A Giant Metrewave Radio Telescope survey for associated H I 21 cm absorption in the Caltech-Jodrell flat-spectrum sample}",
      journal = {\mnras},
         year = 2018,
        month = dec,
       volume = {481},
       number = {2},
        pages = {1578-1596},
          doi = {10.1093/mnras/sty2184},
archivePrefix = {arXiv},
       eprint = {1808.03280},
 primaryClass = {astro-ph.GA},
       adsurl = {https://ui.adsabs.harvard.edu/abs/2018MNRAS.481.1578A}
}

@ARTICLE{murthy2022A&A...659A.185M,
       author = {{Murthy}, Suma and {Morganti}, Raffaella and {Kanekar}, Nissim and {Oosterloo}, Tom},
        title = "{Redshift evolution of the H I detection rate in radio-loud active galactic nuclei}",
      journal = {\aap},
         year = 2022,
        month = mar,
       volume = {659},
          eid = {A185},
        pages = {A185},
          doi = {10.1051/0004-6361/202142550},
archivePrefix = {arXiv},
       eprint = {2201.06625},
 primaryClass = {astro-ph.GA},
       adsurl = {https://ui.adsabs.harvard.edu/abs/2022A&A...659A.185M}
}

@article{dwarakanath1994,
  author={Dwarakanath, K. S. and Owen, F. N. and van Gorkom, J. H.},
  title={A VLA search for neutral hydrogen in central cluster galaxies},
  journal      = {Astrophysical Journal},
  year         = {1994},
        month = Sep,
       volume = {432},
        pages = {469-477},
          doi = {10.1086/174586},
  archivePrefix= {arXiv},
  primaryClass = {astro-ph.GA},
  adsurl       = {https://ui.adsabs.harvard.edu/abs/1994ApJ...432..469D/abstract}
}

@article{Ogle2010,
  author = {Ogle, P. M. and Boulanger, F. and Guillard, P. and Evans, D. A. and Antonucci, R. and Appleton, P. N. and Nesvadba, N. and Leipski, C.},
  title        = {Powerful H$_2$ Line Cooling in Stephan's Quintet II: Warm Molecular Hydrogen in Galaxies},
  journal      = {Astrophysical Journal},
  year         = {2010},
        month = Nov,
       volume = {724},
       number = {2},
        pages = {1193-1209},
          doi = {10.1088/0004-637X/724/2/1193},
  archivePrefix= {arXiv},
  primaryClass = {astro-ph.GA},
  adsurl       = {https://arxiv.org/abs/1009.4533}
}

@ARTICLE{ForsterSchreiber2020,
       author = {{F{\"o}rster Schreiber}, Natascha M. and {Wuyts}, Stijn},
        title = "{Star-Forming Galaxies at Cosmic Noon}",
      journal = {\araa},
         year = 2020,
        month = aug,
       volume = {58},
        pages = {661-725},
          doi = {10.1146/annurev-astro-032620-021910},
archivePrefix = {arXiv},
       eprint = {2010.10171},
 primaryClass = {astro-ph.GA},
       adsurl = {https://ui.adsabs.harvard.edu/abs/2020ARA&A..58..661F}
}

@ARTICLE{Fiore2017,
       author = {{Fiore}, F. and {Feruglio}, C. and {Shankar}, F. and {Bischetti}, M. and
         {Bongiorno}, A. and {Brusa}, M. and {Carniani}, S. and {Cicone}, C. and
         {Duras}, F. and {Lamastra}, A. and {Mainieri}, V. and {Marconi}, A. and
         {Menci}, N. and {Maiolino}, R. and {Piconcelli}, E. and {Vietri}, G. and
         {Zappacosta}, L.},
        title = "{AGN wind scaling relations and the co-evolution of black holes and galaxies}",
      journal = {\aap},
         year = "2017",
        month = "May",
       volume = {601},
          eid = {A143},
        pages = {A143},
          doi = {10.1051/0004-6361/201629478},
archivePrefix = {arXiv},
       eprint = {1702.04507},
 primaryClass = {astro-ph.GA},
       adsurl = {https://ui.adsabs.harvard.edu/abs/2017A&A...601A.143F}
}

@ARTICLE{Jarvis2019,
       author = {{Jarvis}, M.~E. and {Harrison}, C.~M. and {Thomson}, A.~P. and {Circosta}, C. and {Mainieri}, V. and {Alexander}, D.~M. and {Edge}, A.~C. and {Lansbury}, G.~B. and {Molyneux}, S.~J. and {Mullaney}, J.~R.},
        title = "{Prevalence of radio jets associated with galactic outflows and feedback from quasars}",
      journal = {\mnras},
         year = 2019,
        month = may,
       volume = {485},
       number = {2},
        pages = {2710-2730},
          doi = {10.1093/mnras/stz556},
archivePrefix = {arXiv},
       eprint = {1902.07727},
 primaryClass = {astro-ph.GA},
       adsurl = {https://ui.adsabs.harvard.edu/abs/2019MNRAS.485.2710J}
}

@ARTICLE{HerreraCamus2019,
   author = {{Herrera-Camus}, R. and {Tacconi}, L. and {Genzel}, R. and {F{\"o}rster Schreiber}, N. and 
	{Lutz}, D. and {Bolatto}, A. and {Wuyts}, S. and {Renzini}, A. and 
	{Lilly}, S. and {Belli}, S. and {{\"U}bler}, H. and {Shimizu}, T. and 
	{Davies}, R. and {Sturm}, E. and {Combes}, F. and {Freundlich}, J. and 
	{Garc{\'{\i}}a-Burillo}, S. and {Cox}, P. and {Burkert}, A. and 
	{Naab}, T. and {Colina}, L. and {Saintonge}, A. and {Cooper}, M. and 
	{Feruglio}, C. and {Weiss}, A.},
    title = "{Molecular and Ionized Gas Phases of an AGN-driven Outflow in a Typical Massive Galaxy at z{\nbsp}{\ap}{\nbsp}2}",
  journal = {\apj},
archivePrefix = "arXiv",
   eprint = {1807.07074},
     year = 2019,
    month = jan,
   volume = 871,
      eid = {37},
    pages = {37},
      doi = {10.3847/1538-4357/aaf6a7},
   adsurl = {http://adsabs.harvard.edu/abs/2019ApJ...871...37H}
}

@ARTICLE{Banados2023,
       author = {{Ba{\~n}ados}, Eduardo and {Momjian}, Emmanuel and {Connor}, Thomas and {Belladitta}, Silvia and {Decarli}, Roberto and {Mazzucchelli}, Chiara and {Venemans}, Bram P. and {Walter}, Fabian and {Wang}, Feige and {Xie}, Zhang-Liang and {Barth}, Aaron J. and {Eilers}, Anna-Christina and {Fan}, Xiaohui and {Khusanova}, Yana and {Schindler}, Jan-Torge and {Stern}, Daniel and {Yang}, Jinyi and {Andika}, Irham Taufik and {Carilli}, Christopher L. and {Farina}, Emanuele P. and {Fabian}, Andrew and {Hennawi}, Joseph F. and {Pensabene}, Antonio and {Rojas-Ruiz}, Sof{\'\i}a},
        title = "{A blazar in the epoch of reionization}",
      journal = {Nature Astronomy},
         year = 2025,
        month = feb,
       volume = {9},
        pages = {293-301},
          doi = {10.1038/s41550-024-02431-4},
       adsurl = {https://ui.adsabs.harvard.edu/abs/2025NatAs...9..293B}
}

@ARTICLE{Belli2024,
       author = {{Belli}, Sirio and {Park}, Minjung and {Davies}, Rebecca L. and {Mendel}, J. Trevor and {Johnson}, Benjamin D. and {Conroy}, Charlie and {Benton}, Chlo{\"e} and {Bugiani}, Letizia and {Emami}, Razieh and {Leja}, Joel and {Li}, Yijia and {Maheson}, Gabriel and {Mathews}, Elijah P. and {Naidu}, Rohan P. and {Nelson}, Erica J. and {Tacchella}, Sandro and {Terrazas}, Bryan A. and {Weinberger}, Rainer},
        title = "{Star formation shut down by multiphase gas outflow in a galaxy at a redshift of 2.45}",
      journal = {\nat},
         year = 2024,
        month = jun,
       volume = {630},
       number = {8015},
        pages = {54-58},
          doi = {10.1038/s41586-024-07412-1},
archivePrefix = {arXiv},
       eprint = {2308.05795},
 primaryClass = {astro-ph.GA},
       adsurl = {https://ui.adsabs.harvard.edu/abs/2024Natur.630...54B}
}

@ARTICLE{Davies2024,
       author = {{Davies}, Rebecca L. and {Belli}, Sirio and {Park}, Minjung and {Mendel}, J. Trevor and {Johnson}, Benjamin D. and {Conroy}, Charlie and {Benton}, Chlo{\"e} and {Bugiani}, Letizia and {Emami}, Razieh and {Leja}, Joel and {Li}, Yijia and {Maheson}, Gabriel and {Mathews}, Elijah P. and {Naidu}, Rohan P. and {Nelson}, Erica J. and {Tacchella}, Sandro and {Terrazas}, Bryan A. and {Weinberger}, Rainer},
        title = "{JWST reveals widespread AGN-driven neutral gas outflows in massive z   2 galaxies}",
      journal = {\mnras},
         year = 2024,
        month = mar,
       volume = {528},
       number = {3},
        pages = {4976-4992},
          doi = {10.1093/mnras/stae327},
archivePrefix = {arXiv},
       eprint = {2310.17939},
 primaryClass = {astro-ph.GA},
       adsurl = {https://ui.adsabs.harvard.edu/abs/2024MNRAS.528.4976D}
}

@ARTICLE{Nanayakkara2024,
       author = {{Nanayakkara}, Themiya and {Glazebrook}, Karl and {Jacobs}, Colin and {Kawinwanichakij}, Lalitwadee and {Schreiber}, Corentin and {Brammer}, Gabriel and {Esdaile}, James and {Kacprzak}, Glenn G. and {Labbe}, Ivo and {Lagos}, Claudia and {Marchesini}, Danilo and {Marsan}, Z. Cemile and {Oesch}, Pascal A. and {Papovich}, Casey and {Remus}, Rhea-Silvia and {Tran}, Kim-Vy H.},
        title = "{A population of faint, old, and massive quiescent galaxies at 3 <z <4 revealed by JWST NIRSpec Spectroscopy}",
      journal = {Scientific Reports},
         year = 2024,
        month = feb,
       volume = {14},
          eid = {3724},
        pages = {3724},
          doi = {10.1038/s41598-024-52585-4},
archivePrefix = {arXiv},
       eprint = {2212.11638},
 primaryClass = {astro-ph.GA},
       adsurl = {https://ui.adsabs.harvard.edu/abs/2024NatSR..14.3724N}
}

@ARTICLE{Moretti2025,
       author = {{Moretti}, Lorenzo and {Belli}, Sirio and {Rudie}, Gwen C. and {Newman}, Andrew B. and {Park}, Minjung and {Khoram}, Amir H. and {Chartab}, Nima and {Donevski}, Darko},
        title = "{Empirical Calibration of Na I D and Other Absorption Lines as Tracers of High-Redshift Neutral Outflows}",
      journal = {arXiv e-prints},
         year = 2025,
        month = jul,
          eid = {arXiv:2507.07160},
        pages = {arXiv:2507.07160},
          doi = {10.48550/arXiv.2507.07160},
archivePrefix = {arXiv},
       eprint = {2507.07160},
 primaryClass = {astro-ph.GA},
       adsurl = {https://ui.adsabs.harvard.edu/abs/2025arXiv250707160M}
}

@ARTICLE{Zwaan2005,
       author = {{Zwaan}, M.~A. and {van der Hulst}, J.~M. and {Briggs}, F.~H. and {Verheijen}, M.~A.~W. and {Ryan-Weber}, E.~V.},
        title = "{Reconciling the local galaxy population with damped Lyman {\ensuremath{\alpha}} cross-sections and metal abundances}",
      journal = {\mnras},
         year = 2005,
        month = dec,
       volume = {364},
       number = {4},
        pages = {1467-1487},
          doi = {10.1111/j.1365-2966.2005.09698.x},
archivePrefix = {arXiv},
       eprint = {astro-ph/0510127},
 primaryClass = {astro-ph},
       adsurl = {https://ui.adsabs.harvard.edu/abs/2005MNRAS.364.1467Z}
}

@ARTICLE{Roy2021,
       author = {{Roy}, Namrata and {Bundy}, Kevin and {Rubin}, Kate H.~R. and {Rowlands}, Kate and {Westfall}, Kyle and {Riffel}, Rogerio and {Bizyaev}, Dmitry and {Stark}, David V. and {Riffel}, Rogemar A. and {Lacerna}, Ivan and {Nair}, Preethi and {Wu}, Xuanyi and {Drory}, Niv},
        title = "{Signatures of Inflowing Gas in Red Geyser Galaxies Hosting Radio Active Galactic Nuclei}",
      journal = {\apj},
         year = 2021,
        month = oct,
       volume = {919},
       number = {2},
          eid = {145},
        pages = {145},
          doi = {10.3847/1538-4357/ac0f74},
archivePrefix = {arXiv},
       eprint = {2106.14901},
 primaryClass = {astro-ph.GA},
       adsurl = {https://ui.adsabs.harvard.edu/abs/2021ApJ...919..145R}
}

@article{Hamer2014HydraA,
  author       = {Hamer, S. L. and Edge, A. C. and Swinbank, A. M. and Oonk, J. B. R. and Mittal, R. and McNamara, B. R. and Russell, H. R. and Bremer, M. N. and Combes, F. and Fabian, A. C. and Nesvadba, N. P. H. and O'Dea, C. P. and Baum, S. A. and Salomé, P. and Tremblay, G. and Donahue, M. and Ferland, G. J. and Sarazin, C. L.},
  title        = {Cold gas dynamics in Hydra-A: evidence for a rotating disc},
  journal      = {\mnras},
  year         = {2014},
        month = Jan,
       volume = {437},
       number = {1},
        pages = {862–878},
          doi = {10.1093/mnras/stt1949},
  eprint       = {1310.4501},
  archivePrefix= {arXiv},
  primaryClass = {astro-ph.GA},
  adsurl       = {https://arxiv.org/abs/1310.4501}
}

@ARTICLE{chandola2020MNRAS.494.5161C,
       author = {{Chandola}, Yogesh and {Saikia}, D.~J. and {Li}, Di},
        title = "{H I absorption towards radio active galactic nuclei of different accretion modes}",
      journal = {\mnras},
         year = 2020,
        month = jun,
       volume = {494},
       number = {4},
        pages = {5161-5177},
          doi = {10.1093/mnras/staa1029},
archivePrefix = {arXiv},
       eprint = {2004.07074},
 primaryClass = {astro-ph.GA},
       adsurl = {https://ui.adsabs.harvard.edu/abs/2020MNRAS.494.5161C}
}

@ARTICLE{chandola2017MNRAS.465..997C,
       author = {{Chandola}, Yogesh and {Saikia}, D.~J.},
        title = "{H I absorption towards low-luminosity radio-loud active galactic nuclei of different accretion modes and WISE colours}",
      journal = {\mnras},
         year = 2017,
        month = feb,
       volume = {465},
       number = {1},
        pages = {997-1007},
          doi = {10.1093/mnras/stw2705},
archivePrefix = {arXiv},
       eprint = {1607.00841},
 primaryClass = {astro-ph.GA},
       adsurl = {https://ui.adsabs.harvard.edu/abs/2017MNRAS.465..997C}
}

@ARTICLE{Combes2024,
       author = {{Combes}, F. and {Gupta}, N.},
        title = "{Cold molecules in H I 21 cm absorbers across redshifts {\ensuremath{\sim}}0.1-4}",
      journal = {\aap},
         year = 2024,
        month = mar,
       volume = {683},
          eid = {A20},
        pages = {A20},
          doi = {10.1051/0004-6361/202348386},
archivePrefix = {arXiv},
       eprint = {2310.17204},
 primaryClass = {astro-ph.GA},
       adsurl = {https://ui.adsabs.harvard.edu/abs/2024A&A...683A..20C}
}

@ARTICLE{Fabian2012,
       author = {{Fabian}, A.~C.},
        title = "{Observational Evidence of Active Galactic Nuclei Feedback}",
      journal = {\araa},
         year = 2012,
        month = sep,
       volume = {50},
        pages = {455-489},
          doi = {10.1146/annurev-astro-081811-125521},
archivePrefix = {arXiv},
       eprint = {1204.4114},
 primaryClass = {astro-ph.CO},
       adsurl = {https://ui.adsabs.harvard.edu/abs/2012ARA&A..50..455F}
}

@ARTICLE{Rose2024,
       author = {{Rose}, Tom and {McNamara}, B.~R. and {Combes}, F. and {Edge}, A.~C. and {McDonald}, M. and {O'Sullivan}, Ewan and {Russell}, H. and {Fabian}, A.~C. and {Ferland}, G. and {Salom{\'e}}, P. and {Tremblay}, G.},
        title = "{Two distinct molecular cloud populations detected in massive galaxies}",
      journal = {\mnras},
         year = 2024,
        month = sep,
       volume = {533},
       number = {1},
        pages = {771-794},
          doi = {10.1093/mnras/stae1831},
archivePrefix = {arXiv},
       eprint = {2403.03974},
 primaryClass = {astro-ph.GA},
       adsurl = {https://ui.adsabs.harvard.edu/abs/2024MNRAS.533..771R}
}

@ARTICLE{Peroux2020,
  author = {{Péroux}, C. and {Howk}, J.~C.},
  title = "{Cold gas accretion in galaxies}",
  journal = {\araa},
  year = 2020,
  volume = {58},
  pages = {363--403},
  doi = {10.1146/annurev-astro-032620-021859},
  adsurl = {https://ui.adsabs.harvard.edu/abs/2020ARA&A..58..363P}
}

@ARTICLE{Koribalski2004,
       author = {{Koribalski}, B.~S. and {Staveley-Smith}, L. and {Kilborn}, V.~A. and {Ryder}, S.~D. and {Kraan-Korteweg}, R.~C. and {Ryan-Weber}, E.~V. and {Ekers}, R.~D. and {Jerjen}, H. and {Henning}, P.~A. and {Putman}, M.~E. and {Zwaan}, M.~A. and {de Blok}, W.~J.~G. and {Calabretta}, M.~R. and {Disney}, M.~J. and {Minchin}, R.~F. and {Bhathal}, R. and {Boyce}, P.~J. and {Drinkwater}, M.~J. and {Freeman}, K.~C. and {Gibson}, B.~K. and {Green}, A.~J. and {Haynes}, R.~F. and {Juraszek}, S. and {Kesteven}, M.~J. and {Knezek}, P.~M. and {Mader}, S. and {Marquarding}, M. and {Meyer}, M. and {Mould}, J.~R. and {Oosterloo}, T. and {O'Brien}, J. and {Price}, R.~M. and {Sadler}, E.~M. and {Schr{\"o}der}, A. and {Stewart}, I.~M. and {Stootman}, F. and {Waugh}, M. and {Warren}, B.~E. and {Webster}, R.~L. and {Wright}, A.~E.},
        title = "{The 1000 Brightest HIPASS Galaxies: H I Properties}",
      journal = {\aj},
         year = 2004,
        month = jul,
       volume = {128},
       number = {1},
        pages = {16-46},
          doi = {10.1086/421744},
archivePrefix = {arXiv},
       eprint = {astro-ph/0404436},
 primaryClass = {astro-ph},
       adsurl = {https://ui.adsabs.harvard.edu/abs/2004AJ....128...16K}
}

@ARTICLE{koribalski2020,
       author = {{Koribalski}, B{\"a}rbel S. and {Staveley-Smith}, L. and {Westmeier}, T. and {Serra}, P. and {Spekkens}, K. and {Wong}, O.~I. and {Lee-Waddell}, K. and {Lagos}, C.~D.~P. and {Obreschkow}, D. and {Ryan-Weber}, E.~V. and {Zwaan}, M. and {Kilborn}, V. and {Bekiaris}, G. and {Bekki}, K. and {Bigiel}, F. and {Boselli}, A. and {Bosma}, A. and {Catinella}, B. and {Chauhan}, G. and {Cluver}, M.~E. and {Colless}, M. and {Courtois}, H.~M. and {Crain}, R.~A. and {de Blok}, W.~J.~G. and {D{\'e}nes}, H. and {Duffy}, A.~R. and {Elagali}, A. and {Fluke}, C.~J. and {For}, B. -Q. and {Heald}, G. and {Henning}, P.~A. and {Hess}, K.~M. and {Holwerda}, B.~W. and {Howlett}, C. and {Jarrett}, T. and {Jones}, D.~H. and {Jones}, M.~G. and {J{\'o}zsa}, G.~I.~G. and {Jurek}, R. and {J{\"u}tte}, E. and {Kamphuis}, P. and {Karachentsev}, I. and {Kerp}, J. and {Kleiner}, D. and {Kraan-Korteweg}, R.~C. and {L{\'o}pez-S{\'a}nchez}, {\'A}. R. and {Madrid}, J. and {Meyer}, M. and {Mould}, J. and {Murugeshan}, C. and {Norris}, R.~P. and {Oh}, S. -H. and {Oosterloo}, T.~A. and {Popping}, A. and {Putman}, M. and {Reynolds}, T.~N. and {Rhee}, J. and {Robotham}, A.~S.~G. and {Ryder}, S. and {Schr{\"o}der}, A.~C. and {Shao}, Li and {Stevens}, A.~R.~H. and {Taylor}, E.~N. and {van{\^A} der Hulst}, J.~M. and {Verdes-Montenegro}, L. and {Wakker}, B.~P. and {Wang}, J. and {Whiting}, M. and {Winkel}, B. and {Wolf}, C.},
        title = "{WALLABY - an SKA Pathfinder HI survey}",
      journal = {\apss},
         year = 2020,
        month = jul,
       volume = {365},
       number = {7},
          eid = {118},
        pages = {118},
          doi = {10.1007/s10509-020-03831-4},
archivePrefix = {arXiv},
       eprint = {2002.07311},
 primaryClass = {astro-ph.GA},
       adsurl = {https://ui.adsabs.harvard.edu/abs/2020Ap&SS.365..118K}
}

@ARTICLE{Haynes2011,
       author = {{Haynes}, Martha P. and {Giovanelli}, Riccardo and {Martin}, Ann M. and {Hess}, Kelley M. and {Saintonge}, Am{\'e}lie and {Adams}, Elizabeth A.~K. and {Hallenbeck}, Gregory and {Hoffman}, G. Lyle and {Huang}, Shan and {Kent}, Brian R. and {Koopmann}, Rebecca A. and {Papastergis}, Emmanouil and {Stierwalt}, Sabrina and {Balonek}, Thomas J. and {Craig}, David W. and {Higdon}, Sarah J.~U. and {Kornreich}, David A. and {Miller}, Jeffrey R. and {O'Donoghue}, Aileen A. and {Olowin}, Ronald P. and {Rosenberg}, Jessica L. and {Spekkens}, Kristine and {Troischt}, Parker and {Wilcots}, Eric M.},
        title = "{The Arecibo Legacy Fast ALFA Survey: The {\ensuremath{\alpha}}.40 H I Source Catalog, Its Characteristics and Their Impact on the Derivation of the H I Mass Function}",
      journal = {\aj},
         year = 2011,
        month = nov,
       volume = {142},
       number = {5},
          eid = {170},
        pages = {170},
          doi = {10.1088/0004-6256/142/5/170},
archivePrefix = {arXiv},
       eprint = {1109.0027},
 primaryClass = {astro-ph.CO},
       adsurl = {https://ui.adsabs.harvard.edu/abs/2011AJ....142..170H}
}

@ARTICLE{Walter2008,
       author = {{Walter}, Fabian and {Brinks}, Elias and {de Blok}, W.~J.~G. and {Bigiel}, Frank and {Kennicutt}, Jr., Robert C. and {Thornley}, Michele D. and {Leroy}, Adam},
        title = "{THINGS: The H I Nearby Galaxy Survey}",
      journal = {\aj},
         year = 2008,
        month = dec,
       volume = {136},
       number = {6},
        pages = {2563-2647},
          doi = {10.1088/0004-6256/136/6/2563},
archivePrefix = {arXiv},
       eprint = {0810.2125},
 primaryClass = {astro-ph},
       adsurl = {https://ui.adsabs.harvard.edu/abs/2008AJ....136.2563W}
}

@ARTICLE{deBlok2024,
       author = {{de Blok}, W.~J.~G. and {Healy}, J. and {Maccagni}, F.~M. and {Pisano}, D.~J. and {Bosma}, A. and {English}, J. and {Jarrett}, T. and {Marasco}, A. and {Meurer}, G.~R. and {Veronese}, S. and {Bigiel}, F. and {Chemin}, L. and {Fraternali}, F. and {Holwerda}, B.~W. and {Kamphuis}, P. and {Kl{\"o}ckner}, H.~R. and {Kleiner}, D. and {Leroy}, A.~K. and {Mogotsi}, M. and {Oman}, K.~A. and {Schinnerer}, E. and {Verdes-Montenegro}, L. and {Westmeier}, T. and {Wong}, O.~I. and {Zabel}, N. and {Amram}, P. and {Carignan}, C. and {Combes}, F. and {Brinks}, E. and {Dettmar}, R.~J. and {Gibson}, B.~K. and {Jozsa}, G.~I.~G. and {Koribalski}, B.~S. and {McGaugh}, S.~S. and {Oosterloo}, T.~A. and {Spekkens}, K. and {Schr{\"o}der}, A.~C. and {Adams}, E.~A.~K. and {Athanassoula}, E. and {Bershady}, M.~A. and {Beswick}, R.~J. and {Blyth}, S. and {Elson}, E.~C. and {Frank}, B.~S. and {Heald}, G. and {Henning}, P.~A. and {Kurapati}, S. and {Loubser}, S.~I. and {Lucero}, D. and {Meyer}, M. and {Namumba}, B. and {Oh}, S. -H. and {Sardone}, A. and {Sheth}, K. and {Smith}, M.~W.~L. and {Sorgho}, A. and {Walter}, F. and {Williams}, T. and {Woudt}, P.~A. and {Zijlstra}, A.},
        title = "{MHONGOOSE: A MeerKAT nearby galaxy H I survey}",
      journal = {\aap},
         year = 2024,
        month = aug,
       volume = {688},
          eid = {A109},
        pages = {A109},
          doi = {10.1051/0004-6361/202348297},
archivePrefix = {arXiv},
       eprint = {2404.01774},
 primaryClass = {astro-ph.GA},
       adsurl = {https://ui.adsabs.harvard.edu/abs/2024A&A...688A.109D}
}

@ARTICLE{lah2007,
       author = {{Lah}, Philip and {Chengalur}, Jayaram N. and {Briggs}, Frank H. and {Colless}, Matthew and {de Propris}, Roberto and {Pracy}, Michael B. and {de Blok}, W.~J.~G. and {Fujita}, Shinobu S. and {Ajiki}, Masaru and {Shioya}, Yasuhiro and {Nagao}, Tohru and {Murayama}, Takashi and {Taniguchi}, Yoshiaki and {Yagi}, Masafumi and {Okamura}, Sadanori},
        title = "{The HI content of star-forming galaxies at z = 0.24}",
      journal = {\mnras},
         year = 2007,
        month = apr,
       volume = {376},
       number = {3},
        pages = {1357-1366},
          doi = {10.1111/j.1365-2966.2007.11540.x},
archivePrefix = {arXiv},
       eprint = {astro-ph/0701668},
 primaryClass = {astro-ph},
       adsurl = {https://ui.adsabs.harvard.edu/abs/2007MNRAS.376.1357L}
}

@ARTICLE{Meyer2004,
       author = {{Meyer}, M.~J. and {Zwaan}, M.~A. and {Webster}, R.~L. and {Staveley-Smith}, L. and {Ryan-Weber}, E. and {Drinkwater}, M.~J. and {Barnes}, D.~G. and {Howlett}, M. and {Kilborn}, V.~A. and {Stevens}, J. and {Waugh}, M. and {Pierce}, M.~J. and {Bhathal}, R. and {de Blok}, W.~J.~G. and {Disney}, M.~J. and {Ekers}, R.~D. and {Freeman}, K.~C. and {Garcia}, D.~A. and {Gibson}, B.~K. and {Harnett}, J. and {Henning}, P.~A. and {Jerjen}, H. and {Kesteven}, M.~J. and {Knezek}, P.~M. and {Koribalski}, B.~S. and {Mader}, S. and {Marquarding}, M. and {Minchin}, R.~F. and {O'Brien}, J. and {Oosterloo}, T. and {Price}, R.~M. and {Putman}, M.~E. and {Ryder}, S.~D. and {Sadler}, E.~M. and {Stewart}, I.~M. and {Stootman}, F. and {Wright}, A.~E.},
        title = "{The HIPASS catalogue - I. Data presentation}",
      journal = {\mnras},
         year = 2004,
        month = jun,
       volume = {350},
       number = {4},
        pages = {1195-1209},
          doi = {10.1111/j.1365-2966.2004.07710.x},
archivePrefix = {arXiv},
       eprint = {astro-ph/0406384},
 primaryClass = {astro-ph},
       adsurl = {https://ui.adsabs.harvard.edu/abs/2004MNRAS.350.1195M}
}

@ARTICLE{Wolfe1986,
       author = {{Wolfe}, A.~M. and {Turnshek}, D.~A. and {Smith}, H.~E. and {Cohen}, R.~D.},
        title = "{Damped Lyman-Alpha Absorption by Disk Galaxies with Large Redshifts. I. The Lick Survey}",
      journal = {\apjs},
         year = 1986,
        month = jun,
       volume = {61},
        pages = {249},
          doi = {10.1086/191114},
       adsurl = {https://ui.adsabs.harvard.edu/abs/1986ApJS...61..249W}
}

@ARTICLE{Wolfe2005,
       author = {{Wolfe}, Arthur M. and {Gawiser}, Eric and {Prochaska}, Jason X.},
        title = "{Damped Ly {\ensuremath{\alpha}} Systems}",
      journal = {\araa},
         year = 2005,
        month = sep,
       volume = {43},
       number = {1},
        pages = {861-918},
          doi = {10.1146/annurev.astro.42.053102.133950},
archivePrefix = {arXiv},
       eprint = {astro-ph/0509481},
 primaryClass = {astro-ph},
       adsurl = {https://ui.adsabs.harvard.edu/abs/2005ARA&A..43..861W}
}

@ARTICLE{Noterdaeme2012,
       author = {{Noterdaeme}, P. and {Petitjean}, P. and {Carithers}, W.~C. and {P{\^a}ris}, I. and {Font-Ribera}, A. and {Bailey}, S. and {Aubourg}, E. and {Bizyaev}, D. and {Ebelke}, G. and {Finley}, H. and {Ge}, J. and {Malanushenko}, E. and {Malanushenko}, V. and {Miralda-Escud{\'e}}, J. and {Myers}, A.~D. and {Oravetz}, D. and {Pan}, K. and {Pieri}, M.~M. and {Ross}, N.~P. and {Schneider}, D.~P. and {Simmons}, A. and {York}, D.~G.},
        title = "{Column density distribution and cosmological mass density of neutral gas: Sloan Digital Sky Survey-III Data Release 9}",
      journal = {\aap},
         year = 2012,
        month = nov,
       volume = {547},
          eid = {L1},
        pages = {L1},
          doi = {10.1051/0004-6361/201220259},
archivePrefix = {arXiv},
       eprint = {1210.1213},
 primaryClass = {astro-ph.CO},
       adsurl = {https://ui.adsabs.harvard.edu/abs/2012A&A...547L...1N}
}

@ARTICLE{Neeleman2016,
       author = {{Neeleman}, Marcel and {Prochaska}, J. Xavier and {Ribaudo}, Joseph and {Lehner}, Nicolas and {Howk}, J. Christopher and {Rafelski}, Marc and {Kanekar}, Nissim},
        title = "{The H I Content of the Universe Over the Past 10 GYRS}",
      journal = {\apj},
         year = 2016,
        month = feb,
       volume = {818},
       number = {2},
          eid = {113},
        pages = {113},
          doi = {10.3847/0004-637X/818/2/113},
archivePrefix = {arXiv},
       eprint = {1601.01691},
 primaryClass = {astro-ph.GA},
       adsurl = {https://ui.adsabs.harvard.edu/abs/2016ApJ...818..113N}
}

@ARTICLE{Delhaize2013,
       author = {{Delhaize}, J. and {Meyer}, M.~J. and {Staveley-Smith}, L. and {Boyle}, B.~J.},
        title = "{Detection of H I in distant galaxies using spectral stacking}",
      journal = {\mnras},
         year = 2013,
        month = aug,
       volume = {433},
       number = {2},
        pages = {1398-1410},
          doi = {10.1093/mnras/stt810},
archivePrefix = {arXiv},
       eprint = {1305.1968},
 primaryClass = {astro-ph.CO},
       adsurl = {https://ui.adsabs.harvard.edu/abs/2013MNRAS.433.1398D}
}

@ARTICLE{Drouart:20,
       author = {{Drouart}, Guillaume and {Seymour}, Nick and {Galvin}, Tim J. and {Afonso}, Jose and {Callingham}, Joseph R. and {De Breuck}, Carlos and {Johnston-Hollitt}, Melanie and {Kapi{\'n}ska}, Anna D. and {Lehnert}, Matthew D. and {Vernet}, Jo{\"e}l},
        title = "{The GLEAMing of the first supermassive black holes}",
      journal = {\pasa},
         year = 2020,
        month = jul,
       volume = {37},
          eid = {e026},
        pages = {e026},
          doi = {10.1017/pasa.2020.6},
       adsurl = {https://ui.adsabs.harvard.edu/abs/2020PASA...37...26D}
}

@ARTICLE{Rhee2018,
       author = {{Rhee}, Jonghwan and {Lah}, Philip and {Briggs}, Frank H. and {Chengalur}, Jayaram N. and {Colless}, Matthew and {Willner}, Steven P. and {Ashby}, Matthew L.~N. and {Le F{\`e}vre}, Olivier},
        title = "{Neutral hydrogen (H I) gas content of galaxies at z {\ensuremath{\approx}} 0.32}",
      journal = {\mnras},
         year = 2018,
        month = jan,
       volume = {473},
       number = {2},
        pages = {1879-1894},
          doi = {10.1093/mnras/stx2461},
archivePrefix = {arXiv},
       eprint = {1709.07596},
 primaryClass = {astro-ph.GA},
       adsurl = {https://ui.adsabs.harvard.edu/abs/2018MNRAS.473.1879R}
}

@ARTICLE{Chowdhury2022,
       author = {{Chowdhury}, Aditya and {Kanekar}, Nissim and {Chengalur}, Jayaram N.},
        title = "{Atomic Gas Scaling Relations of Star-forming Galaxies at z {\ensuremath{\approx}} 1}",
      journal = {\apjl},
         year = 2022,
        month = dec,
       volume = {941},
       number = {1},
          eid = {L6},
        pages = {L6},
          doi = {10.3847/2041-8213/ac9d8a},
archivePrefix = {arXiv},
       eprint = {2212.06176},
 primaryClass = {astro-ph.GA},
       adsurl = {https://ui.adsabs.harvard.edu/abs/2022ApJ...941L...6C}
}

@ARTICLE{Chowdhury2020,
       author = {{Chowdhury}, Aditya and {Kanekar}, Nissim and {Chengalur}, Jayaram N. and {Sethi}, Shiv and {Dwarakanath}, K.~S.},
        title = "{H I 21-centimetre emission from an ensemble of galaxies at an average redshift of one}",
      journal = {\nat},
         year = 2020,
        month = oct,
       volume = {586},
       number = {7829},
        pages = {369-372},
          doi = {10.1038/s41586-020-2794-7},
archivePrefix = {arXiv},
       eprint = {2010.06617},
 primaryClass = {astro-ph.GA},
       adsurl = {https://ui.adsabs.harvard.edu/abs/2020Natur.586..369C}
}

@ARTICLE{Bera2023,
       author = {{Bera}, Apurba and {Kanekar}, Nissim and {Chengalur}, Jayaram N. and {Bagla}, Jasjeet S.},
        title = "{Atomic Hydrogen Scaling Relations at z {\ensuremath{\approx}} 0.35}",
      journal = {\apjl},
         year = 2023,
        month = jun,
       volume = {950},
       number = {2},
          eid = {L18},
        pages = {L18},
          doi = {10.3847/2041-8213/acd0b3},
archivePrefix = {arXiv},
       eprint = {2305.01389},
 primaryClass = {astro-ph.CO},
       adsurl = {https://ui.adsabs.harvard.edu/abs/2023ApJ...950L..18B}
}

@ARTICLE{Kanekar2014b,
       author = {{Kanekar}, N. and {Prochaska}, J.~X. and {Smette}, A. and {Ellison}, S.~L. and {Ryan-Weber}, E.~V. and {Momjian}, E. and {Briggs}, F.~H. and {Lane}, W.~M. and {Chengalur}, J.~N. and {Delafosse}, T. and {Grave}, J. and {Jacobsen}, D. and {de Bruyn}, A.~G.},
        title = "{The spin temperature of high-redshift damped Lyman {\ensuremath{\alpha}} systems}",
      journal = {\mnras},
         year = 2014,
        month = mar,
       volume = {438},
       number = {3},
        pages = {2131-2166},
          doi = {10.1093/mnras/stt2338},
archivePrefix = {arXiv},
       eprint = {1312.3640},
 primaryClass = {astro-ph.CO},
       adsurl = {https://ui.adsabs.harvard.edu/abs/2014MNRAS.438.2131K}
}

@ARTICLE{Braun2012,
       author = {{Braun}, Robert},
        title = "{Cosmological Evolution of Atomic Gas and Implications for 21 cm H I Absorption}",
      journal = {\apj},
         year = 2012,
        month = apr,
       volume = {749},
       number = {1},
          eid = {87},
        pages = {87},
          doi = {10.1088/0004-637X/749/1/87},
archivePrefix = {arXiv},
       eprint = {1202.1840},
 primaryClass = {astro-ph.CO},
       adsurl = {https://ui.adsabs.harvard.edu/abs/2012ApJ...749...87B}
}

@ARTICLE{Field1958,
       author = {{Field}, George B.},
        title = "{Excitation of the Hydrogen 21-CM Line}",
      journal = {Proceedings of the IRE},
         year = 1958,
        month = jan,
       volume = {46},
        pages = {240-250},
          doi = {10.1109/JRPROC.1958.286741},
       adsurl = {https://ui.adsabs.harvard.edu/abs/1958PIRE...46..240F}
}

@ARTICLE{Liszt2001,
       author = {{Liszt}, H.},
        title = "{The spin temperature of warm interstellar H I}",
      journal = {\aap},
         year = 2001,
        month = may,
       volume = {371},
        pages = {698-707},
          doi = {10.1051/0004-6361:20010395},
archivePrefix = {arXiv},
       eprint = {astro-ph/0103246},
 primaryClass = {astro-ph},
       adsurl = {https://ui.adsabs.harvard.edu/abs/2001A&A...371..698L}
}

@ARTICLE{Murthy2024,
       author = {{Murthy}, Suma and {Morganti}, Raffaella and {Oosterloo}, Tom and {Schulz}, Robert and {Paragi}, Zsolt},
        title = "{Turbulent circumnuclear disc and cold gas outflow in the newborn radio source 4C 31.04}",
      journal = {\aap},
         year = 2024,
        month = aug,
       volume = {688},
          eid = {A84},
        pages = {A84},
          doi = {10.1051/0004-6361/202450233},
archivePrefix = {arXiv},
       eprint = {2405.17389},
 primaryClass = {astro-ph.GA},
       adsurl = {https://ui.adsabs.harvard.edu/abs/2024A&A...688A..84M}
}

@ARTICLE{Schulz2018,
       author = {{Schulz}, R. and {Morganti}, R. and {Nyland}, K. and {Paragi}, Z. and {Mahony}, E.~K. and {Oosterloo}, T.},
        title = "{Mapping the neutral atomic hydrogen gas outflow in the restarted radio galaxy 3C 236}",
      journal = {\aap},
         year = 2018,
        month = sep,
       volume = {617},
          eid = {A38},
        pages = {A38},
          doi = {10.1051/0004-6361/201833108},
archivePrefix = {arXiv},
       eprint = {1806.06653},
 primaryClass = {astro-ph.GA},
       adsurl = {https://ui.adsabs.harvard.edu/abs/2018A&A...617A..38S}
}

@ARTICLE{Murthy2021,
       author = {{Murthy}, Suma and {Morganti}, Raffaella and {Oosterloo}, Tom and {Maccagni}, Filippo M.},
        title = "{The H I absorption zoo: JVLA extension to z {\ensuremath{\sim}} 0.4}",
      journal = {\aap},
         year = 2021,
        month = oct,
       volume = {654},
          eid = {A94},
        pages = {A94},
          doi = {10.1051/0004-6361/202141566},
archivePrefix = {arXiv},
       eprint = {2108.08122},
 primaryClass = {astro-ph.GA},
       adsurl = {https://ui.adsabs.harvard.edu/abs/2021A&A...654A..94M}
}

@ARTICLE{Morganti2013,
       author = {{Morganti}, Raffaella and {Fogasy}, Judit and {Paragi}, Zsolt and {Oosterloo}, Tom and {Orienti}, Monica},
        title = "{Radio Jets Clearing the Way Through a Galaxy: Watching Feedback in Action}",
      journal = {Science},
         year = 2013,
        month = sep,
       volume = {341},
       number = {6150},
        pages = {1082-1085},
          doi = {10.1126/science.1240436},
archivePrefix = {arXiv},
       eprint = {1309.1240},
 primaryClass = {astro-ph.CO},
       adsurl = {https://ui.adsabs.harvard.edu/abs/2013Sci...341.1082M}
}

@ARTICLE{Aditya2019,
       author = {{Aditya}, J.~N.~H.~S.},
        title = "{uGMRT detections of H I 21-cm absorption associated with intermediate redshift galaxies}",
      journal = {\mnras},
         year = 2019,
        month = feb,
       volume = {482},
       number = {4},
        pages = {5597-5605},
          doi = {10.1093/mnras/sty3062},
archivePrefix = {arXiv},
       eprint = {1811.03048},
 primaryClass = {astro-ph.GA},
       adsurl = {https://ui.adsabs.harvard.edu/abs/2019MNRAS.482.5597A}
}

@Article{vanBreugel1999,
  author = 	 {W.~van~Breugel and C.~{De~Breuck} and S.~A.~Stanford and D.~Stern and H.~R\"ottgering and G.~Miley},
  title =        "A Radio Galaxy at $z = 5.19$",
  journal = 	 {\apjl},
  year = 	 1999,
  volume =	 518,
  pages =	 "L61"
}

@ARTICLE{Capettie2024,
       author = {{Capetti}, Alessandro and {Balmaverde}, Barbara},
        title = "{The radio properties of z > 3.5 quasars: Are most high-redshift radio-loud active galactic nuclei obscured?}",
      journal = {\aap},
         year = 2024,
        month = sep,
       volume = {689},
          eid = {A174},
        pages = {A174},
          doi = {10.1051/0004-6361/202449676},
archivePrefix = {arXiv},
       eprint = {2407.01288},
 primaryClass = {astro-ph.GA},
       adsurl = {https://ui.adsabs.harvard.edu/abs/2024A&A...689A.174C}
}

@ARTICLE{Vermeulen2003,
       author = {{Vermeulen}, R.~C. and {Pihlstr{\"o}m}, Y.~M. and {Tschager}, W. and {de Vries}, W.~H. and {Conway}, J.~E. and {Barthel}, P.~D. and {Baum}, S.~A. and {Braun}, R. and {Bremer}, M.~N. and {Miley}, G.~K. and {O'Dea}, C.~P. and {R{\"o}ttgering}, H.~J.~A. and {Schilizzi}, R.~T. and {Snellen}, I.~A.~G. and {Taylor}, G.~B.},
        title = "{Observations of H I absorbing gas in compact radio sources at cosmological redshifts}",
      journal = {\aap},
         year = 2003,
        month = jun,
       volume = {404},
        pages = {861-870},
          doi = {10.1051/0004-6361:20030468},
archivePrefix = {arXiv},
       eprint = {astro-ph/0304291},
 primaryClass = {astro-ph},
       adsurl = {https://ui.adsabs.harvard.edu/abs/2003A&A...404..861V}
}

@ARTICLE{Gupta2006,
       author = {{Gupta}, Neeraj and {Saikia}, D.~J.},
        title = "{Unification scheme and the distribution of neutral gas in compact radio sources}",
      journal = {\mnras},
         year = 2006,
        month = aug,
       volume = {370},
       number = {2},
        pages = {738-742},
          doi = {10.1111/j.1365-2966.2006.10498.x},
archivePrefix = {arXiv},
       eprint = {astro-ph/0605399},
 primaryClass = {astro-ph},
       adsurl = {https://ui.adsabs.harvard.edu/abs/2006MNRAS.370..738G}
}

@ARTICLE{Kanekar2004,
       author = {{Kanekar}, N. and {Briggs}, F.~H.},
        title = "{21-cm absorption studies with the Square Kilometer Array}",
      journal = {\nar},
         year = 2004,
        month = dec,
       volume = {48},
       number = {11-12},
        pages = {1259-1270},
          doi = {10.1016/j.newar.2004.09.030},
archivePrefix = {arXiv},
       eprint = {astro-ph/0409169},
 primaryClass = {astro-ph},
       adsurl = {https://ui.adsabs.harvard.edu/abs/2004NewAR..48.1259K}
}

@ARTICLE{Morganti2015,
       author = {{Morganti}, Raffaella and {Sadler}, Elaine M. and {Curran}, Stephen J.},
        title = "{Cool Outflows and HI Absorption}",
      journal = {AASKA14},
         year = 2015,
        month = apr,
       volume = {AASKA14},
        pages = {134},
          doi = {10.22323/1.215.0134},
       adsurl = {https://ui.adsabs.harvard.edu/abs/2015aska.confE.134M}
}

@ARTICLE{Prochaska2009,
       author = {{Prochaska}, J. Xavier and {Wolfe}, Arthur M.},
        title = "{On the (Non)Evolution of H I Gas in Galaxies Over Cosmic Time}",
      journal = {\apj},
         year = 2009,
        month = may,
       volume = {696},
       number = {2},
        pages = {1543-1547},
          doi = {10.1088/0004-637X/696/2/1543},
archivePrefix = {arXiv},
       eprint = {0811.2003},
 primaryClass = {astro-ph},
       adsurl = {https://ui.adsabs.harvard.edu/abs/2009ApJ...696.1543P}
}

@ARTICLE{Dutta2019,
       author = {{Dutta}, R. and {Srianand}, R. and {Gupta}, N.},
        title = "{Prevalence of neutral gas in centres of merging galaxies-II: nuclear H I and multiwavelength properties}",
      journal = {\mnras},
         year = 2019,
        month = oct,
       volume = {489},
       number = {1},
        pages = {1099-1109},
          doi = {10.1093/mnras/stz2178},
archivePrefix = {arXiv},
       eprint = {1908.02291},
 primaryClass = {astro-ph.GA},
       adsurl = {https://ui.adsabs.harvard.edu/abs/2019MNRAS.489.1099D}
}

@ARTICLE{Chandola2013,
       author = {{Chandola}, Yogesh and {Gupta}, Neeraj and {Saikia}, D.~J.},
        title = "{Associated 21-cm absorption towards the cores of radio galaxies}",
      journal = {\mnras},
         year = 2013,
        month = mar,
       volume = {429},
       number = {3},
        pages = {2380-2391},
          doi = {10.1093/mnras/sts499},
archivePrefix = {arXiv},
       eprint = {1211.6852},
 primaryClass = {astro-ph.CO},
       adsurl = {https://ui.adsabs.harvard.edu/abs/2013MNRAS.429.2380C}
}

@ARTICLE{Ellison2008,
       author = {{Ellison}, Sara L. and {York}, Brian A. and {Pettini}, Max and {Kanekar}, Nissim},
        title = "{A search for damped Lyman {\ensuremath{\alpha}} systems towards radio-loud quasars I: the optical survey*}",
      journal = {\mnras},
         year = 2008,
        month = aug,
       volume = {388},
       number = {3},
        pages = {1349-1360},
          doi = {10.1111/j.1365-2966.2008.13482.x},
archivePrefix = {arXiv},
       eprint = {0805.2940},
 primaryClass = {astro-ph},
       adsurl = {https://ui.adsabs.harvard.edu/abs/2008MNRAS.388.1349E}
}

@ARTICLE{Pontzen2009,
       author = {{Pontzen}, Andrew and {Pettini}, Max},
        title = "{Dust biasing of damped Lyman alpha systems: a Bayesian analysis}",
      journal = {\mnras},
         year = 2009,
        month = feb,
       volume = {393},
       number = {2},
        pages = {557-568},
          doi = {10.1111/j.1365-2966.2008.14193.x},
archivePrefix = {arXiv},
       eprint = {0810.3236},
 primaryClass = {astro-ph},
       adsurl = {https://ui.adsabs.harvard.edu/abs/2009MNRAS.393..557P}
}

@ARTICLE{Krogager2019,
       author = {{Krogager}, Jens-Kristian and {Fynbo}, Johan P.~U. and {M{\o}ller}, Palle and {Noterdaeme}, Pasquier and {Heintz}, Kasper E. and {Pettini}, Max},
        title = "{The effect of dust bias on the census of neutral gas and metals in the high-redshift Universe due to SDSS-II quasar colour selection}",
      journal = {\mnras},
         year = 2019,
        month = jul,
       volume = {486},
       number = {3},
        pages = {4377-4397},
          doi = {10.1093/mnras/stz1120},
archivePrefix = {arXiv},
       eprint = {1904.06966},
 primaryClass = {astro-ph.GA},
       adsurl = {https://ui.adsabs.harvard.edu/abs/2019MNRAS.486.4377K}
}

@ARTICLE{Krogager2016,
       author = {{Krogager}, J.-K. and {Fynbo}, J.~P.~U. and {Heintz}, K.~E. and {Geier}, S. and {Ledoux}, C. and {M{\o}ller}, P. and {Noterdaeme}, P. and {Venemans}, B.~P. and {Vestergaard}, M.},
        title = "{The Extended High A(V) Quasar Survey: Searching for Dusty Absorbers toward Mid-infrared-selected Quasars}",
      journal = {\apj},
         year = 2016,
        month = nov,
       volume = {832},
       number = {1},
          eid = {49},
        pages = {49},
          doi = {10.3847/0004-637X/832/1/49},
archivePrefix = {arXiv},
       eprint = {1608.08404},
 primaryClass = {astro-ph.GA},
       adsurl = {https://ui.adsabs.harvard.edu/abs/2016ApJ...832...49K}
}

@ARTICLE{Dutta2017qgp,
       author = {{Dutta}, R. and {Srianand}, R. and {Gupta}, N. and {Momjian}, E. and {Noterdaeme}, P. and {Petitjean}, P. and {Rahmani}, H.},
        title = "{H I 21-cm absorption survey of quasar-galaxy pairs: distribution of cold gas around z < 0.4 galaxies}",
      journal = {\mnras},
         year = 2017,
        month = feb,
       volume = {465},
       number = {1},
        pages = {588-618},
          doi = {10.1093/mnras/stw2689},
archivePrefix = {arXiv},
       eprint = {1610.05316},
 primaryClass = {astro-ph.GA},
       adsurl = {https://ui.adsabs.harvard.edu/abs/2017MNRAS.465..588D}
}

@ARTICLE{Uzan2011,
       author = {{Uzan}, Jean-Philippe},
        title = "{Varying Constants, Gravitation and Cosmology}",
      journal = {Living Reviews in Relativity},
         year = 2011,
        month = mar,
       volume = {14},
       number = {1},
          eid = {2},
        pages = {2},
          doi = {10.12942/lrr-2011-2},
archivePrefix = {arXiv},
       eprint = {1009.5514},
 primaryClass = {astro-ph.CO},
       adsurl = {https://ui.adsabs.harvard.edu/abs/2011LRR....14....2U}
}

@ARTICLE{Dutta2016,
       author = {{Dutta}, R. and {Gupta}, N. and {Srianand}, R. and {O'Meara}, J.~M.},
        title = "{Mapping kiloparsec-scale structures in the extended H I disc of the galaxy UGC 000439 by H I 21-cm absorption}",
      journal = {\mnras},
         year = 2016,
        month = mar,
       volume = {456},
       number = {4},
        pages = {4209-4218},
          doi = {10.1093/mnras/stv2980},
archivePrefix = {arXiv},
       eprint = {1601.00971},
 primaryClass = {astro-ph.GA},
       adsurl = {https://ui.adsabs.harvard.edu/abs/2016MNRAS.456.4209D}
}

@ARTICLE{Keeney2005,
       author = {{Keeney}, Brian A. and {Momjian}, Emmanuel and {Stocke}, John T. and {Carilli}, Chris L. and {Tumlinson}, Jason},
        title = "{Absorption-Line Study of Halo Gas in NGC 3067 toward the Background Quasar 3C 232}",
      journal = {\apj},
         year = 2005,
        month = mar,
       volume = {622},
       number = {1},
        pages = {267-278},
          doi = {10.1086/427899},
archivePrefix = {arXiv},
       eprint = {astro-ph/0409448},
 primaryClass = {astro-ph},
       adsurl = {https://ui.adsabs.harvard.edu/abs/2005ApJ...622..267K}
}

@ARTICLE{Srianand2010,
       author = {{Srianand}, R. and {Gupta}, N. and {Petitjean}, P. and {Noterdaeme}, P. and {Ledoux}, C.},
        title = "{Detection of 21-cm, H$_{2}$ and deuterium absorption at z > 3 along the line of sight to J1337+3152}",
      journal = {\mnras},
         year = 2010,
        month = jul,
       volume = {405},
       number = {3},
        pages = {1888-1900},
          doi = {10.1111/j.1365-2966.2010.16574.x},
archivePrefix = {arXiv},
       eprint = {1002.4620},
 primaryClass = {astro-ph.CO},
       adsurl = {https://ui.adsabs.harvard.edu/abs/2010MNRAS.405.1888S}
}

@ARTICLE{Kanekar2001,
       author = {{Kanekar}, Nissim and {Chengalur}, Jayaram N.},
        title = "{Variable 21-cm absorption at z=0.3127}",
      journal = {\mnras},
         year = 2001,
        month = aug,
       volume = {325},
       number = {2},
        pages = {631-635},
          doi = {10.1046/j.1365-8711.2001.04424.x},
archivePrefix = {arXiv},
       eprint = {astro-ph/0102424},
 primaryClass = {astro-ph},
       adsurl = {https://ui.adsabs.harvard.edu/abs/2001MNRAS.325..631K}
}

@ARTICLE{Kanekar2008,
       author = {{Kanekar}, Nissim and {Chengalur}, Jayaram N.},
        title = "{Outflowing atomic and molecular gas at z \raisebox{-0.5ex}\textasciitilde 0.67 towards 1504 + 377}",
      journal = {\mnras},
         year = 2008,
        month = feb,
       volume = {384},
       number = {1},
        pages = {L6-L10},
          doi = {10.1111/j.1745-3933.2007.00410.x},
archivePrefix = {arXiv},
       eprint = {0710.3224},
 primaryClass = {astro-ph},
       adsurl = {https://ui.adsabs.harvard.edu/abs/2008MNRAS.384L...6K}
}

@ARTICLE{Combes2023,
       author = {{Combes}, F. and {Gupta}, N. and {Muller}, S. and {Balashev}, S. and {Deka}, P.~P. and {Emig}, K.~L. and {Kl{\"o}ckner}, H.-R. and {Klutse}, D. and {Knowles}, K. and {Mohapatra}, A. and {Momjian}, E. and {Noterdaeme}, P. and {Petitjean}, P. and {Salas}, P. and {Srianand}, R. and {Wagenveld}, J.~D.},
        title = "{PKS 1413+135: OH and H I at z = 0.247 with MeerKAT}",
      journal = {\aap},
         year = 2023,
        month = mar,
       volume = {671},
          eid = {A43},
        pages = {A43},
          doi = {10.1051/0004-6361/202245482},
archivePrefix = {arXiv},
       eprint = {2211.09355},
 primaryClass = {astro-ph.GA},
       adsurl = {https://ui.adsabs.harvard.edu/abs/2023A&A...671A..43C}
}

\end{document}